\documentclass[pdflatex,sn-chicago]{sn-jnl}

\usepackage{natbib}

\jyear{2022}%

\theoremstyle{thmstyleone}%
\theoremstyle{thmstyletwo}%

\theoremstyle{thmstylethree}%

\newcommand{\vecY}{{\mathbf{Y}}}
\newcommand{\bphi}{\mbox{\boldmath${\phi}$}}
\newcommand{\matr}[1]{\mathbf{#1}}
\newcommand{\vect}[1]{\mathbf{#1}}
\newcommand{\I}{ {\rm{I}}}
\newcommand{\addeq}{ \overset{\textrm{\tiny{+}}}{\approx}}
\newcommand{\EZi}{\E_{q^{*}(Z_i)}\big(\I(Z_i=k)\big)}
\newcommand{\vecX}{{\mathbf{X}}}

\def\negr#1{\mbox{\boldmath$#1$}}
\def\E{{\mathbb E}}

\usepackage{graphicx} 
\usepackage{caption}
\usepackage{subcaption}
\usepackage{multirow}

\begin{document}

\title[Clustering Functional Data via Variational Inference]{Clustering Functional Data via Variational Inference}


\author*[1]{\fnm{Chengqian} \sur{Xian}}\email{cxian3@uwo.ca}

\author[1]{\fnm{Camila} \sur{P. E. de Souza}}\email{camila.souza@uwo.ca}

\author[2]{\fnm{John} \sur{Jewell}}\email{jjewell6@uwo.ca}

\author[3]{\fnm{Ronaldo} \sur{Dias}}\email{dias@unicamp.br}

\affil*[1]{\orgdiv{Department of Statistical and Actuarial Sciences}, \orgname{University of Western Ontario}, \orgaddress{\street{1151 Richmond Street}, \city{London}, \postcode{N6A 5B7}, \state{Ontario}, \country{Canada}}}

\affil[2]{\orgdiv{Department of Computer Science}, \orgname{University of Western Ontario}, \orgaddress{\street{1151 Richmond Street}, \city{London}, \postcode{N6A 5B7}, \state{Ontario}, \country{Canada}}}

\affil[3]{\orgdiv{Department of Statistics}, \orgname{University of Campinas}, \orgaddress{\street{Barão Geraldo}, \city{Campinas}, \postcode{13083-970}, \state{São Paulo}, \country{Brazil}}}


\abstract{Functional data analysis deals with data recorded densely over time (or any other continuum) with one or more observed curves per subject. Conceptually, functional data are continuously defined, but in practice, they are usually observed at discrete points. Among different kinds of functional data analyses, clustering analysis aims to determine underlying groups of curves in the dataset when there is no information on the group membership of each individual curve. In this work, we propose a new model-based approach for clustering and smoothing functional data simultaneously via variational inference. We derive coordinate ascent mean-field variational Bayes algorithms to approximate the posterior distribution of our model parameters by finding the variational distribution with the smallest Kullback-Leibler divergence to the posterior. The performance of our proposed method is evaluated using simulated data and publicly available datasets.}

\keywords{Functional data analysis, Model-based clustering, Bayesian variational inference}



\maketitle

\section{Introduction}\label{sec1}

Functional data analysis (FDA), term first coined by \citet{Ramsay_1991}, deals with the analysis of data that are defined on some continuum such as time. Theoretically, data are in the form of functions, but in practice they are observed as a series of discrete points representing an underlying curve. \citet{ramsay_2005} establish a foundation for FDA on topics including smoothing functional data, functional principal components analysis and functional linear models. \citet{ramsay_hooker_graves_2009} provide a guide for analyzing functional data in R and Matlab using publicly available datasets. \citet{wang_2016} present a comprehensive review of FDA, in which clustering and classification methods for functional data are also discussed. Functional data analysis has been applied to various research areas such as energy consumption \citep{Lenzi_2017, de_2017, Franco_2021}, rainfall data visualization \citep{Hael_2020}, income distribution \citep{Hu_2021}, spectroscopy \citep{Dias_2015, Yang_2021, Frizzarin_2021}, and Covid-19 pandemic \citep{Boschi_2021, sousa2022bayesian, Collazos2023}, to mention a few. 

Cluster analysis of functional data aims to determine underlying groups in a set of observed curves when there is no information on the group label of each curve. As described in \cite{Jacques_2014}, there are three main types of methods used for functional data clustering: dimension reduction-based (or filtering) methods, distance-based methods, and model-based methods. Functional data generally belongs to the infinite-dimensional space, making those clustering methods for finite-dimensional data ineffective. Therefore, dimension reduction-based methods have been proposed to solve this problem. Before clustering, a dimension reduction step (also called \textit{filtering} in \citealp{James_2003}) is carried out by the techniques including spline basis function expansion \citep{Tarpey_2003} and functional principal component analysis \citep{Jones_1992}. Clustering is then performed using the basis expansion coefficients or the principal component scores, resulting in a two-stage clustering procedure. Distance-based methods are the most well-known and popular approaches for clustering functional data since no parametric assumptions are necessary for these algorithms. Nonparametric clustering techniques, including $k$-means clustering \citep{Hartigan_1979} and hierarchical clustering \citep{Ward_1963}, are usually applied using specific distances or dissimilarities between curves \citep{Delaigle_2019, Martino_2019, Li_2020}. It is important to note that distance-based methods are sometimes equivalent to dimension reduction-based methods if, for example, distances are computed using the basis expansion coefficients. Another widely-used approach is model-based clustering, where functional data are assumed to arise from a mixture of underlying probability distributions. For example, in Bayesian hierarchical clustering, a common methodology is to assume that the set of coefficients in the basis expansion representing functional data follow a mixture of Gaussian distributions \citep{wang_2016}. Inference in model-based clustering for functional data is generally conducted via the Expectation-Maximization algorithm \citep{Same_2011, Jacques_2013, Giacofci_2013, Chamroukhi_2016} or Markov Chain Monte Carlo (MCMC) sampling techniques \citep{Ray_2006}. 

More recently, \citet{Zambom_2019} proposed a method for clustering functional data via hypothesis testing where they replaced the classical distance computation step in the $k$-means algorithm with hypothesis tests to decide to which cluster a curve belongs. However, these authors first smooth the curves and then apply the test-based clustering procedure, which falls in the scope of two-stage clustering.

In this paper, we propose a new model-based approach for clustering functional data via Bayesian variational inference. We model the raw data obtaining clustering assignments and cluster-specific smooth mean curves simultaneously. To our knowledge, this is the first attempt to use variational inference in the context of functional data clustering. Our proposed method is implemented in R and codes are available at \url{https://github.com/chengqianxian/funclustVI}.

The remainder of the paper is organized as follows. Section \ref{sec.method} presents an overview of variational inference, our two model settings and proposed algorithms. In Section \ref{sec.sim}, we conduct simulation studies to assess the performance of our methods under various scenarios. In Section \ref{sec.real}, we apply our proposed methodology to real datasets. A conclusion of our study and a discussion on the proposed method are provided in Section \ref{sec.con.dis}.

\section{Methodology}\label{sec.method}

\subsection{Overview of Variational Inference}

Variational inference (VI) is a method from machine learning that approximates the posterior density in a Bayesian model through optimization \citep{Jordan_1999, Wainwright_2008}. \cite{David_2017} provide an interesting review of VI from a statistical perspective, including some guidance on when to use MCMC or VI. For example, one may apply VI to large datasets and scenarios where the interest is to develop probabilistic models. In contrast, one may apply MCMC to small datasets for more precise samples but with a higher computational cost. In Bayesian inference, our goal is to find the posterior density, denoted by $p(\cdot \vert y)$, where $y$ corresponds to the observed data. One can apply Bayes' theorem to find the posterior, but this might not be easy if there are many parameters and non-conjugate prior distributions. Therefore, one can aim to find an approximation to the posterior. To be specific, one wants to find $q^*$ coming from a family of possible densities $Q$ to approximate $p(\cdot \vert y)$, which can be solved in terms of an optimization problem with criterion $f$ as follows:
$$q^* = \underset{q \in Q}{\mathrm{argmin}}\,f(q(\cdot), p(\cdot \vert y)).$$

The criterion $f$ measures the closeness between the possible densities $q$ in the family $Q$ and the exact posterior density $p$. When we consider the Kullback-Leibler (KL) divergence \citep{Kullback_1951} as criterion $f$, i.e.,
\begin{eqnarray}
 q^* = \underset{q \in Q}{\mathrm{argmin}}\,\mbox{KL}(q(\cdot) \Vert p(\cdot \vert y)),   
 \label{eq:KL}
\end{eqnarray}
this optimization-based technique to approximate the posterior density is called Variational Bayes (VB). \cite{Jordan_1999} and \citet{David_2017} show that minimizing the KL divergence is equivalent to maximizing the so-called evidence lower bound (ELBO). Let $\theta$ be a set of latent model variables, the KL divergence is defined as $$\mbox{KL}(q(\cdot) \Vert p(\cdot \vert y)) := \int q(\theta)\log\frac{q(\theta)}{p(\theta \vert y)}d\theta,$$
and it can be shown that $$\int q(\theta)\log\frac{q(\theta)}{p(\theta \vert y)}d\theta=\log p(y)-\int q(\theta)\log\frac{p(\theta, y)}{q(\theta)}d\theta$$
where the last term is the ELBO. Since $\log p(y)$ is a constant with respect to $q(\theta)$, this changes the problem in (\ref{eq:KL}) to
\begin{eqnarray}
q^* = \underset{q \in Q}{\mathrm{argmax}}\,\mbox{ELBO}(q).
\label{eq:maxELBO}
\end{eqnarray}

We, therefore, derive a VB algorithm for clustering functional data. We consider the mean-field variational family in which the latent variables are mutually independent, and a distinct factor governs each of them in the variational density. Finally, we apply the coordinate ascent variational inference algorithm \citep{Bishop_2006} to solve the optimization problem in (\ref{eq:maxELBO}).

\subsection{Assumptions and model settings}
Let $\vecY_i$, $\{i=1,\ldots,N\}$, denote the observed data from $N$ curves, and for each curve $i$ there are $n$ evaluation points, $t_1,..., t_n$, so that $\vecY_i =(Y_i(t_1),\ldots,Y_i(t_n))^T$. Let $Z_i$ be a hidden variable taking values in $\{ 1,\ldots,K\}$ that determines which cluster $\vecY_i$ belongs to. We assume $Z_1,\ldots,Z_N$ are independent and identically distributed with $P(Z_i=k) = \pi_k, \, k=1,...,K$, and $\sum_{k=1}^K \pi_k =1$. 
For each cluster $k$, there is a smooth function $f_k$ evaluated at $t_1,..., t_n$ so that $f_k(\mathbf{t}) = (f_k(t_1),\ldots,f_k(t_n))^T$. Given that $Z_i=k$, we consider two different models for  $\vecY_i$ based on the correlation structure of the errors. In Model 1, described in Section \ref{sec:model:one}, we assume independent errors, and in Model 2, described in Section \ref{sec:model:two}, we add a random intercept to induce a correlation between observations within each curve.

\subsubsection{Model 1}\label{sec:model:one}

Let us assume that
\begin{eqnarray}
\vecY_i\, \vert \,(Z_i = k) = f_k(\mathbf{t}) + \sigma_k\negr{\epsilon_i}
    \label{Model:one}
\end{eqnarray}
with independent errors $\negr{\epsilon_1},..., \negr{\epsilon_N}$ and $\negr{\epsilon_i} \sim MVN (\mathbf{0}, \I_n), i=1,...,N$, where $\I_n$ is an identity matrix of size $n$. 
The functions $f_1,\ldots,f_K$ can be written as a linear combination of $M$ known B-spline basis functions, that is, $f_k(t_j) = \sum_{m=1}^M B_m(t_j)\phi_{km}, \; j=1,...,n$, such that
$f_k(\mathbf{t}) = \matr{B}_{(n\times M)}\bphi_{k(M\times 1)}, k=1,...,K$. Therefore, $$\vecY_i \, \vert \,Z_i =k \sim MVN (\matr{B}\bphi_{k}, \sigma_k^2 \I_n), \; i=1,...,N, \;k=1,...,K.$$

Let $\vect{Z}=(Z_1,\ldots,Z_N)^T$, $\negr{\phi}=\{\bphi_1,\ldots,\bphi_K\}$, $\negr{\pi} = (\pi_1,\ldots,\pi_K)^T $ and $\negr{\tau} = (\tau_1,\ldots,\tau_K)^T $, where $\tau_k = 1/\sigma^2_k$ is the precision parameter. We develop a Bayesian approach to infer $\negr{Z}$, $\bphi$, $\negr{\pi}$ and $\negr{\tau}$  via VB; that is, we derive a procedure that, for given data, approximates the posterior distribution by finding the variational distribution (VD), $ q(\vect{Z},\negr{\pi},\bphi,\negr{\tau})$,
with smallest Kullback-Leibler divergence to the posterior distribution $p(\vect{Z},\negr{\pi},\negr{\phi},\negr{\tau}\vert \vecY )$, which is equivalent to maximizing the evidence lower bound (ELBO) given by
\begin{eqnarray}
    \mbox{ELBO}(q) = \E \big[ \log p(\vecY ,\vect{Z},\negr{\pi},\negr{\phi},\negr{\tau}) \big] - \E \big[ \log q(\vect{Z},\negr{\pi},\bphi,\negr{\tau}) \big].
    \label{Eq:elbo}
\end{eqnarray}

We assume the following prior distributions for parameters in Model 1:

\begin{itemize}
\setlength\itemsep{0.2cm}
\item $\negr{\pi}  \sim \mbox{Dirichlet}(\vect{d}^0)$;
\item $\bphi_k \sim MVN(\vect{m}_k^0,s^0\matr{I})$ with precision $v^0 = 1/s^0$ and $\matr{I}$ an $M \times M$ identity matrix;
\item $\tau_k = 1/\sigma^2_k \sim \mbox{Gamma}(a^0,r^0), \; k=1,...,K$. 
\end{itemize}

\subsubsection{Model 2}\label{sec:model:two}
We extend the model in Section \ref{sec:model:one} by adding a curve-specific random intercept $a_i$ which induces correlation among observations within each curve. The model now becomes:
\begin{eqnarray}
Y_{ij}\, \vert \,(Z_i = k)=a_i+f_k(t_j)+\sigma_k\epsilon_{ij}
    \label{Model:two}
\end{eqnarray}
where $\epsilon_{ij}\sim N(0, 1)$ and $a_i\sim N(0, \sigma_a^2)$ with $a_i$ and $\epsilon_{ij}$ independent for all $i$ and $j$. We can write Model 2 in a vector form as
$$ \vecY_i\, \vert \,(Z_i = k) = a_i\vect{1}_n + f_k(\mathbf{t}) + \sigma_k\negr{\epsilon_i},\, i=1,2,...,N,$$
in which $\vect{1}_n$ is a vector of length $n$ with all elements equal to 1, and further assume that
$\negr{\epsilon_i}\sim MVN(\vect{0}, \I_{n})$ and $a_i\sim N(0, \sigma_a^2)$.

This model can be rewritten as a two-step model:
$$ \vecY_i\, \vert \,(Z_i = k, a_i)\sim MVN (\matr{B}\bphi_{k} + a_i\vect{1}_n, \sigma_k^2 \I_n)$$
and $a_i\sim N(0, \sigma_a^2), i=1,2,...,N$.

Let $\vect{a}=(a_1,\ldots,a_N)^T$ and $\tau_a=1/\sigma^2_a$. As in Model 1, we build a VB algorithm to infer $\negr{Z}$, $\bphi$, $\negr{\pi}$, $\negr{\tau}$, $\vect{a}$ and $\tau_a$. The ELBO under Model 2 is given by
\begin{eqnarray}
\mbox{ELBO}(q) = \E_{q^*} \big[ \log p(\vecY ,\vect{Z},\negr{\pi},\negr{\phi},\negr{\tau}, \vect{a},\tau_a)\big] - \E_{q^*} \big[ \log q(\vect{Z},\negr{\pi},\bphi,\negr{\tau},\vect{a},\tau_a) \big].
    \label{Eq:elbo:ext}
\end{eqnarray}
We assume the following prior distributions for the parameters in Model 2:
\begin{itemize}
\setlength\itemsep{0.2cm}
    \item $\negr{\pi}  \sim \mbox{Dirichlet}(\vect{d}^0)$;
    \item $\bphi_k \sim MVN(\vect{m}_k^0,s^0\matr{I})$ with precision $v^0 = 1/s^0$;
    \item $\tau_k = 1/\sigma^2_k \sim \mbox{Gamma}(b^0,r^0), \; k=1,...,K$;
    \item $\tau_a = 1/\sigma^2_a \sim \mbox{Gamma}(\alpha^0,\beta^0)$;
    \item $a_i \sim N(0, \sigma_a^2)$ with $\tau_a=1/\sigma^2_a$. 
\end{itemize}

\subsection{Steps of the VB algorithm}

This section describes the main steps of the VB algorithm under Model 2 for inferring $\negr{Z}$, $\bphi$, $\negr{\pi}$, $\negr{\tau}$, $\vect{a}$ and $\tau_a$. The proposed VB is summarized in Algorithm \ref{VBalgorithm:ext}. The VB algorithm's main steps and the ELBO calculation for Model 1 can be found in Appendix A.

First, we assume that the variational distribution belongs to the mean-field variational family, where $\negr{Z}$, $\bphi$, $\negr{\pi}$ $\negr{\tau}$, $\vect{a}$ and $\tau_a$ are
mutually independent and each governed by a distinct factor in the variational density, that is:
\begin{eqnarray}
 q(\vect{Z},\negr{\pi},\negr{\phi},\negr{\tau}, \vect{a}, \tau_a) 
&=& \prod_{i=1}^N q(Z_i) \times \prod_{k=1}^K q(\bphi_k) \times \prod_{k=1}^K 
 q(\tau_k)  \nonumber \\ && 
\times q(\negr{\pi}) \times  \prod_{i=1}^N q(a_i) \times q(\tau_a). \label{eq:factorization:ext}
\end{eqnarray}
We then derive a coordinate ascent algorithm to obtain the  VD \citep{Jordan_1999,David_2017}. That is, we derive an update equation for each term in the factorization (\ref{eq:factorization:ext}) by calculating the expectation of $\log p(\vect{Y},\vect{Z},\negr{\pi},\negr{\phi},\negr{\tau}, \vect{a}, \tau_a)$ (the joint distribution of the observed data $\vect{Y}$, hidden variables $\vect{Z}$ and parameters $\negr{\pi},\negr{\phi},\negr{\tau}, \vect{a}, \tau_a$, which is also called complete-data log-likelihood) over the VD of all random variables except the one of interest, where
\begin{eqnarray}
 \log p(\vect{Y},\vect{Z},\negr{\pi},\negr{\phi},\negr{\tau}, \vect{a}, \tau_a) 
&=& \log p(\vecY \vert \vect{Z},\negr{\phi},\negr{\tau}, \vect{a}) \; + \; \log p(\vect{Z} \vert   \negr{\pi}) +  \nonumber \\ &&   \log p(\negr{\phi}) + \log p(\negr{\tau}) \;+\; \log p(\negr{\pi})  \; + \nonumber \\ && \log p(\vect{a} \vert \tau_a)  \; + \; \log p(\tau_a).
\label{eq:complete:ext}
\end{eqnarray}
So, for example, the optimal update equation for $q(\negr{\pi})$, $q^*(\negr{\pi})$, is given by calculating 
\begin{eqnarray}
        \log q^*(\negr{\pi}) = \E_{-\negr{\pi}} \big( \log p(\vect{Y},\vect{Z},\negr{\pi},\negr{\phi},\negr{\tau}, \vect{a}, \tau_a) \big) \;+\; \mbox{constant} , \nonumber 
\end{eqnarray}
where $-\negr{\pi}$ indicates that the expectation is taken with respect to the VD of all other latent variables but $\negr{\pi}$, i.e., $\vect{Z},\negr{\phi}$, $\negr{\tau}$, $\vect{a}$ and $\tau_a$. In what follows we derive the update equation for each component in our model. 
For convenience, we use $\addeq$ to denote equality up to a constant additive factor.

\subsubsection{VB update equations}\label{sec:updatesVB}

\textit{i) Update equation for $q(\negr{\pi}$})

\vspace{0.2cm}
Since only the second term, $\log p(\vect{Z} \vert \negr{\pi})$, and the fifth term, $\log p(\negr{\pi})$, in (\ref{eq:complete:ext}) depend on $\negr{\pi}$, the update equation $q^*(\negr{\pi})$ can be derived as follows.
\begin{eqnarray}
{\log q^*(\negr{\pi})}   &\addeq& \E_{-\negr{\pi}} \big( \log p(\vect{Y},\vect{Z},\negr{\pi},\negr{\phi},\negr{\tau}, \vect{a}, \tau_a) \big) \nonumber \\
 &\addeq & \E_{-\negr{\pi}} \big( \log p(\vect{Z}\vert \negr{\pi}) \big)\; +\; \E_{-\negr{\pi}} \big(\log p(\negr{\pi})\big)  \nonumber \\
 &= & \E_{-\negr{\pi}} \Big[ \sum_{i=1}^N \sum_{k=1}^{K} \I(Z_i =k) \log \pi_k \Big]+  \log p(\negr{\pi})  \nonumber \\
 &\addeq & \sum_{k=1}^{K} \log \pi_k \Big[\sum_{i=1}^N \EZi \Big] + \sum_{k=1}^{K} [d^0_k -1 ]\log \pi_k\nonumber \\
 &=& \sum_{k=1}^{K} \log \pi_k \Big[ \Big( \sum_{i=1}^N \EZi + d^0_k   \Big) -1  \Big] .\nonumber 
\end{eqnarray}
Therefore, $q^*(\negr{\pi})$ is a Dirichlet distribution with parameters $\vect{d}^*=(d_1^*,\ldots,d_K^*)$, where  
\begin{eqnarray}d^{*}_{k}= d^0_k + \sum_{i=1}^N \EZi. 
\label{Eq:qstarPi:ext}
\end{eqnarray}
\vspace{0.4cm}

\noindent\textit{ii) Update equation for $q(Z_i)$} 
\begin{eqnarray}
\log q^*(Z_i) &\addeq& \E_{-Z_i} \big(\log p(\vect{Y},\vect{Z},\negr{\pi},\negr{\phi},\negr{\tau}, \vect{a}, \tau_a) \big) \nonumber \\
&\addeq&  \E_{-Z_i} \big( \log p(\vecY \vert \vect{Z},\negr{\phi},\negr{\tau}, \vect{a})\big)\; +\; \E_{-Z_i} \big(\log p(\vect{Z} \vert \negr{\pi})\big)
\label{eq:logqZi:ext}
\end{eqnarray}
Note that we can write $\log p(\vecY \vert \vect{Z},\negr{\phi},\negr{\tau}, \vect{a})$ and $\log p(\vect{Z} \vert \negr{\pi})$ into two parts, one that depends on $Z_i$ and one that does not, that is: 
\begin{eqnarray}
\log p(\vect{Y},\vect{Z},\negr{\phi},\negr{\tau}, \vect{a})&=& \sum_{k=1}^K \I(Z_i = k) \log p(\vecY_i \vert Z_i =k, \bphi_k,\tau_k, a_i) \nonumber \\
&& +\, \sum_{l:l\neq i} \sum_{k=1}^K \I(Z_l=k) \log p(\vecY_l \vert Z_l=k,\negr{\phi}_k,\tau_k, a_l)  \nonumber
\end{eqnarray}
\begin{eqnarray}
\log p(\vect{Z} \vert \negr{\pi}) &=& \sum_{k=1}^K \I (Z_i=k) \log \pi_k + \sum_{l:l\neq i} \sum_{k=1}^K \I (Z_l = k) \log \pi_k . \nonumber
\end{eqnarray}
Now when taking the expectation in (\ref{eq:logqZi:ext}), the parts that do not depend on $Z_i$ in $\log p(\vecY \vert \vect{Z},\negr{\phi},\negr{\tau},\vect{a})$ and $\log p(\vect{Z} \vert \negr{\pi})$ will be added as a constant in the expectation. So, we obtain
\begin{eqnarray}
\log q^*(Z_i) &\addeq& \sum_{k=1}^K \I(Z_i =k)\Big \{\frac{n}{2}\E_{q^*(\tau_k)}(\log \tau_k)   
 \nonumber \\
&& -\frac{1}{2}\E_{q^*(\tau_k)}(\tau_k)\E_{q^*(\negr{\phi}_k)\cdot q^*(a_i)}\big[ (\vecY_i - \matr{B}\negr{\phi}_k-a_i\vect{1}_n)^T(\vecY_i - \matr{B}\negr{\phi}_k-a_i\vect{1}_n) \big]   \nonumber \\
&& \, + \,\E_{q^*(\negr{\pi})} (\log \pi_k)
\Big\}  \nonumber     
\end{eqnarray}
Therefore, $q^*(Z_i)$ is a categorical distribution with parameters 
\begin{eqnarray}
    p^*_{ik} = \frac{e^{\alpha_{ik}}}{\sum_{k=1}^Ke^{\alpha_{ik}}},
    \label{eq:pikstar:ext}
\end{eqnarray} 
\noindent where 
\begin{eqnarray}
\alpha_{ik} &=& \frac{n}{2}\E_{q^*(\tau_k)}(\log \tau_k)  \nonumber \\
&& -\frac{1}{2}\E_{q^*(\tau_k)}(\tau_k)\E_{q^*(\negr{\phi}_k) q^*(a_i)}\big[ (\vecY_i - \matr{B}\negr{\phi}_k-a_i\vect{1}_n)^T(\vecY_i - \matr{B}\negr{\phi}_k-a_i\vect{1}_n) \big]  \nonumber \\
&& +\, \E_{q^*(\negr{\pi})} (\log \pi_k). \nonumber
\end{eqnarray}

Note that all expectations involved in the VB update equations are calculated in Section \ref{sec:expectations}. 

\vspace{0.2cm}
\textit{iii) Update equation for $q(\negr{\phi}_k)$} 

Only the first term, $\log p(\vecY\vert\vect{Z},\negr{\phi},\negr{\tau}, \vect{a})$, and the third term, $\log p(\negr{\phi})$, in (\ref{eq:complete:ext}) depend on $\bphi_k$. In addition, similarly to the previous case for $q^*(Z_i)$, we can write $\log p(\vecY\vert\vect{Z},\negr{\phi},\negr{\tau}, \vect{a})$ and $\log p(\negr{\phi})$ in two parts, one that depends on $\bphi_k$ and the other that does not. Therefore, we obtain
\begin{eqnarray}
\log q^*(\bphi_k)  &\addeq& \E_{-\bphi_k} \big(\log p(\vecY\vert\vect{Z},\negr{\phi},\negr{\tau}, \vect{a})\big) +\E_{-\bphi_k}\log p(\negr{\phi}) \nonumber \\
&\addeq& \frac{n}{2}\E_{q^*(\tau_k)}(\log \tau_k) \sum_{i=1}^N \E_{q^*(Z_i)}[\I(Z_i=k)] \nonumber \\
&& \; - \; \frac{1}{2} \E_{q^*(\tau_k)}(\tau_k) \sum_{i=1}^N \Big\{\E_{q^*(Z_i)}[\I(Z_i=k)] \nonumber \\
&& \times \,\E_{q^*(a_i)}[(\vecY_i \; - \, \matr{B}\negr{\phi}_k-a_i\vect{1}_n)^T(\vecY_i - \matr{B}\negr{\phi}_k-a_i\vect{1}_n)]\Big\} \label{eq:quad1:ext} \\
&& + \frac{M}{2} \log v^0 \; - \;\frac{1}{2}v^0 (\bphi_k -\vect{m}_k^0)^T(\bphi_k -\vect{m}_k^0)
\label{eq:quad2:ext}
\end{eqnarray}

All expectations are defined in Section \ref{sec:expectations}, but note that, for example, $\E_{q^*(Z_i)}[\I(Z_i=k)] = p^*_{ik}$ and
\begin{eqnarray}
&&{\E_{q^*(a_i)}[(\vecY_i \; - \, \matr{B}\negr{\phi}_k-a_i\vect{1}_n)^T(\vecY_i - \matr{B}\negr{\phi}_k-a_i\vect{1}_n)]} \nonumber \\ 
&\addeq& (\vecY_i \; - \, \matr{B}\negr{\phi}_k-\mu_{a_i}^*\vect{1}_n)^T(\vecY_i - \matr{B}\negr{\phi}_k-\mu_{a_i}^*\vect{1}_n) \nonumber
\end{eqnarray}
where $\mu_{a_i}^*$ is the posterior mean of $q^*(a_i)$ which is derived later. We focus on the quadratic forms that appear in (\ref{eq:quad1:ext}) and (\ref{eq:quad2:ext}). Let $\vecY_i^*=\vecY_i-\mu_{a_i}^*\vect{1}_n$, we can write:
\begin{eqnarray}
\log q^*(\bphi_k)  &\addeq& -\frac{1}{2} \E_{q^*(\tau_k)}(\tau_k) \sum_{i=1}^N p^*_{ik} (\vecY_i^* \; - \, \matr{B}\negr{\phi}_k)^T(\vecY_i^* - \matr{B}\negr{\phi}_k) \nonumber \\
&&- \;\frac{1}{2}v^0 (\bphi_k -\vect{m}_k^0)^T(\bphi_k -\vect{m}_k^0) \nonumber \\
&=&  - \; \frac{1}{2} \E_{q^*(\tau_k)}(\tau_k) \sum_{i=1}^N p^*_{ik}\big[\vecY_i^{*T} \vecY_i^* - 2\vecY_i^{*T}\matr{B}\bphi_k + \bphi_k^T\matr{B}^T\matr{B}\bphi_k \big] \nonumber \\
&& \; -\frac{1}{2}v^0\big[ \bphi^T_k\bphi_k  - 2(\vect{m}_k^0)^T\bphi_k + (\vect{m}_k^0)^T\vect{m}_k^0 \big]\nonumber \\
&\addeq&-\frac{1}{2} \bphi_k^T \Big[v^0\matr{I} + \E_{q^*(\tau_k)}(\tau_k) \sum_{i=1}^N p^*_{ik} \matr{B}^T\matr{B} \Big] \bphi_k \nonumber \\
&& + \Big [v^0(\vect{m}_k^0)^T + \E_{q^*(\tau_k)}(\tau_k) \sum_{i=1}^N p^*_{ik} \vecY_i^{*T} \matr{B}  \Big]\bphi_k .
\label{eq:quad_form:ext}
\end{eqnarray}
Now let 
\begin{eqnarray}
    \matr{\Sigma}^*_k = \Big[v^0\matr{I} + \E_{q^*(\tau_k)}(\tau_k) \sum_{i=1}^N p^*_{ik} \matr{B}^T\matr{B}   \Big]^{-1}.
    \label{eq:sigma_star:ext}
\end{eqnarray} 
\noindent We can then rewrite (\ref{eq:quad_form:ext}) as

$$ -\frac{1}{2} \bphi_k^T \matr{\Sigma}^{*-1}_k \bphi_k -\frac{1}{2}(-2)\Big [v^0(\vect{m}_k^0)^T + \E_{q^*(\tau_k)}(\tau_k) \sum_{i=1}^N p^*_{ik} \vecY_i^{*T} \matr{B} \Big]\matr{\Sigma}^*_k\matr{\Sigma}^{*-1}_k \bphi_k.$$  

\noindent Therefore, $q^*(\bphi_k)$ is $MVN(\vect{m}^*_k,\matr{\Sigma}^*_k)$ with  $\matr{\Sigma}^*_k$ as in (\ref{eq:sigma_star:ext}) and mean vector
\begin{eqnarray}
 \vect{m}^*_k = \big [v^0(\vect{m}_k^0)^T + \E_{q^*(\tau_k)}(\tau_k) \sum_{i=1}^N p^*_{ik} \vecY_i^{*T} \matr{B}  \big]\matr{\Sigma}^*_k. 
 \label{eq:mkstar:ext}
\end{eqnarray}

\vspace{0.4cm}
\noindent\textit{iv) Update equation for $q(\tau_k)$}

Similarly to the calculations in iii) we can write
\begin{eqnarray}
 \log q^*(\tau_k) &\addeq& \frac{n}{2}\log \tau_k \sum_{i=1}^N p^*_{ik} \nonumber \\
 && -\frac{1}{2}\tau_k\sum_{i=1}^N p^*_{ik}\E_{q^*(\negr{\phi}_k)\cdot q^*(a_i)}\big[ (\vecY_i - \matr{B}\negr{\phi}_k-a_i\vect{1}_n)^T(\vecY_i - \matr{B}\negr{\phi}_k-a_i\vect{1}_n) \big]  \nonumber \\
&& + \,(b^0 -1)\log \tau_k - r^0\tau_k  \nonumber
\end{eqnarray}
Therefore, $q^*(\tau_k)$ is a Gamma distribution with parameters
\begin{eqnarray}
A^*_k = b^0 + \frac{n}{2} \sum_{i=1}^N p^*_{ik} 
\label{eq:Akstar:ext}
\end{eqnarray}
\noindent and
\begin{eqnarray}
R^*_{k} &=& r^0 + \frac{1}{2} \sum_{i=1}^N\Big\{ p^*_{ik}  \nonumber \\ && \times\,\E_{q^*(\negr{\phi}_k)\cdot q^*(a_i)}\big[ (\vecY_i - \matr{B}\negr{\phi}_k-a_i\vect{1}_n)^T(\vecY_i - \matr{B}\negr{\phi}_k-a_i\vect{1}_n) \big]\Big\}.
\label{eq:Rkstar:ext}
\end{eqnarray}

\noindent\textit{v) Update equation for $q(a_i)$} 
\begin{eqnarray}
 \log q^*(a_i) &\addeq& \E_{-a_i} \big(\log p(\vect{Y},\vect{Z},\negr{\pi},\negr{\phi},\negr{\tau}, \vect{a}, \tau_a) \big)  \nonumber \\
&\addeq&  \E_{-a_i} \big( \log p(\vecY \vert \vect{Z},\negr{\phi},\negr{\tau}, \vect{a})\big)\; +\; \E_{-a_i} \big(\log p(\vect{a} \vert \tau_a)\big) \nonumber \\
&\addeq& \E_{-a_i}\big[\sum_{k=1}^K \I(Z_i = k) \log p(\vecY_i \vert Z_i =k, \bphi_k,\tau_k, a_i)\big] \nonumber \\
&& + \,\E_{-a_i}\big[\sum_{k=1}^K \I(Z_i = k) \log p(a_i \vert\tau_a)\big]  \nonumber \\
&\addeq& \sum_{k=1}^K p^*_{ik} \big\{ \frac{n}{2}\E_{q^*(\tau_k)}\log \tau_k \nonumber \\
&& -\frac{1}{2}\E_{q^*(\tau_k)}\tau_k \E_{q^*(\negr{\phi}_k)}\big[ (\vecY_i - \matr{B}\negr{\phi}_k-a_i\vect{1}_n)^T(\vecY_i - \matr{B}\negr{\phi}_k-a_i\vect{1}_n) \big] \nonumber \\
&&-\frac{1}{2}a_i^2\E_{q^*(\tau_a)}\tau_a \big\} \nonumber \\
&\addeq& \sum_{k=1}^K p^*_{ik} \big\{-\frac{1}{2}\E_{q^*(\tau_k)}\tau_k \big[ (\vecY_i - \matr{B}\vect{m}^*_k-a_i\vect{1}_n)^T(\vecY_i - \matr{B}\vect{m}^*_k-a_i\vect{1}_n) \big] \nonumber \\
&&-\frac{1}{2}a_i^2\E_{q^*(\tau_a)}\tau_a \big\} \nonumber
\end{eqnarray}
Let $\vecY_{ik}^*=\vecY_i-\matr{B}\vect{m}^*_k$, then
\begin{eqnarray}
 \log q^*(a_i) &\addeq& \sum_{k=1}^K p^*_{ik} \big\{-\frac{1}{2}\E_{q^*(\tau_k)}\tau_k \big[ (\vecY_{ik}^*-a_i\vect{1}_n)^T(\vecY_{ik}^*-a_i\vect{1}_n) \big] -\frac{1}{2}a_i^2\E_{q^*(\tau_a)}\tau_a \big\} \nonumber \\
 &\addeq& -\frac{n}{2} a_i^2 \sum_{k=1}^K p^*_{ik}\E_{q^*(\tau_k)}\tau_k + a_i \sum_{k=1}^K p^*_{ik}\E_{q^*(\tau_k)}\tau_k \vect{1}_n^T\vecY_{ik}^* -\frac{1}{2}a_i^2 \E_{q^*(\tau_a)}\tau_a \nonumber \\
 &=& -\frac{1}{2}a_i^2 \big[n\sum_{k=1}^K p^*_{ik}\E_{q^*(\tau_k)}\tau_k + \E_{q^*(\tau_a)}\tau_a\big] + a_i \sum_{k=1}^K p^*_{ik}\E_{q^*(\tau_k)}\tau_k \vect{1}_n^T\vecY_{ik}^* \nonumber
\end{eqnarray}
Let 
\begin{eqnarray}
    {\sigma}^{2*}_{a_i} = \big( n\sum_{k=1}^K p^*_{ik}\E_{q^*(\tau_k)}\tau_k + \E_{q^*(\tau_a)}\tau_a\big)^{-1}
    \label{eq:sigma_a_star:ext}
\end{eqnarray} 
and
\begin{eqnarray}
 {\mu}^{*}_{a_i} = {\sigma}^{2*}_{a_i}\sum_{k=1}^K p^*_{ik}\E_{q^*(\tau_k)}\tau_k \vect{1}_n^T \vecY_{ik}^* 
 \label{eq:mu_a_star:ext}
\end{eqnarray}
Then $q^*(a_i)$ is $N({\mu}^{*}_{a_i},{\sigma}^{*2}_{a_i})$.

\vspace{0.4cm}
\noindent\textit{vi) Update equation for $q(\tau_a)$}
\begin{eqnarray}
 \log q^*(\tau_a) &\addeq& \E_{-\tau_a} \big(\log p(\vect{a} \vert \tau_a) + \log p(\tau_a)\big)  \nonumber \\
&\addeq& \E_{-\tau_a}\Big(\sum_{i=1}^N \log p(a_i \vert \tau_a)\Big) + (\alpha^0-1)\log \tau_a - \beta^0 \tau_a \nonumber \\
&\addeq& \frac{N}{2} \log \tau_a -\frac{1}{2}\tau_a \sum_{i=1}^N \E_{q^*(a_i)} a_i^2 + (\alpha^0-1)\log \tau_a - \beta^0 \tau_a \nonumber \\
&=& \big(\alpha^0+\frac{N}{2} -1\big)\log \tau_a -\Big(\beta^0 +\frac{1}{2} \sum_{i=1}^N \E_{q^*(a_i)} a_i^2 \Big)\tau_a \nonumber
\end{eqnarray}
Let 
$$\alpha^* = \alpha^0+\frac{N}{2}$$
and
\begin{eqnarray}
 \beta^*=\beta^0 +\frac{1}{2}\sum_{i=1}^N \E_{q^*(a_i)} a_i^2
 \label{eq:beta_star_ext}
\end{eqnarray}
$q^*(\tau_a)$ is Gamma($\alpha^*, \beta^*$). 

\subsubsection{Expectations}\label{sec:expectations}

In this section, we calculate the expectations in the update equations derived in Section \ref{sec:updatesVB} for each component in the VD.
Let  $ \negr{\Psi}$ be the digamma function defined as
\begin{eqnarray}
 \negr{\Psi}(x)=\frac{d}{dx}\log \Gamma(x), 
 \label{eq:digamma:ext}
\end{eqnarray}
\noindent which can be easily calculated via numerical approximation. The values of the expectations taken with respect to the approximated distributions are given as follows.
\begin{eqnarray}\E_{q^*(Z_i)}[\I(Z_i=k)] = p^*_{ik}
\label{eq:EqZstar:ext}
\end{eqnarray}
\begin{eqnarray}\E_{q^*(\tau_k)}(\tau_{k}) = \frac{A^{*}_{k}}{R^{*}_{k}}
\label{eq:Eqtaustar:ext}
\end{eqnarray}
\begin{eqnarray}
\E_{q^*(\tau_k)}(\log \tau_{k}) =  \negr{\Psi}(A^{*}_k) - \log R^{*}_{k}
\label{eq:Eqlogtau:ext}
\end{eqnarray}
\begin{eqnarray}
\E_{q^*(\negr{\pi})}(\log \pi_{k}) =  \negr{\Psi}(d^{*}_{k}) -  \negr{\Psi}\Big(\sum_{k=1}^K d^{*}_{k}\Big)
\label{eq:Eqlogpi:ext}
\end{eqnarray}
\begin{eqnarray}
\E_{q^*(\tau_a)}(\tau_{a}) = \frac{
\alpha^*}{\beta^*} 
\label{eq:Eqtau:a:ext}
\end{eqnarray}
\begin{eqnarray}
\E_{q^*(\tau_a)}(\log \tau_{a}) = \negr{\Psi}(\alpha^{*}) - \log \beta^{*}
\label{eq:Eqlogtau:a:ext}
\end{eqnarray}
\begin{eqnarray}
\E_{q^*(a_i)} a_i^2 =  \sigma_{a_i}^{*2} + \mu_{a_i}^{*2}
\label{eq:Eq:a2:ext}
\end{eqnarray}

\noindent In addition, using the fact that $\E(\vecX^T \vecX) = \mbox{trace}[\mbox{Var}(\vecX)] + \E(\vecX)^T\E(\vecX)$, we obtain
\begin{eqnarray}
&&{\E_{q^*(\negr{\phi}_k)}\big[ (\vecY_i - \matr{B}\negr{\phi}_k-a_i\vect{1}_n)^T(\vecY_i - \matr{B}\negr{\phi}_k-a_i\vect{1}_n) \big]} \nonumber \\
&=& \mbox{trace}\big(\matr{B} \matr{\Sigma}^*_k \matr{B}^T \big) \nonumber \\ && + \,(\vecY_i - \matr{B}\vect{m}^*_k-a_i\vect{1}_n)^T(\vecY_i - \matr{B}\vect{m}^*_k-a_i\vect{1}_n),
 \label{eq:Eqphistar:ext}
\end{eqnarray}
and 
\begin{eqnarray}
&&{\E_{q^*(\negr{\phi}_k)\cdot q^*(a_i)}\big[ (\vecY_i - \matr{B}\negr{\phi}_k-a_i\vect{1}_n)^T(\vecY_i - \matr{B}\negr{\phi}_k-a_i\vect{1}_n) \big]} \nonumber \\
  &=& \E_{q^*(a_i)}\Big[\E_{q^*(\negr{\phi}_k)}\big[ (\vecY_i - \matr{B}\negr{\phi}_k-a_i\vect{1}_n)^T(\vecY_i - \matr{B}\negr{\phi}_k-a_i\vect{1}_n) \big]\Big] \nonumber \\
  &=& \E_{q^*(a_i)}\Big[\mbox{trace}\big(\matr{B} \matr{\Sigma}^*_k \matr{B}^T \big) + (\vecY_i - \matr{B}\vect{m}^*_k-a_i\vect{1}_n)^T(\vecY_i - \matr{B}\vect{m}^*_k-a_i\vect{1}_n)\Big] \nonumber\\
  &=& \mbox{trace}\big(\matr{B} \matr{\Sigma}^*_k \matr{B}^T \big) + n\sigma_{a_i}^{*2} \nonumber\\
  && +\, (\vecY_i - \matr{B}\vect{m}^*_k-\mu_{a_i}^*\vect{1}_n)^T(\vecY_i - \matr{B}\vect{m}^*_k-\mu_{a_i}^*\vect{1}_n).
\label{eq:Eqphi:qa:star:ext}
\end{eqnarray}

\subsection{ELBO calculation}\label{sec:elbo_derivation}

In this section, we show how to calculate the ELBO under Model 2, which is the convergence criterion of our proposed VB algorithm and is updated at the end of each iteration until convergence. Equation (\ref{Eq:elbo:ext}) gives the ELBO:
$$
    \mbox{ELBO}(q) = \E_{q^*} \big[ \log p(\vecY ,\vect{Z},\negr{\pi},\negr{\phi},\negr{\tau}, \vect{a},\tau_a)\big] - \E_{q^*} \big[ \log q(\vect{Z},\negr{\pi},\bphi,\negr{\tau},\vect{a},\tau_a) \big],
$$
\noindent where
\begin{eqnarray}
    \E_{q^*} \big[ \log p(\vecY ,\vect{Z},\negr{\pi},\negr{\phi},\negr{\tau}, \vect{a},\tau_a) \big] &=& 
    \E_{q^*} \big[ \log p(\vecY \vert \vect{Z},\negr{\phi},\negr{\tau}, \vect{a}) \big] + 
    \E_{q^*} \big[ \log p(\vect{Z} \vert \negr{\pi} \big)]  \nonumber \\
    && +\, \E_{q^*} \big[ \log p(\negr{\phi})] + 
    \E_{q^*} \big[ \log p(\negr{\tau})] \nonumber \\
    && +\,\E_{q^*} \big[ \log p(\negr{\phi})] + \E_{q^*} \big[ \log p(\vect{a} \vert \tau_a \big)] \nonumber \\
    && +\, \E_{q^*} \big[ \log p(\tau_a)] ,  \nonumber
\end{eqnarray}
\noindent and 
\begin{eqnarray}
    \E_{q^*} \big[ \log q(\vect{Z},\negr{\pi},\bphi,\negr{\tau}, \vect{a},\tau_a) \big] &=& \E_{q^*} \big[ \log q(\vect{Z}) \big] +  
    \E_{q^*} \big[ \log q(\bphi) \big] + 
    \E_{q^*} \big[ \log q(\negr{\pi}) \big] \nonumber \\ 
    && +\,\E_{q^*} \big[ \log q(\negr{\tau}) \big] + \E_{q^*} \big[ \log q(\vect{a}) \big] 
    + \E_{q^*} \big[ \log q(\tau_a) \big].\nonumber
\end{eqnarray}
\noindent Therefore, we can write the ELBO as the summation of 7 terms: 
\begin{eqnarray}
    \mbox{ELBO}(q) &=& \E_{q^*} \big[ \log p(\vecY \vert \vect{Z},\negr{\phi},\negr{\tau}, \vect{a}) \big] + diff_{\vect{Z}} +  diff_{\negr{\phi}} \nonumber
\\ && + \,diff_{\negr{\tau}} + diff_{\negr{\pi}} + diff_{\vect{a}} + diff_{\tau_a}   
    \label{eq:elbo_calc:ext}
\end{eqnarray}
\noindent where, 
$$diff_{\vect{Z}} = \E_{q^*}\big[\log p (\vect{Z} \vert\negr{\pi})\big ]-\E_{q^*}\big[\log q(\vect{Z})\big].$$
\noindent Specifically,
\begin{eqnarray}
 diff_{\vect{Z}} = \sum_{i=1}^{N}\sum_{k=1}^K p^{*}_{ik} \E_{q^*(\negr{\pi})}(\log \pi_k) - \sum_{i=1}^{N}\sum_{k=1}^K p^{*}_{ik} \log p^{*}_{ik}.
\label{eq:diff_Z:ext}
\end{eqnarray}
The other terms in (\ref{eq:elbo_calc:ext}) are calculated as follows:
$$diff_{\negr{\phi}} = -\frac{1}{2}\sum_{k=1}^K v_k^0\{\mbox{trace}\big(\matr{\Sigma}^*_k \big) + (\vect{m}^*_k-\vect{m}^0_k)^T(\vect{m}^*_k-\vect{m}^0_k)\} +\frac{1}{2}\sum_{k=1}^K \log \vert\matr{\Sigma}^*_k\vert,$$
\begin{eqnarray}
diff_{\negr{\tau}} &=& \sum_{k=1}^K \{(b^0-1)\E_{q^*(\tau_k)}(\log \tau_{k})-r^0\E_{q^*(\tau_k)}(\tau_{k})\} \nonumber
\\&& - \, \sum_{k=1}^K\{A^*_k \log R_k^{*} - \log \Gamma(A^*_k)
\nonumber
\\&& +\,(A^*_k-1)\E_{q^*(\tau_k)}(\log \tau_{k}) - R_k^{*}\E_{q^*(\tau_k)}(\tau_{k}) \}
\label{diff.tau:ext},
\end{eqnarray}
$$diff_{\negr{\pi}} \equiv \sum_{k=1}^K (d_k^0-d_k^{*})\E_{q^*(\negr{\pi})}(\log \pi_{k}),$$
$$diff_{\vect{a}}=-\frac{1}{2}\E_{q^*(\tau_a)}\tau_a \sum_{i=1}^{N}\E_{q^*(a_i)} a_i^2+\sum_{i=1}^{N}\log \sigma_{a_i}^*,$$
\begin{eqnarray}
    diff_{\tau_a}&=&(\alpha^0-1)\E_{q^*(\tau_a)}(\log\tau_{a})-\beta^0 \E_{q^*(\tau_a)}\tau_a \nonumber \\ 
    && - \, \alpha^*\log\beta^*-(\alpha^*-1)\E_{q^*(\tau_a)}(\log\tau_{a})+\beta^* \E_{q^*(\tau_a)}\tau_a\nonumber \\
    &=& (\alpha^0-\alpha^*)\E_{q^*(\tau_a)}(\log\tau_{a}) - (\beta^0 -\beta^*) \E_{q^*(\tau_a)}\tau_a -\alpha^*\log\beta^* \nonumber
\end{eqnarray}
\noindent and
\begin{eqnarray}
&&{\E_{q^*} \big[ \log p(\vecY \vert \vect{Z},\negr{\phi},\negr{\tau}, \vect{a}) \big]}\nonumber \\&=&\sum_{i=1}^{N}\sum_{k=1}^K p^{*}_{ik}\Big\{\frac{n}{2}\E_{q^*(\tau_k)}(\log \tau_{k}) \nonumber \\ 
    && -\frac{1}{2}\frac{A_k^{*}}{R_k^{*}}\E_{q^*(\negr{\phi}_k)\cdot q^*(a_i)}\big[ (\vecY_i - \matr{B}\negr{\phi}_k-a_i\vect{1}_n)^T(\vecY_i - \matr{B}\negr{\phi}_k-a_i\vect{1}_n) \big]\Big\}. \nonumber
\end{eqnarray}

\noindent Therefore, at iteration $c$, we calculate $\mbox{ELBO}^{(c)}$ using all parameters obtained at the end of iteration $c$. Convergence of the algorithm is achieved if $\mbox{ELBO}^{(c)}-\mbox{ELBO}^{(c-1)}$ is smaller than a given threshold. It is important to note that we use the fact that $\displaystyle \lim_{p^{*}_{ik} \to 0} p^{*}_{ik}\log p^{*}_{ik}=0$ to avoid numerical issues when calculating (\ref{eq:diff_Z:ext}). Numerical issues also exist in calculating the term $\{A^*_k \log R_k^{*} - \log \Gamma(A^*_k)
+(A^*_k-1)\E_{q^*(\tau_k)}(\log \tau_{k}) - R_k^{*}\E_{q^*(\tau_k)}(\tau_{k}) \}$ in (\ref{diff.tau:ext}), so we will approximate it by the following digamma and log-gamma approximations. Note that we use (\ref{eq:Eqtaustar:ext}) and (\ref{eq:Eqlogtau:ext}) for $\E_{q^*(\tau_k)}(\tau_{k})$ and $\E_{q^*(\tau_k)}(\log \tau_{k})$, respectively.
\vspace{0.2cm}

\noindent (1) digamma approximation based on asymptotic expansion: $$\negr{\Psi}(A_k^{*})\approx\log A_k^{*} - 1/(2A_k^{*}).$$

\noindent (2) log-gamma Stirling's series approximation: $$\log \Gamma(A_k^{*})\approx A_k^{*}\log (A_k^{*}) - A_k^{*} - \frac{1}{2}\log(A_k^{*}).$$

Therefore, plugging in these two approximations, we obtain
\begin{eqnarray}
&& A^*_k \log R_k^{*} - \log \Gamma(A^*_k)
+(A^*_k-1)\E_{q^*(\tau_k)}(\log \tau_{k}) - R_k^{*}\E_{q^*(\tau_k)}(\tau_{k}) \nonumber \\ 
&=& A^*_k \log R_k^{*} - \log \Gamma(A^*_k)
+(A^*_k-1)(\negr{\Psi}(A^{*}_k) - \log R^{*}_{k}) - R_k^{*}\frac{A^{*}_{k}}{R^{*}_{k}} \nonumber \\ 
&\approx& \frac{1}{2}\log A_k^{*} +\frac{1}{2A^*_k}-\frac{1}{2}\nonumber \\
&\addeq& \frac{1}{2}\log A_k^{*} +\frac{1}{2A^*_k} = \frac{1}{2}(\log A_k^{*} + \frac{1}{A^*_k})\nonumber   
\end{eqnarray}

\begin{algorithm}[!ht]
  \KwData{$N$ original curves with $n$ evaluation points; number of clusters $K$; values of hyperparameters: $\vect{d}^0$, $\vect{m}_k^0, k=1,...,K$, $s^0$, $b^0$, $r^0$, $\alpha^0$, $\beta^0$; convergence threshold and maximum number of iterations}
  \KwResult{VB estimated mean curves for each cluster and the cluster index for each original curve}
  \textbf{Initialization}: initialize $R_{k}^*$, $\mu^*_a$ and $\beta^*$ with arbitrary values (e.g., $R_{k}^*= r^0$, $\mu^*_a=0$, \,$\beta^*=\beta^0$) and $p^{*}_{ik}$ from $k$-means, and set $c=0$\;
  \While{$c<$ maximum number of iterations and difference of ELBO $>$ convergence threshold}{
  $\alpha^*=\alpha^0+\frac{N}{2}$\;
    \Repeat{maximum iteration is achieved or the ELBO converges}{
      $c=c+1$\;
      update $A^{*(c)}_k$ using $p^{*(c-1)}_{1k},\ldots,p^{*(c-1)}_{Nk}$ with equation (\ref{eq:Akstar:ext})\;
      update $\matr{\Sigma}^{*(c)}_k$ using $A^{*(c)}_k$, $R_{k}^{*(c-1)}$ and $p^{*(c-1)}_{1k},\ldots,p^{*(c-1)}_{Nk}$ with equations (\ref{eq:sigma_star:ext}) and (\ref{eq:Eqtaustar:ext})\;
      update $\vect{m}^{*(c)}_k$ using $\matr{\Sigma}^{*(c)}_k$, $A^{*(c)}_k$, $R_{k}^{*(c-1)}$, $\mu_a^{*(c-1)}$ and $p^{*(c-1)}_{1k},\ldots,p^{*(c-1)}_{Nk} $ with equations (\ref{eq:mkstar:ext}) and (\ref{eq:Eqtaustar:ext})\;
      update $\sigma_{a_i}^{*2(c)}$ using $A^{*(c)}_k$, $R_{k}^{*(c-1)}$, $\alpha^*$, $\beta^{*(c-1)}$ and $p^{*(c-1)}_{ik},\ldots,p^{*(c-1)}_{iK}$ with equations (\ref{eq:sigma_a_star:ext}), (\ref{eq:Eqtaustar:ext}) and (\ref{eq:Eqtau:a:ext}) \;
      update $\mu_{a_i}^{*(c)}$ using $\sigma_{a_i}^{*2(c)}$, $A^{*(c)}_k$, $R_{k}^{*(c-1)}$ and $p^{*(c-1)}_{ik},\ldots,p^{*(c-1)}_{iK}$ with equations (\ref{eq:mu_a_star:ext}) and (\ref{eq:Eqtaustar:ext})\;
      update $R_k^{*(c)}$ using $\vect{m}^{*(c)}_k$, $\matr{\Sigma}^{*(c)}_k$, $\sigma_{a_i}^{*2(c)}$, $\mu_{a_i}^{*(c)}$ and $p^{*(c-1)}_{1k},\ldots,p^{*(c-1)}_{Nk} $ with equations (\ref{eq:Rkstar:ext}) and (\ref{eq:Eqphi:qa:star:ext})\;
      update $\beta^{*(c)}$ using $\sigma_{a_i}^{*2(c)}$ and $\mu_{a_i}^{*(c)}$ with equations (\ref{eq:beta_star_ext}) and (\ref{eq:Eq:a2:ext})\;
      update $\negr{d}^{*(c)}$ using $p^{*(c-1)}_{1k},\ldots,p^{*(c-1)}_{Nk} $ with equations (\ref{Eq:qstarPi:ext}) and (\ref{eq:EqZstar:ext})\;
      update $p^{*(c)}_{1k},\ldots,p^{*(c)}_{Nk} $ using $A^{*(c)}_k$, $R_k^{*(c)}$ , $\negr{d}^{*(c)}$, $\sigma_{a_i}^{*2(c)}$, $\mu_{a_i}^{*(c)}$, $\vect{m}^{*(c)}_k$ and $\matr{\Sigma}^{*(c)}_k$ with equations (\ref{eq:pikstar:ext}), (\ref{eq:Eqtaustar:ext}), (\ref{eq:Eqlogtau:ext}), (\ref{eq:Eqlogpi:ext}) and (\ref{eq:Eqphi:qa:star:ext})\;
      calculate the current ELBO, $\text{ELBO}^{(c)}$ using equation (\ref{eq:elbo_calc:ext}) \;
      calculate difference of ELBO $=\text{ELBO}^{(c)}-\text{ELBO}^{(c-1)}$\;
    }
  }
  \caption{Clustering functional data via variational inference with random intercepts}
  \label{VBalgorithm:ext}
\end{algorithm}

\section{Simulation studies}\label{sec.sim}

In Section \ref{sec:perf_metrics}, we present the metrics used to evaluate the performance our proposed methodology. Sections \ref{sec:sim_model_1} and \ref{sec:sim_model_2} present the simulation scenarios and results for Model 1 and Model 2, respectively.


\subsection{Performance metrics}\label{sec:perf_metrics}

We evaluate the clustering performance of our proposed algorithm by two metrics: mismatches \citep{Zambom_2019} and V-measure \citep{Rosenberg_2007}. Mismatch rate is the proportion of subjects misclassified by the clustering procedure. In our case, each subject corresponds to a curve in our functional dataset. V-measure, a score between zero and one, evaluates the subject-to-cluster assignments and indicates the homogeneity and completeness of a clustering procedure result. Homogeneity is satisfied if the clustering procedure assigns only those subjects that are members of a single group to a single cluster. Completeness is symmetrical to homogeneity, and it is satisfied if all those subjects that are members of a single group are assigned to a single cluster. The V-measure is one when all subjects are assigned to their correct groups by the clustering procedure.

To further evaluate the performance of the proposed VB algorithm in terms of the estimated mean curves, we calculate the empirical mean integrated squared error (EMISE) for each cluster in each simulation scenario. The EMISE is obtained as follows: 
\begin{eqnarray}
 \text{EMISE}_k=\frac{T}{n}\sum_{j=1}^{n}\text{EMSE}_k(t_j),
 \label{eq:EMISE}
\end{eqnarray}
where $T$ is the curve evaluation interval length, $n$ is total number of observed evaluation points, and the empirical mean squared error (EMSE) at point $t_j$ for cluster $k$, $\text{EMSE}_k(t_j)$, is given by
$$\text{EMSE}_k(t_j)=\frac{1}{S}\sum_{s=1}^{S}\big[f_k(t_j)-\hat{f}_k^s(t_j)\big]^2,$$
in which $s$ corresponds to the $s$th simulated dataset among $S$ datasets in total, $f_k(t_j)$ is the value of the true mean function in cluster $k$ evaluated at point $t_j$ and $\hat{f}_k^s(t_j)$ is its corresponding estimated value for the $s$th simulated dataset. The estimated value $\hat{f}_k^s(t_j)$ is calculated using the B-spline basis expansion with coefficients corresponding the to posterior mean (\ref{eq:mkstar:ext}) obtained at the convergence of the VB algorithm.

\subsection{Simulation study on Model 1} \label{sec:sim_model_1}
\subsubsection{Simulation scenarios}\label{sec:scenarios}

Our simulation study for Model 1 comprises six different scenarios, five of which have three clusters ($K=3$) while the last scenario has four clusters ($K=4$). For each simulation scenario, we generate 50 datasets and apply the proposed VB algorithm to each dataset, considering the number of basis functions to be six except for Scenario 5, which uses 12 basis functions. The ELBO convergence threshold is 0.01, with a maximum of 100 iterations. For comparison purposes, we also investigate the performance of the $k$-means algorithm applied to the raw curves. In addition, we use the clustering results of $k$-means to initialize $p^{*}_{ik}$ in our VB algorithm.

Scenarios 1 and 2 are adopted from \citet{Zambom_2019}. Each dataset is generated from 3 possible clusters ($k=1,2,3$)  with $N=50$ curves per cluster. For each curve, we assume there are $n=100$ observed values across a grid of equally spaced points in the interval $[0, \pi/3]$. \vspace{0.5cm}

\noindent \textit{\textbf{Scenario 1, $K = 3$:}}
$$Y_{ik}(t_j)=a_i+b_k+c_k \sin(1.3t_j)+t_j^3+\delta_{ij}; i=1,...,50; j=1,...,100;  k=1,2,3,$$

\noindent where $Y_{ik}(t_j)$ denotes the value at point $t_j$ of the $i$th curve from cluster $k$, $a_i\sim U(-1/4,1/4)$, $\delta_{ij}\sim N(0, 0.4^2)$, $b_1=0.3$, $b_2=1$, $b_3=0.2$, $c_1=1/1.3$, $c_2=1/1.2$, and $c_3=1/4$. \vspace{0.5cm}

\noindent \textit{\textbf{Scenario 2, $K = 3$:}}
$$Y_{ik}(t_j)=a_i+b_k \exp(c_kt_j)-t_j^3+\delta_{ij}; i=1,...,50; j=1,...,100; k=1,2,3,$$

\noindent where $Y_{ik}(t_j)$ denotes the value at point $t_j$ of the $i$th curve from cluster $k$, $a_i\sim U(-1/4,1/4)$, $\delta_{ij}\sim N(0, 0.3^2)$, $b_1=1/1.8$, $b_2=1/1.7$, $b_3=1/1.5$, $c_1=1.1$, $c_2=1.4$, and $c_3=1.5$. 
\vspace{0.5cm}

In Scenarios 3 and 4, each dataset is also generated considering three clusters ($k=1,2,3$) with 50 curves each. The mean curve of the functional data in each cluster is generated from a pre-specified linear combination of B-spline basis functions. The number of basis functions is the same across clusters but the coefficients of the linear combination are different, one set per cluster (see Table \ref{Scenario_34_coe}). We apply the function \textit{create.bspline.basis} in the R package \textit{fda} to generate six B-spline basis functions of order 4, $B_l(\cdot)$, $l=1,...,6,$ evaluated on equally spaced points, $t_j$, $j=1,...,100$, in the interval $[0, 1]$.
\vspace{0.5cm}

\begin{figure}[!ht]
\centering
\begin{subfigure}{.5\textwidth}
  \centering
  \includegraphics[height = 5cm]{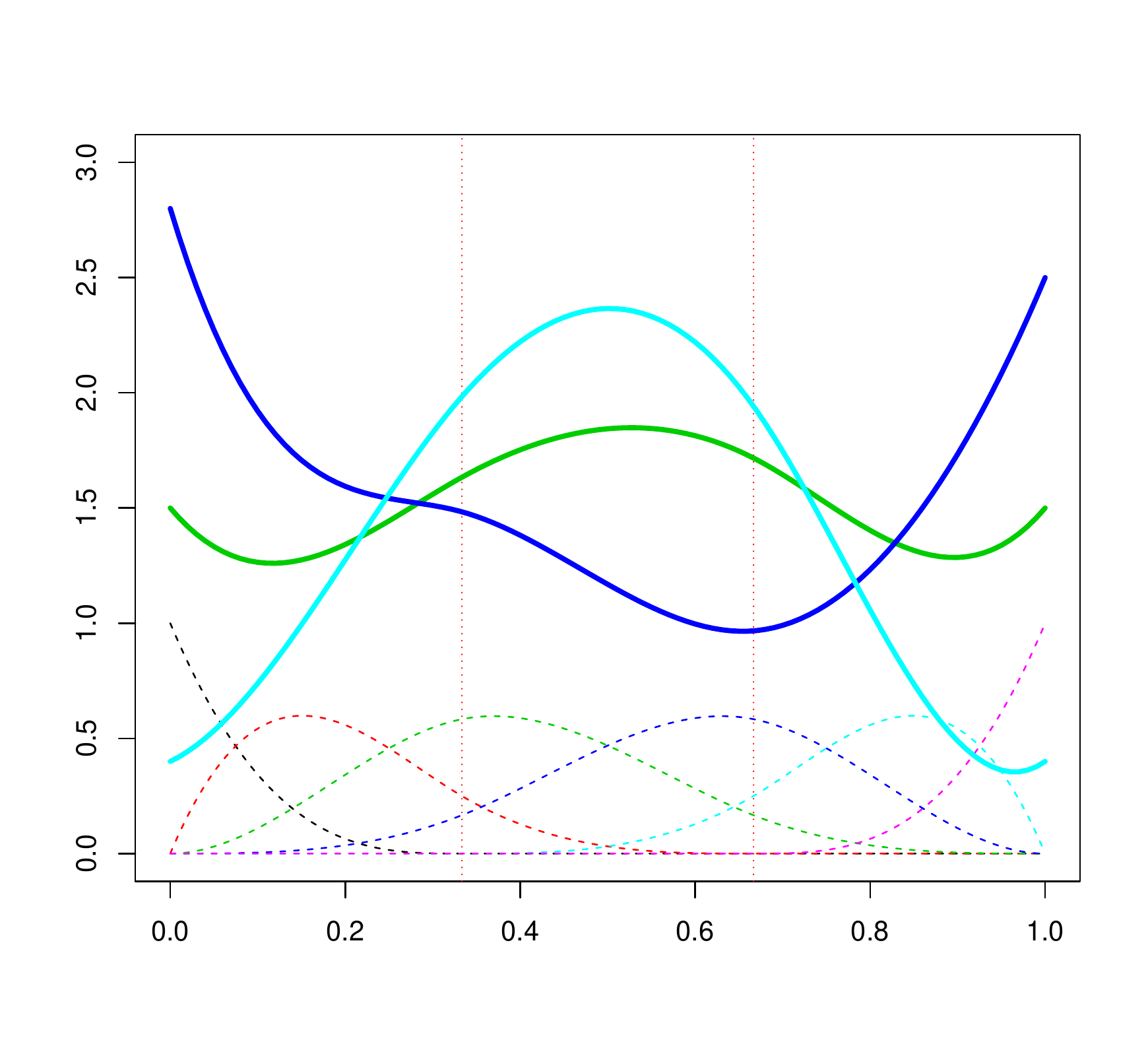}
\end{subfigure}%
\begin{subfigure}{.5\textwidth}
  \centering
  \includegraphics[height = 5cm]{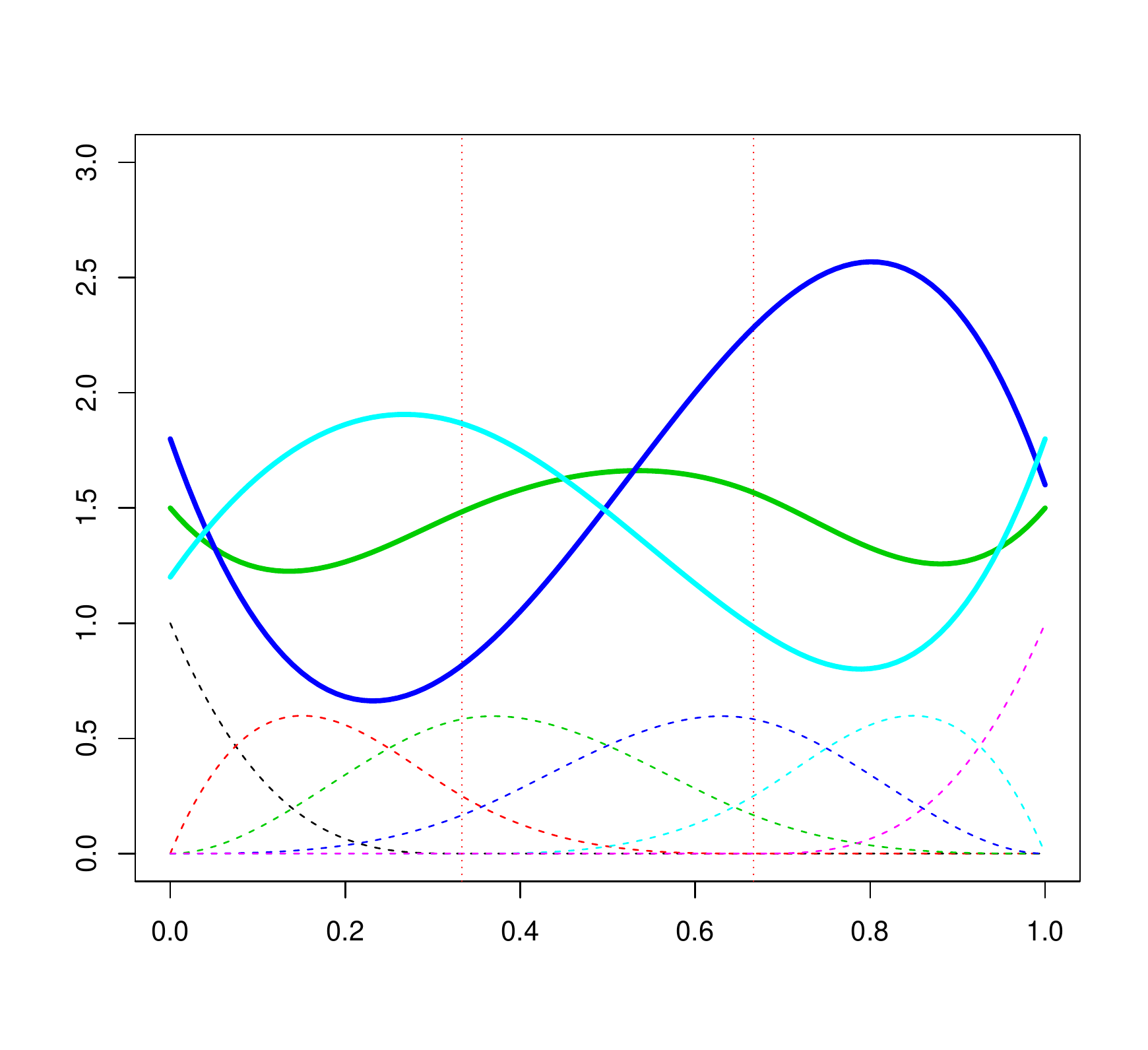}
\end{subfigure}
\caption{Cluster true mean curves (solid curves) and their corresponding six B-splines basis functions (dashed curves) for simulation scenarios 3 (left) and 4 (right).}
\label{fig:scen.3.4}
\end{figure}

\noindent\textit{\textbf{Scenarios 3 and 4, $K = 3$:}}
$$Y_{ik}(t_j)=\sum_{l=1}^{6}B_l(t_j)\phi_{kl}+\delta_{ij}; i=1,...,50; j=1,...,100; k=1,2,3,$$

\noindent where $Y_{ik}(t_j)$ denotes the value at point $t_j$ of the $i$th curve from cluster $k$ and $\delta_{ij}\sim N(0, 0.4^2)$. Table \ref{Scenario_34_coe} presents the vector of coefficients for each cluster $k$, $\bphi_k = (\phi_{k1},\ldots,\phi_{k6})^T$, used in Scenarios 3 and 4. Figure \ref{fig:scen.3.4} illustrates the true mean curves for the three clusters and their corresponding basis functions for Scenarios 3 and 4.

\begin{table}[h]
\begin{center}
\begin{minipage}{\textwidth}
\caption{Coefficient vectors of six B-spline basis functions for each cluster in Scenarios 3 and 4}\label{Scenario_34_coe}
\begin{tabular*}{\textwidth}{@{\extracolsep{\fill}}ccccccccccccc@{\extracolsep{\fill}}}
\toprule%
& \multicolumn{6}{@{}c@{}}{Scenario 3} & \multicolumn{6}{@{}c@{}}{Scenario 4} \\\cmidrule{2-7}\cmidrule{8-13}%
$\bphi_k$  & $\phi_{k1}$  & $\phi_{k2}$   & $\phi_{k3}$ & $\phi_{k4}$   & $\phi_{k5}$   & $\phi_{k6}$ &  $\phi_{k1}$  & $\phi_{k2}$   & $\phi_{k3}$ & $\phi_{k4}$   & $\phi_{k5}$   & $\phi_{k6}$  \\
\midrule
$k=1$  & 1.5 & 1   & 1.8 & 2   & 1   & 1.5 & 1.5 & 1   & 1.6 & 1.8 & 1   & 1.5  \\
$k=2$ & 2.8 & 1.4 & 1.8 & 0.5 & 1.5 & 2.5 & 1.8 & 0.6 & 0.4 & 2.6 & 2.8 & 1.6 \\
$k=3$  & 0.4 & 0.6 & 2.4 & 2.6 & 0.1 & 0.4 & 1.2 & 1.8 & 2.2 & 0.8 & 0.6 & 1.8\\
\botrule
\end{tabular*}
\end{minipage}
\end{center}
\end{table}
Scenario 5 ($K=3$) is based on one of the simulation scenarios used in \citet{Dias_2009} in which the curves mimic the energy consumption of different types of consumers in Brazil. There are 50 curves per cluster and for each curve we generate 96 points based on equally spaced time points, $t_j, \; j=1,...,96$ in the interval $[0, 24]$ (corresponding to one observation every 15 minutes over a 24-hour period). 

\vspace{0.5cm}

\noindent\textit{\textbf{Scenario 5, $K = 3$:}}
\begin{eqnarray}
Y_{i1}(t_j)&=&0.1(0.4 + \exp(-(t_j-6)^2/3) \nonumber \\ 
&& +\, 0.2 \exp(-(t_j-12)^2/25) \nonumber \\
&& +\,0.5 \exp(-(t_j-19)^2/4))+\delta_{ij} \nonumber   
\end{eqnarray}
\begin{eqnarray}
Y_{i2}(t_j)&=&0.1(0.2 + \exp(-(t_j-5)^2/4) \nonumber \\ 
&& +\, 0.25 \exp(-(t_j-18)^2/5))+\delta_{ij}\nonumber 
\end{eqnarray}
\begin{eqnarray}
Y_{i3}(t_j)&=&0.1(0.2 + \exp(-(t_j-3)^2/4) \nonumber \\ 
&& +\, 0.25 \exp(-(t_j-16)^2/5))+\delta_{ij}\nonumber 
\end{eqnarray}
\noindent where $Y_{ik}(t_j)$ denotes the value at time $t_j$ of the $i$th curve from cluster $k$, $i=1,...,50$, $j=1,...,96$,  $k=1,2,3$, and $\delta_{ij}\sim N(0, 0.012^2)$.

\vspace{0.5cm}

Scenario 6 also corresponds to one of the simulation scenarios considered by \citet{Zambom_2019}, where there are $K=4$ clusters with 50 curves each. Each curve has 100 observed values based on equally spaced points, $t_j$, $j=1,...,100$, in the interval $[0, \pi/3]$. 

\vspace{0.5cm}
\noindent\textit{\textbf{Scenario 6, $K = 4$:}}
$$Y_{ik}(t_j)=a_i+b_k - \sin(c_k\pi t_j) + t_j^3+\delta_{ij}; i=1,...,50; j=1,...,100;  k=1,2,3,4,$$

\noindent where $Y_{ik}(t_j)$ denotes the value at point $t_j$ of the $i$th curve from cluster $k$, $a_i\sim U(-1/3,1/3)$, $\delta_{ij}\sim N(0, 0.4^2)$, $b_1=0.2$, $b_2=0.5$, $b_3=0.7$, $b_3=1.3$, $c_1=1.1$, $c_2=1.4$, $c_3=1.6$ and $c_4=1.8$. 

\subsubsection{Simulation results for Model 1}

\begin{figure}[!ht]
\centering
\begin{subfigure}{.5\textwidth}
  \centering
  \includegraphics[height = 5cm, width = 5.8cm]{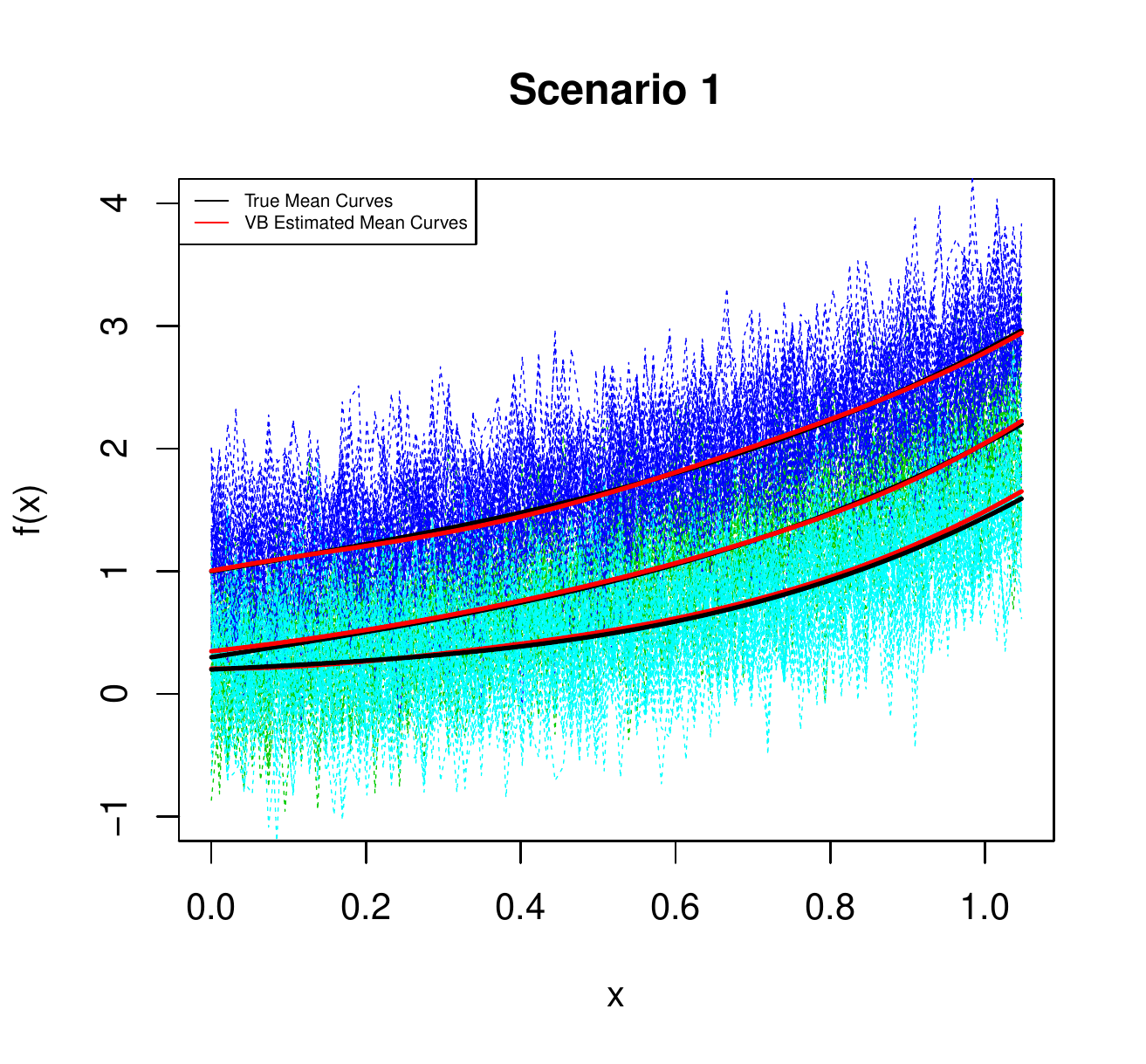}
\end{subfigure}%
\begin{subfigure}{.5\textwidth}
  \centering
  \includegraphics[height = 5cm, width = 5.8cm]{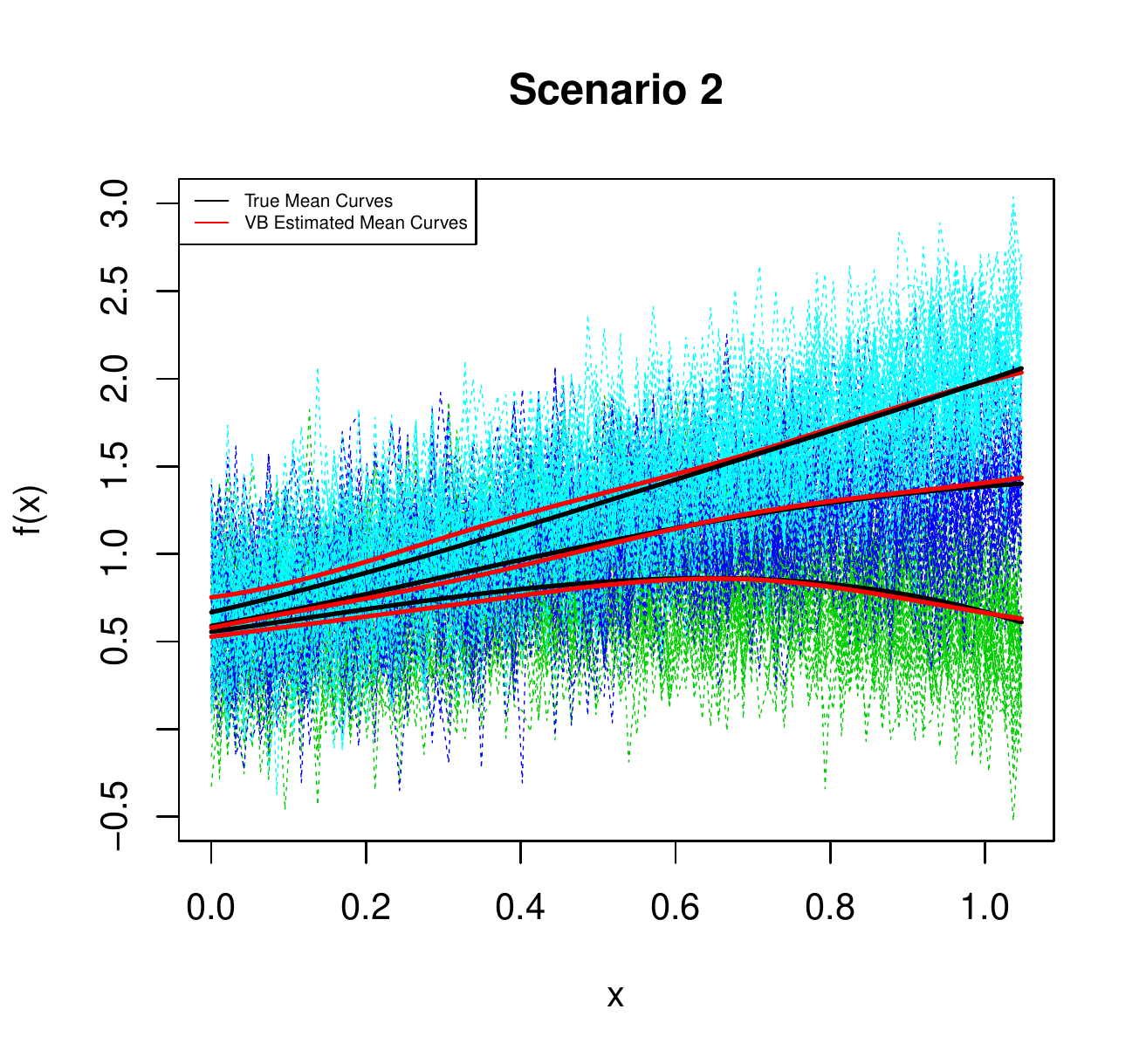}
\end{subfigure} 
\medskip
\begin{subfigure}{.5\textwidth}
  \centering
  \includegraphics[height = 5cm, width = 5.8cm]{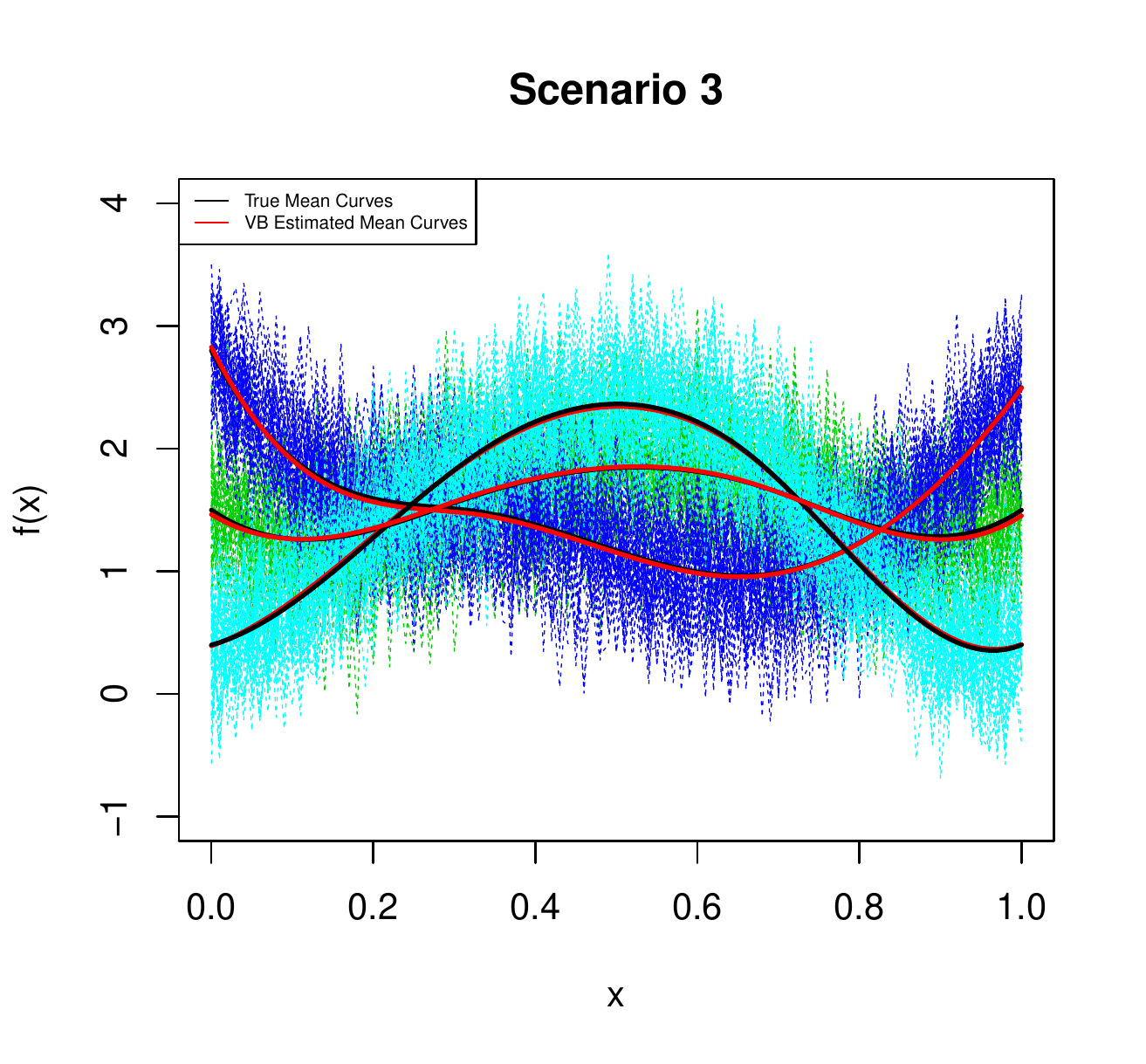}
\end{subfigure}%
\begin{subfigure}{.5\textwidth}
  \centering
  \includegraphics[height = 5cm, width = 5.8cm]{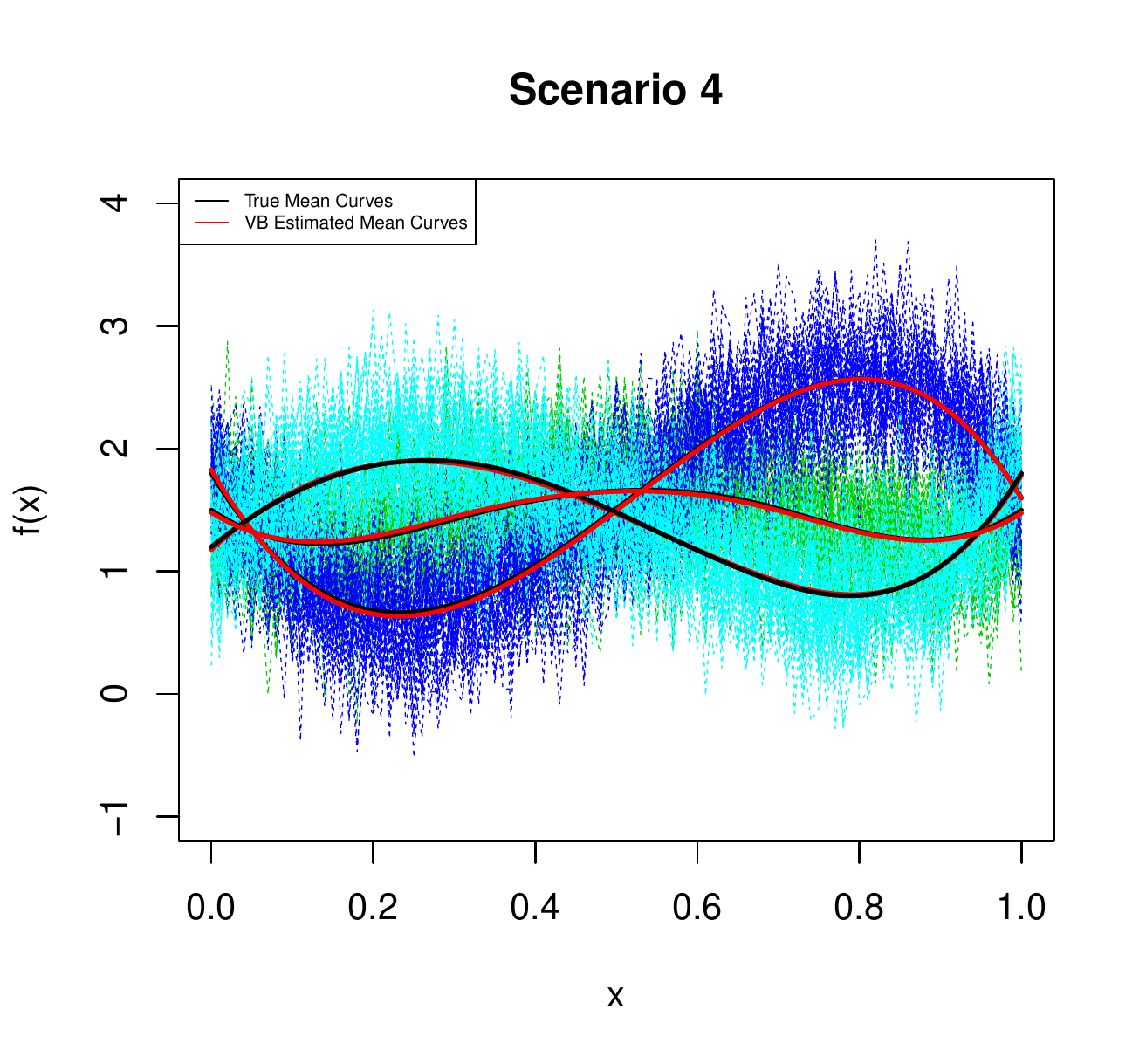}
\end{subfigure} 
\medskip
\begin{subfigure}{.5\textwidth}
  \centering
  \includegraphics[height = 5cm, width = 5.8cm]{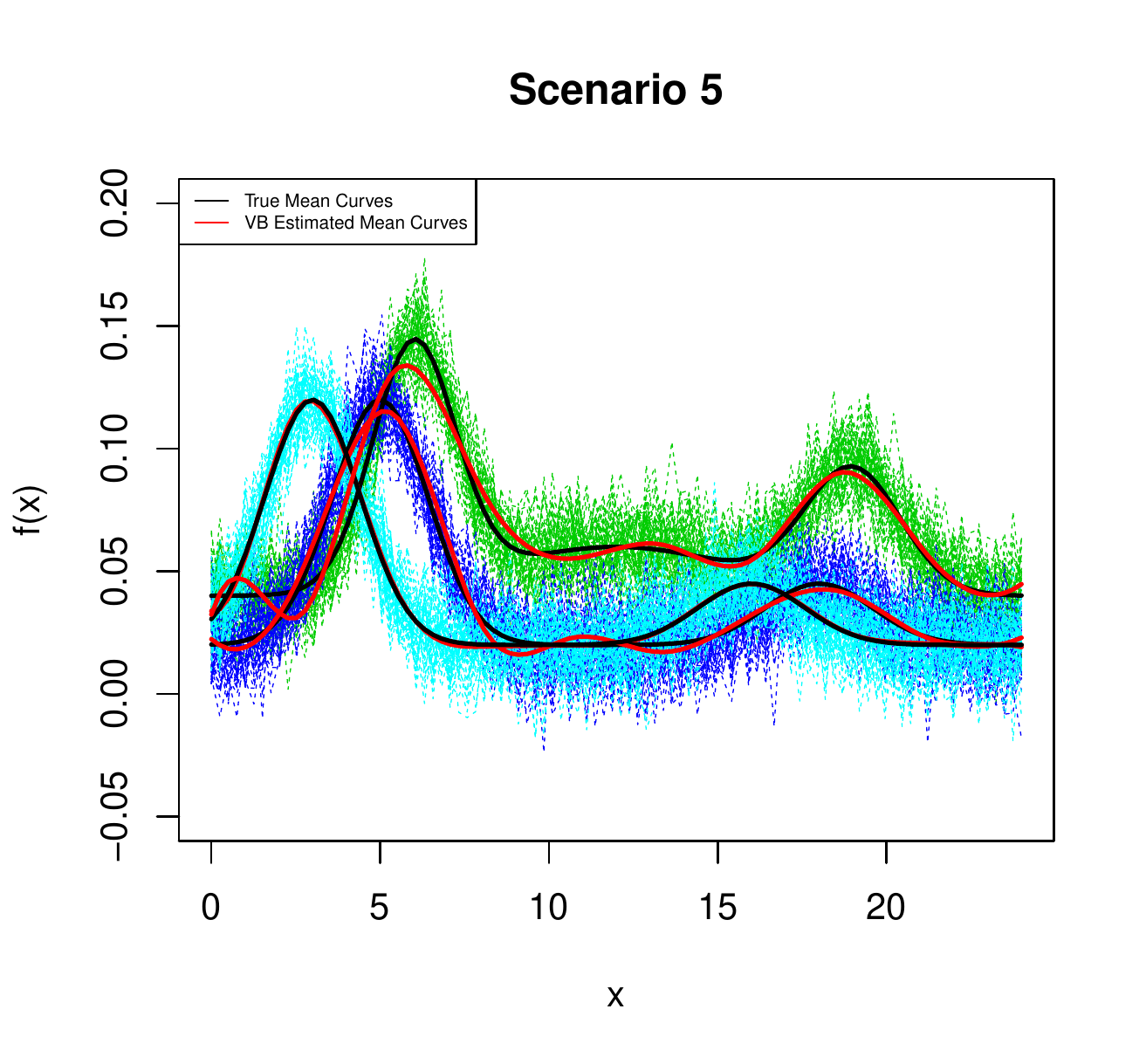}
\end{subfigure}%
\begin{subfigure}{.5\textwidth}
  \centering
  \includegraphics[height = 5cm, width = 5.8cm]{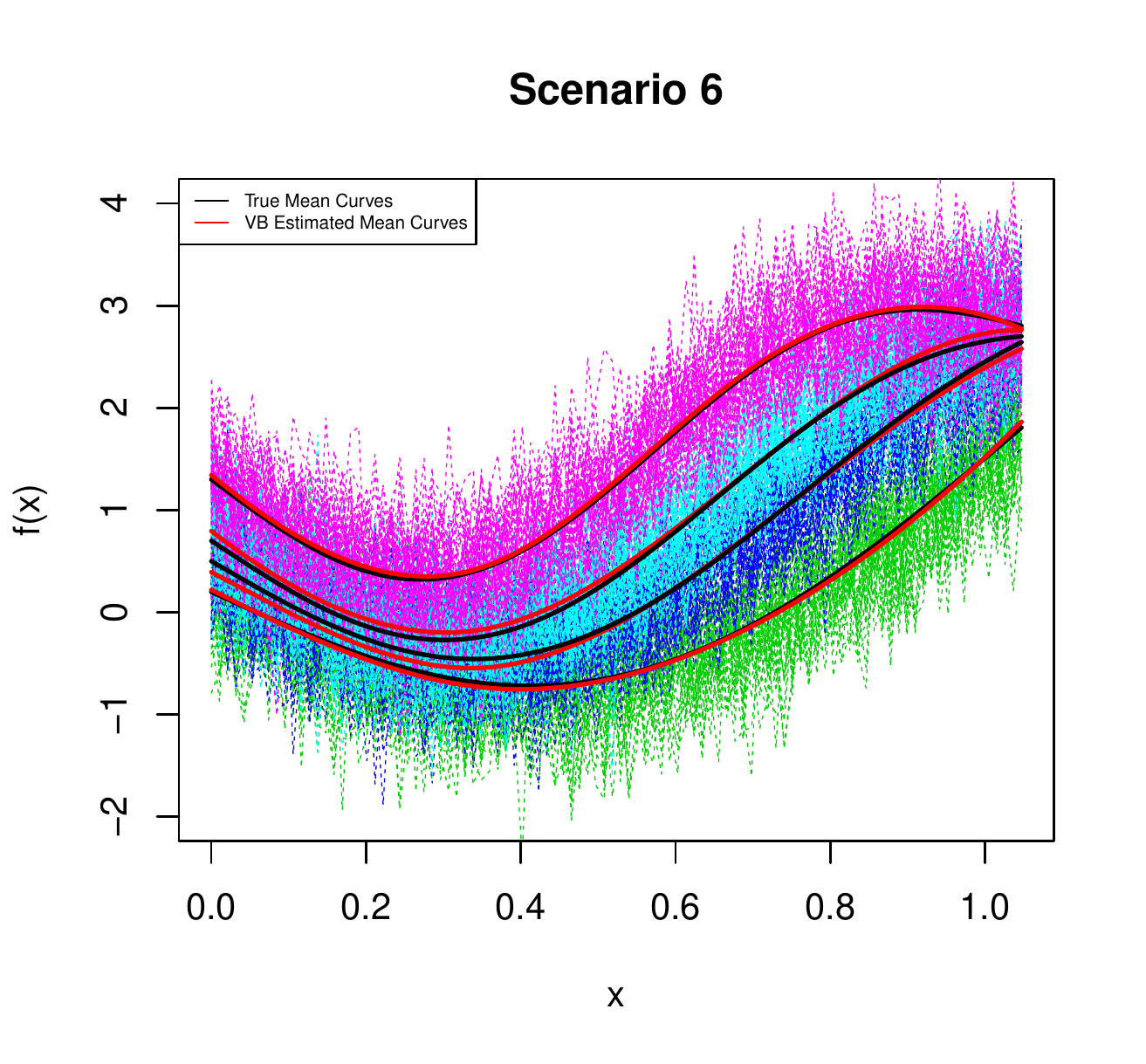}
\end{subfigure}
\caption{\textit{Simulation results for Model 1.} Example of simulated data under each proposed scenario. Raw curves (different colors correspond to different clusters), cluster-specific true mean curves (in black) and corresponding estimated mean curves (in red).}
\label{Performance_fig}
\end{figure}

Figure \ref{Performance_fig} shows the raw curves (color-coded by cluster) from one of the 50 generated datasets for each simulation scenario. In addition, the true mean curves ($f_k(\vect{t})$, $k=1,\ldots,K$) and the estimated smoothed mean curves ($\hat{f}_k(\vect{t})=\matr{B}\vect{m}^*_k$, $k=1,\ldots,K$) are shown in black and red, respectively. We can observe that the true and estimated mean curves almost coincide within each cluster in all scenarios. 

Table \ref{Performance_tab} presents the mean and standard deviation of the mismatch rates and V-measure values over the 50 simulated datasets under each scenario for our proposed VB algorithm and $k$-means. Our VB algorithm performs better than $k$-means in Scenarios 1, 2, and 6, consisting of relatively parallel curves without crossing each other, but there are areas of overlap. In Scenarios 3 and 4, where we simulate data via a linear combination of six predefined basis functions, the proposed method has a better clustering performance than $k$-means, particularly in Scenario 3 with no mismatched curves, in contrast to a 17.15\% mismatch rate from $k$-means. These results are expected since, in these scenarios, the raw data are generated with the same structure as the proposed model. In Scenario 5, where we simulate data based on energy consumption, the VB algorithm also shows satisfactory performance, with only 2\% of curves being misclassified compared to 10.53\% for $k$-means. 

In terms of V-measure, Table \ref{Performance_tab} shows that the proposed method results in average V-measure values above 0.8 in all scenarios except Scenario 2. Mean V-measure values are higher for the proposed VB algorithm than $k$-means in all scenarios. We can observe that V-measure is more sensitive than mismatch rate, and a small increase in misclassification of the curves can make a marked drop in V-measure. These results show that our proposed VB algorithm improves the clustering results upon the naive $k$-means initial clustering. 

For the computational cost, the run times of the proposed VB algorithm of Model 1 for 50 simulated datasets from each scenario are 1.97 min, 5.41 min, 1.41 min, 1.61 min, 3.60 min and 5.32 min, respectively. The algorithm was implemented in R version 3.6.3 on a computer using
the Mac OS X operating system with a 1.6 GHz processor and 8 GBytes of random access memory, same for the simulation study for Model 2 in Section \ref{sec:sim_model_2}.

\begin{table}[h]
\begin{center}
\begin{minipage}{\textwidth}
\caption{\textit{Simulation results for Model 1.} Mismatch rate and V-measure values for each simulation scenario.}\label{Performance_tab}
\begin{tabular*}{\textwidth}{@{\extracolsep{\fill}}ccccc@{\extracolsep{\fill}}}
\toprule%
& \multicolumn{2}{@{}c@{}}{VB algorithm} & \multicolumn{2}{@{}c@{}}{$k$-means} \\\cmidrule{2-3}\cmidrule{4-5}%
Scenario & M\footnotemark[1] (sd\footnotemark[2]) & V\footnotemark[3] (sd) & M (sd)  & V (sd)  \\
\midrule
1 & 0.0409 (0.0153) & 0.8654 (0.0350) & 0.0488 (0.0181) & 0.8594 (0.0388) \\
2 & 0.1416 (0.0334) & 0.6300 (0.0655) & 0.1739 (0.0517) & 0.6188 (0.0650) \\
3 & 0.0000 (0.0000) & 1.0000 (0.0000) & 0.1715 (0.2312) & 0.8738 (0.1700)\\
4 & 0.0000 (0.0000) & 1.0000 (0.0000) & 0.0559 (0.1531) & 0.9581 (0.1145) \\
5 & 0.0200 (0.0800) & 0.9840 (0.0639) & 0.1053 (0.2005) & 0.9227 (0.1469)\\
6 & 0.1054 (0.0197) & 0.8043 (0.0262) & 0.1398 (0.0655) & 0.7819 (0.0546)\\
\botrule
\end{tabular*}
\footnotetext[1]{M: mean mismatch rate from 50 runs.}
\footnotetext[2]{sd: standard deviation.}
\footnotetext[3]{V: mean V-measure from 50 runs.}
\end{minipage}
\end{center}
\end{table}

\begin{figure}[!ht]
\centering
  \includegraphics[height = 6cm, width = 8cm]{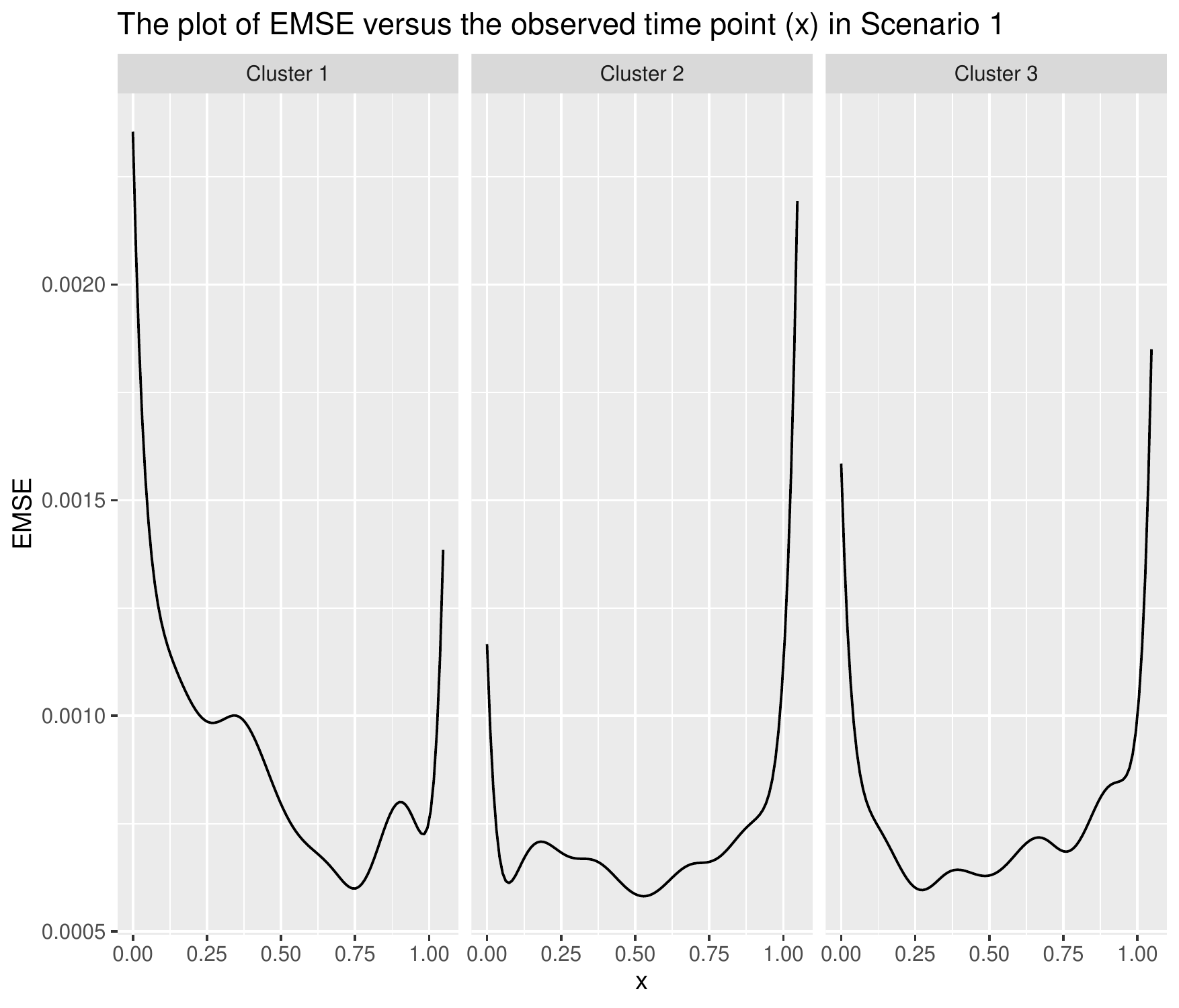}
\caption{\textit{Simulation results for Model 1.} Empirical mean squared error (EMSE) versus each evaluation point $x$ for each cluster in Scenario 1.}
\label{EMSE_Scenario_1}
\end{figure}

Table \ref{EMISE_table} presents the EMISE for each cluster in each Scenario. We can observe small EMISE values, which are consistent with the results shown in Figure \ref{Performance_fig}, where there is a small difference between the red curves (i.e., the estimated mean functions) and the black curves (i.e., the true mean functions). A plot of EMSE values versus observed points for each cluster in Scenario 1 is presented in Figure \ref{EMSE_Scenario_1} while plots of EMSE values for Scenarios 2, 3, 4, 5 and 6 are provided in Figure \ref{EMSE_others} in Appendix \ref{secA2}.

\begin{table}[h]
\caption{\textit{Simulation results for Model 1.} The empirical mean integrated squared error (EMISE) for the estimated mean curve in each cluster in each scenario.}
\label{EMISE_table}
\begin{center}
\begin{minipage}{250pt}
\begin{tabular}{cccccc}
\toprule
Scenario           & Cluster & EMISE   & Scenario           & Cluster & EMISE   \\ \hline
\multirow{3}{*}{1} & 1       & 0.00096 & \multirow{3}{*}{2} & 1       & 0.00164 \\ \cline{2-3} \cline{5-6} 
                   & 2       & 0.00077 &                    & 2       & 0.00246 \\ \cline{2-3} \cline{5-6} 
                   & 3       & 0.00080 &                    & 3       & 0.00169 \\ \cline{2-6} 
\multirow{3}{*}{3} & 1       & 0.00031 & \multirow{3}{*}{4} & 1       & 0.00023 \\ \cline{2-3} \cline{5-6} 
                   & 2       & 0.00045 &                    & 2       & 0.00034 \\ \cline{2-3} \cline{5-6} 
                   & 3       & 0.00042 &                    & 3       & 0.00033 \\ \cline{2-6} 
\multirow{4}{*}{5} & 1       & 0.00001 & \multirow{4}{*}{6} & 1       & 0.00076 \\ \cline{2-3} \cline{5-6} 
                   & 2       & 0.00114 &                    & 2       & 0.00419 \\ \cline{2-3} \cline{5-6} 
                   & 3       & 0.00022 &                    & 3       & 0.00472 \\ \cline{2-3} \cline{5-6} 
                   &         &         &                    & 4       & 0.00130 \\
\botrule
\end{tabular}
\end{minipage}
\end{center}
\end{table}

\subsubsection{Prior sensitivity analysis}
In Bayesian analysis, it is important to assess the effects of different prior settings in the posterior estimation. In this section, we carry out a sensitivity analysis on how different prior settings may affect the results of our proposed VB algorithm. Our sensitivity analysis focuses on the prior distribution of the coefficients $\bphi_k$ of the B-spline basis expansion of each cluster-specific mean curve. We assume $\bphi_k$ follows a  multivariate normal prior distribution with a mean vector $\mathbf{m}_k^0 $ and $s^0\matr{I}$ as the covariance matrix. We simulated data according to Scenario 3 in Section \ref{sec:scenarios} and four different prior settings as follows:
\begin{itemize}
\setlength\itemsep{0.2cm}
    \item Setting 1: use the true coefficients as the prior mean vector and consider a small variance ($s^0=0.01$). 
    \item Setting 2: use the true coefficients as the prior mean vector but consider a larger variance than in Setting 1 ($s^0=1$).
    \item Setting 3: use a prior mean vector that is different than the true vector of coefficients with a small variance ($s^0=0.01$).
    \item Setting 4: set the prior mean vector of coefficients to a vector of zeros with a small variance ($s^0=0.01$).
\end{itemize}

Setting 1 has the strongest prior information among these four prior settings, while setting 4 is the most non-informative prior case. In setting 3, the prior mean vector of coefficients is generated from sampling from a multivariate normal distribution with a mean vector corresponding to the true coefficients and  covariance matrix $\sigma^2\mathbf{I}$, with $\sigma^2 =0.5$. For each prior setting, we simulate 50 datasets as in Scenario 3, obtaining the average mismatch rate and V-measure, which are displayed in Table \ref{sensitivity_ana}. First, we can observe that all the curves are correctly clustered under Setting 1, which has the strongest prior information. Then, as we relax the prior assumptions in two possible directions (i.e., more considerable variance or less informative mean vector), the mismatch rate increases, and the V-measure decreases. However, the clustering performance does not decrease much, only 4.67\% higher in mismatches and 3.73\% lower in V-measure. 

\begin{table}[h]
\begin{center}
\begin{minipage}{\textwidth}
\caption{\textit{Simulation results for Model 1.} Mean mismatch rate and V-measure value from prior sensitivity analysis in Scenario 3}\label{sensitivity_ana}
\begin{tabular*}{\textwidth}
{@{\extracolsep{\fill}}ccccc@{\extracolsep{\fill}}}
\toprule%
Setting & 1  & 2 & 3 & 4\\
\midrule
M\footnotemark[1]  & 0.0000 & 0.0067 & 0.0067 & 0.0467 \\ 
V\footnotemark[2]  & 1.0000 & 0.9947 & 0.9947 & 0.9627 \\
\botrule
\end{tabular*}
\footnotetext[1]{M: mean mismatch rate from 50 runs.}
\footnotetext[2]{V: mean V-measure from 50 runs.}
\end{minipage}
\end{center}
\end{table}

\subsection{Simulation study on Model 2}\label{sec:sim_model_2}
\subsubsection{Simulation scenarios}\label{sec:scenarios:2}
We also investigate the performance of our proposed VB algorithm under Model 2 using simulated data. We consider the simulation schemes of Scenario 1 and Scenario 3 in Section \ref{sec:scenarios}, but add a random intercept to each curve, to construct four different scenarios namely Scenario 7, Scenario 8, Scenario 9, and Scenario 10. 
\vspace{0.5cm}

\noindent \textit{\textbf{Scenario 7, $K = 3$:}}

Scenario 7 is constructed based on Scenario 1. The data are simulated as follows.
$$Y_{ik}(t_j)=a_{ik}+b_k+c_k \sin(1.3t_j)+t_j^3+\delta_{ij}; i=1,...,50; j=1,...,100; k=1,2,3,$$
\noindent where $Y_{ik}(t_j)$ denotes the value at point $t_j$ of the $i$th curve from cluster $k$, $a_{ik}\sim N(0, 0.4^2)$, $\delta_{ij}\sim N(0, 0.2^2)$, $b_1=-0.25$, $b_2=1.25$, $b_3=2.50$, $c_1=1/1.3$, $c_2=1/1.2$, and $c_3=1/4$.
\vspace{0.5cm}

\noindent \textit{\textbf{Scenario 8, $K = 3$:}}

Scenario 8 is developed based on Scenario 3. In this scenario, we consider a very small variance for the random intercept which almost resembles the case without a random intercept. Data are generated as follows.
$$Y_{ik}(t_j)=a_{ik} + \sum_{l=1}^{6}B_l(t_j)\phi_{kl}+\delta_{ij}; i=1,...,50; j=1,...,100; k=1,2,3,$$
\noindent where $Y_{ik}(t_j)$ denotes the value at point $t_j$ of the $i$th curve from cluster $k$, $a_{ik}\sim N(0, 0.05^2)$, $\delta_{ij}\sim N(0, 0.4^2)$. The B-spline coefficients, $\phi_{kl}$, remain the same and are presented in Table \ref{Scenario_34_coe}, which are also used in Scenarios 9 and 10. 
\vspace{0.5cm}

\noindent \textit{\textbf{Scenario 9, $K = 3$:}}

Scenario 9 is similar to Scenario 8, but with larger variance for the random intercept but smaller variance for the random error. Data are generated as follows.
$$Y_{ik}(t_j)=a_{ik} + \sum_{l=1}^{6}B_l(t_j)\phi_{kl}+\delta_{ij};  i=1,...,50; j=1,...,100;  k=1,2,3,$$
\noindent where $Y_{ik}(t_j)$ denotes the value at point $t_j$ of the $i$th curve from cluster $k$, $a_{ik}\sim N(0, 0.3^2)$, $\delta_{ij}\sim N(0, 0.15^2)$.
\vspace{0.5cm}

\noindent \textit{\textbf{Scenario 10, $K = 3$:}}

Scenario 10 is similar to Scenario 8, but with larger variance for the random intercept. In this scenario, we use larger variance for the random error compared with that in Scenario 9, indicating a more complex case. Data are generated as follows.
$$Y_{ik}(t_j)=a_{ik} + \sum_{l=1}^{6}B_l(t_j)\phi_{kl}+\delta_{ij}; i=1,...,50; j=1,...,100; k=1,2,3,$$
\noindent where $Y_{ik}(t_j)$ denotes the value at point $t_j$ of the $i$th curve from cluster $k$, $a_{ik}\sim N(0, 0.6^2)$, $\delta_{ij}\sim N(0, 0.4^2)$.
\vspace{0.5cm}

\subsubsection{Simulation results for Model 2}

\begin{figure}[!ht]
\centering
\begin{subfigure}{.5\textwidth}
  \centering
  \includegraphics[height = 5.5cm, width = 6cm]{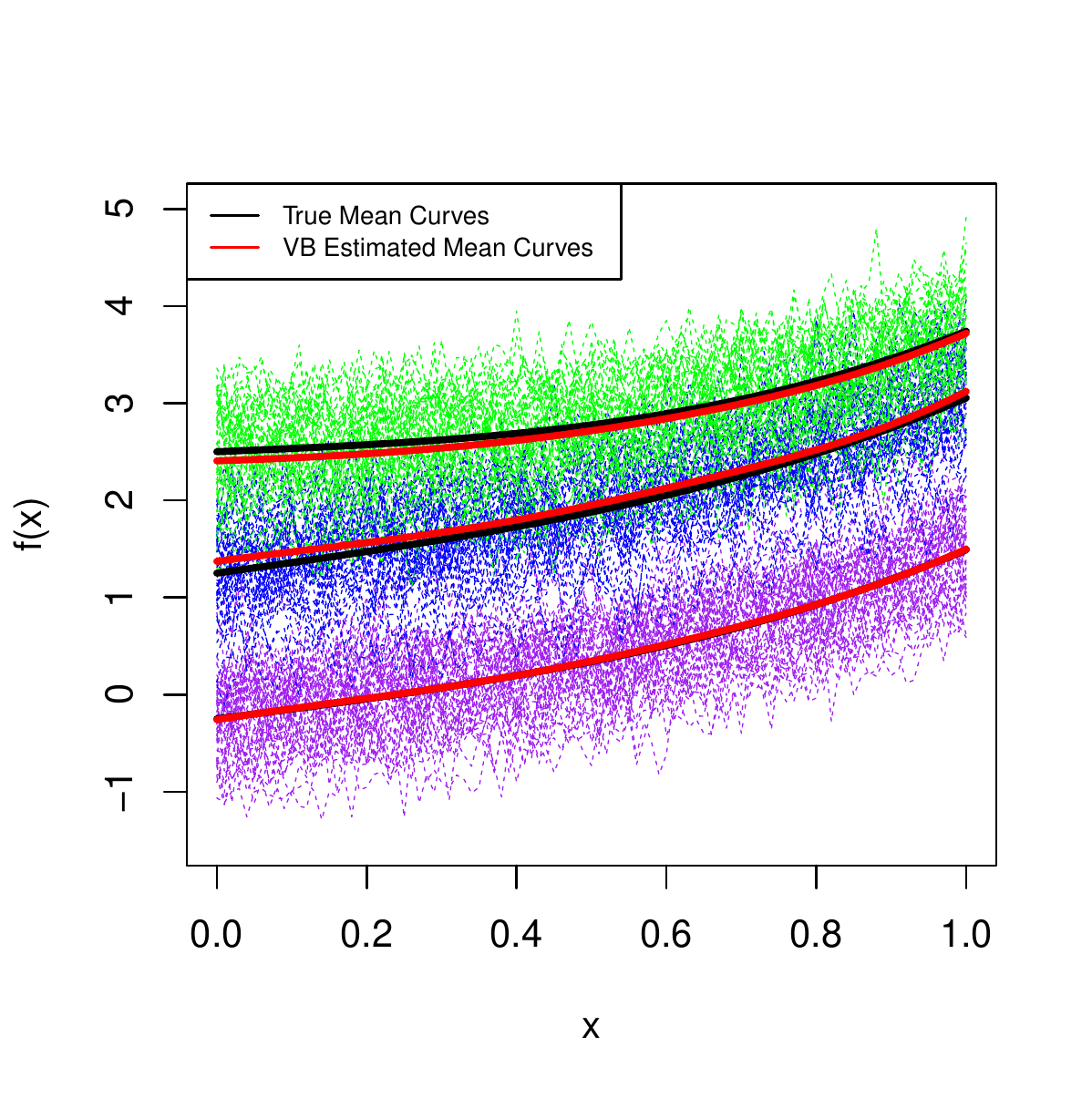}
\end{subfigure}%
\begin{subfigure}{.5\textwidth}
  \centering
  \includegraphics[height = 5.5cm, width = 6cm]{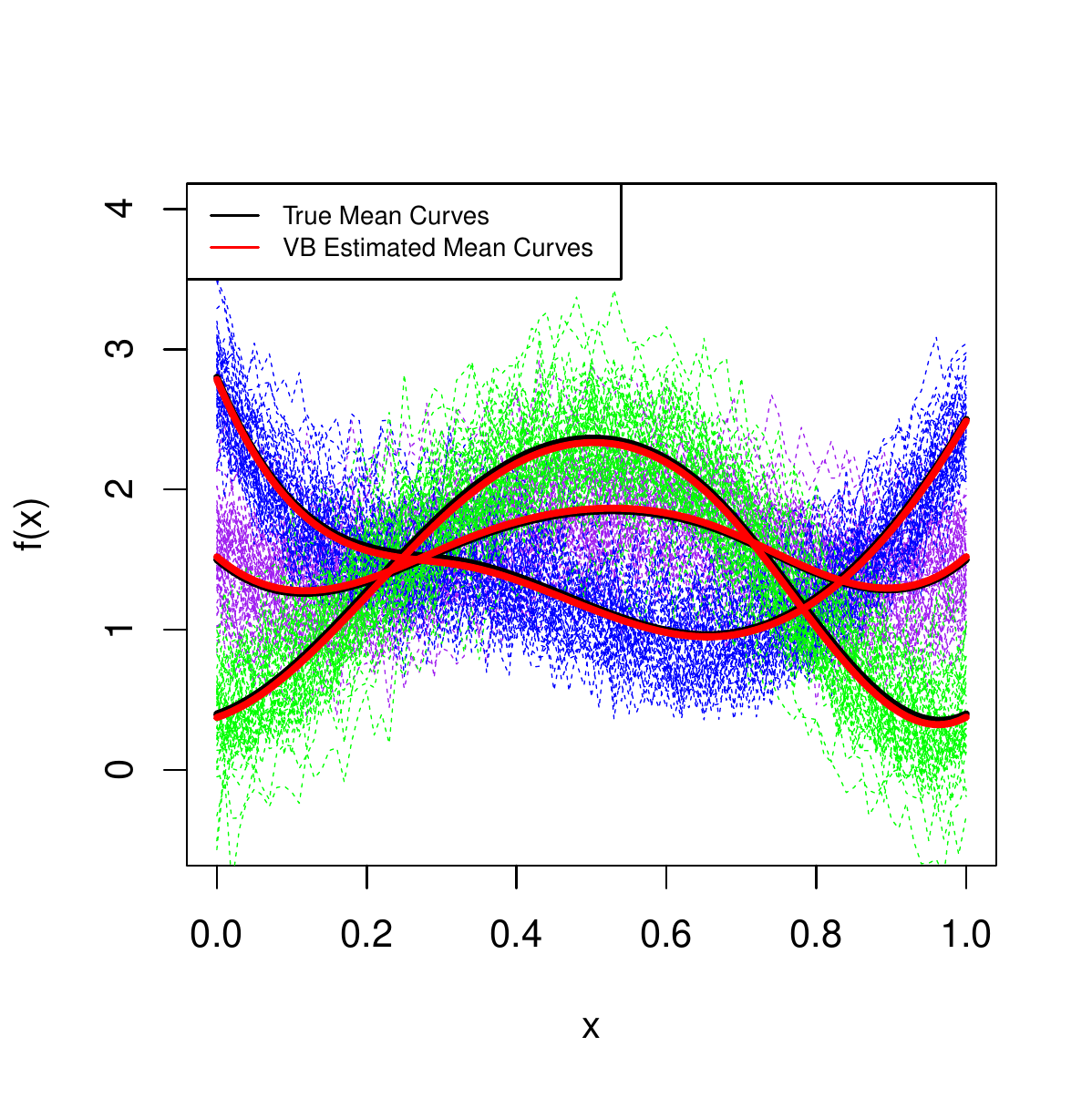}
\end{subfigure} 
\caption{\textit{Simulation results for Model 2.} Example of simulated data under Scenario 7 (left) and Scenario 9 (right). Raw curves (different colors correspond to different clusters), cluster-specific true mean curves (in black) and corresponding estimated mean curves (in red).}
\label{Performance_fig_ext}
\end{figure}

Figure \ref{Performance_fig_ext} shows the curves from one of the 50 simulated datasets for Scenarios 7 and 9. Due to the similarity among Scenarios 8, 9 and 10, the curves for Scenarios 8 and 10 are presented in Figure \ref{Scenario.8.10} of Appendix B. In Figure \ref{Performance_fig_ext}, we can observe a slight difference between each cluster's true mean curve and the estimated mean curve. Furthermore, more variation occurs after adding the random intercept. Especially in Scenario 10, with large variances, there is a more substantial overlap among curves from different clusters, resulting in a more complex scenario for clustering than the corresponding Scenario 3 in Section \ref{sec:sim_model_1}. 

Table \ref{Performance_tab_ext} presents the numerical results, including the mean mismatch rate and the mean V-measure with their corresponding standard deviations from the 50 different simulated datasets under each scenario considered. In Scenario 7, where the true mean curves are relatively parallel, we do not observe a large difference in the mean mismatch rate and the mean V-measure between our model and $k$-means. However, our method still provides a lower mismatch rate and a higher V-measure than $k$-means. In Scenario 8, when the true mean curves cross each other, the mean mismatch rate using our proposed model is 0.0299, much lower than that of using $k$-means, which is 0.1404, while the mean V-measure is 0.9767, higher than that of using $k$-means (0.8937). When the random intercept variance becomes larger in Scenario 9, even with a smaller random error variance, clustering curves via our proposed model becomes more challenging. The mean mismatch rate increases to 0.1453 from 0.0299, while the mean V-measure drops to 0.7923 from 0.9767 in Scenario 8. However, our model still outperforms $k$-means with a difference in mismatch rate of 1.18\% and a difference in mean V-measure of 3.43\%. In Scenario 10, we can see that although the standard deviations of mismatch rates and V-measures are larger than the ones obtained by $k$-means, our extended model provides a much lower mean mismatch rate (lower by 13.31\%) and a much higher mean V-measure (higher by 23.04\%). The larger standard deviations compared to $k$-means happen because, among the 50 different runs, there are 11 runs where our method can 100\% correctly assign each curve to the cluster it belongs to, resulting in a mismatch rate of zero and a V-measure of one. At the same time, there is no run where $k$-means provides such perfect clustering results. Besides, among 50 different runs, there are 41 runs where our method provides lower mismatch rates and higher V-measures than $k$-means.

Table \ref{EMISE_ext_table} shows the EMISE for each cluster in Scenarios 7, 8, 9 and 10 based on Model 2. Small EMISE values once again indicate that the true mean curves and the corresponding curves have a small difference. We also find that compared with Table \ref{EMISE_table} based on Model 1, the EMISE values based on Model 2 are larger. This is in our expectation since adding a random intercept to each curve will bring more variation to the curves, and as a result, more variation in the estimated mean curves, in Scenario 10 especially when we have a larger variance for generating random intercepts. Plots of EMSE values in Scenarios 7, 8, 9, and 10 based on Model 2 are provided in Figure \ref{EMSE_others_ext} in Appendix \ref{secA2}.

For the computational cost, the run times of the proposed VB algorithm of Model 2 for 50 simulated datasets from Scenarios 7, 8, 9, and 10 are 40.96 min, 1.52 min, 10.46 min, and 11.52 min, respectively. We can see that Scenarios 8, 9 and 10 have lower computational cost than that in Scenario 7, since we generate the data with the same structure specified in the proposed model in these scenarios, that is, the data are generated from a linear combination of known B-spline basis functions in Scenarios 8, 9 and 10.

\begin{table}[h]
\begin{center}
\begin{minipage}{\textwidth}
\caption{\textit{Simulation results for Model 2.} Mean mismatch rate and V-measure value under each scenario and compared with $k$-means}\label{Performance_tab_ext}
\begin{tabular*}{\textwidth}{@{\extracolsep{\fill}}ccccc@{\extracolsep{\fill}}}
\toprule%
& \multicolumn{2}{@{}c@{}}{VB algorithm} & \multicolumn{2}{@{}c@{}}{$k$-means} \\\cmidrule{2-3}\cmidrule{4-5}%
Scenario & M\footnotemark[1] (sd\footnotemark[2]) & V\footnotemark[3] (sd) & M (sd)  & V (sd)  \\
\midrule
7 & 0.1045 (0.0265) & 0.7077 (0.0565) & 0.1069 (0.0259) & 0.7033 (0.0505) \\
8 & 0.0299 (0.1040) & 0.9767 (0.0804) & 0.1404 (0.2169) & 0.8937 (0.1641) \\
9 & 0.1453 (0.1485) & 0.7923 (0.1865) & 0.1571 (0.1400) & 0.7580 (0.1793)\\
10 & 0.2493 (0.1416) & 0.6078 (0.2285) & 0.3824 (0.0367) & 0.3774 (0.0581) \\
\botrule
\end{tabular*}
\footnotetext[1]{M: mean mismatch rate from 50 runs.}
\footnotetext[2]{sd: standard deviation.}
\footnotetext[3]{V: mean V-measure from 50 runs.}
\end{minipage}
\end{center}
\end{table}

\begin{table}[h]
\caption{\textit{Simulations results for Model 2}. The empirical mean integrated squared error (EMISE) for the estimated mean curve in each cluster in each scenario.}
\label{EMISE_ext_table}
\begin{center}
\begin{minipage}{250pt}
\begin{tabular}{cccccc}
\toprule
Scenario           & Cluster & EMISE   & Scenario           & Cluster & EMISE   \\ \hline
\multirow{3}{*}{7} & 1       & 0.07666 & \multirow{3}{*}{8} & 1       & 0.00498 \\ \cline{2-3} \cline{5-6} 
                   & 2       & 0.03109 &                    & 2       & 0.00203 \\ \cline{2-3} \cline{5-6} 
                   & 3       & 0.06953 &                    & 3       & 0.00316 \\ \cline{2-6} 
\multirow{3}{*}{9} & 1       & 0.05171 & \multirow{3}{*}{10} & 1       & 0.25312 \\ \cline{2-3} \cline{5-6} 
                   & 2       & 0.01938 &                    & 2       & 0.13287 \\ \cline{2-3} \cline{5-6} 
                   & 3       & 0.02638 &                    & 3       & 0.12465 \\
\botrule
\end{tabular}
\end{minipage}
\end{center}
\end{table}

\section{Application to real data}\label{sec.real}

In this section, we apply our proposed method in Section 2 to the growth and the Canadian weather datasets, which are both publicly available in the R package \textit{fda}. 

\begin{figure}[!ht]
\centering
\begin{subfigure}{.5\textwidth}
  \centering
  \includegraphics[height = 5.5cm, width = 6cm]{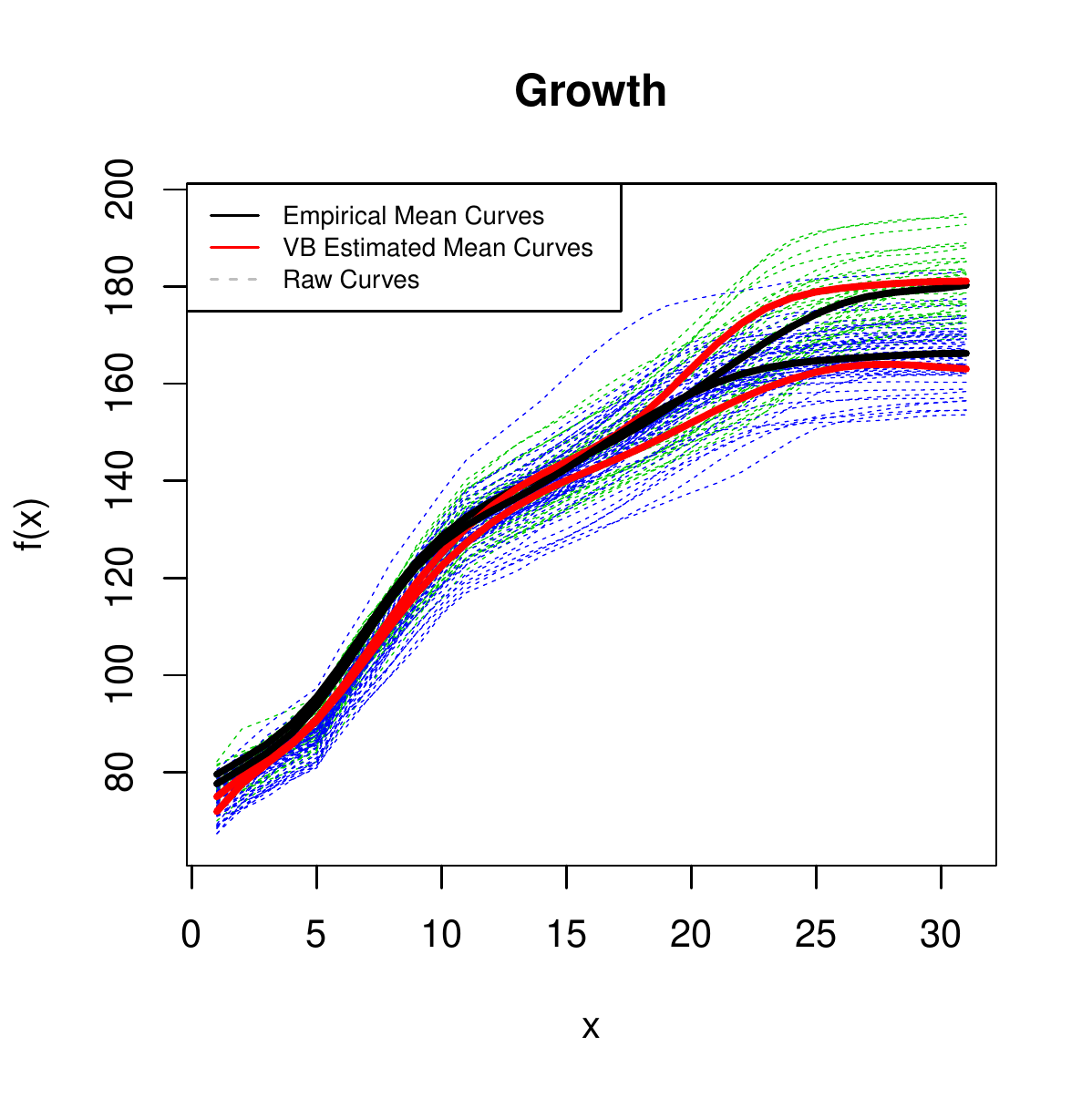}
\end{subfigure}%
\begin{subfigure}{.5\textwidth}
  \centering
  \includegraphics[height = 5.5cm, width = 6cm]{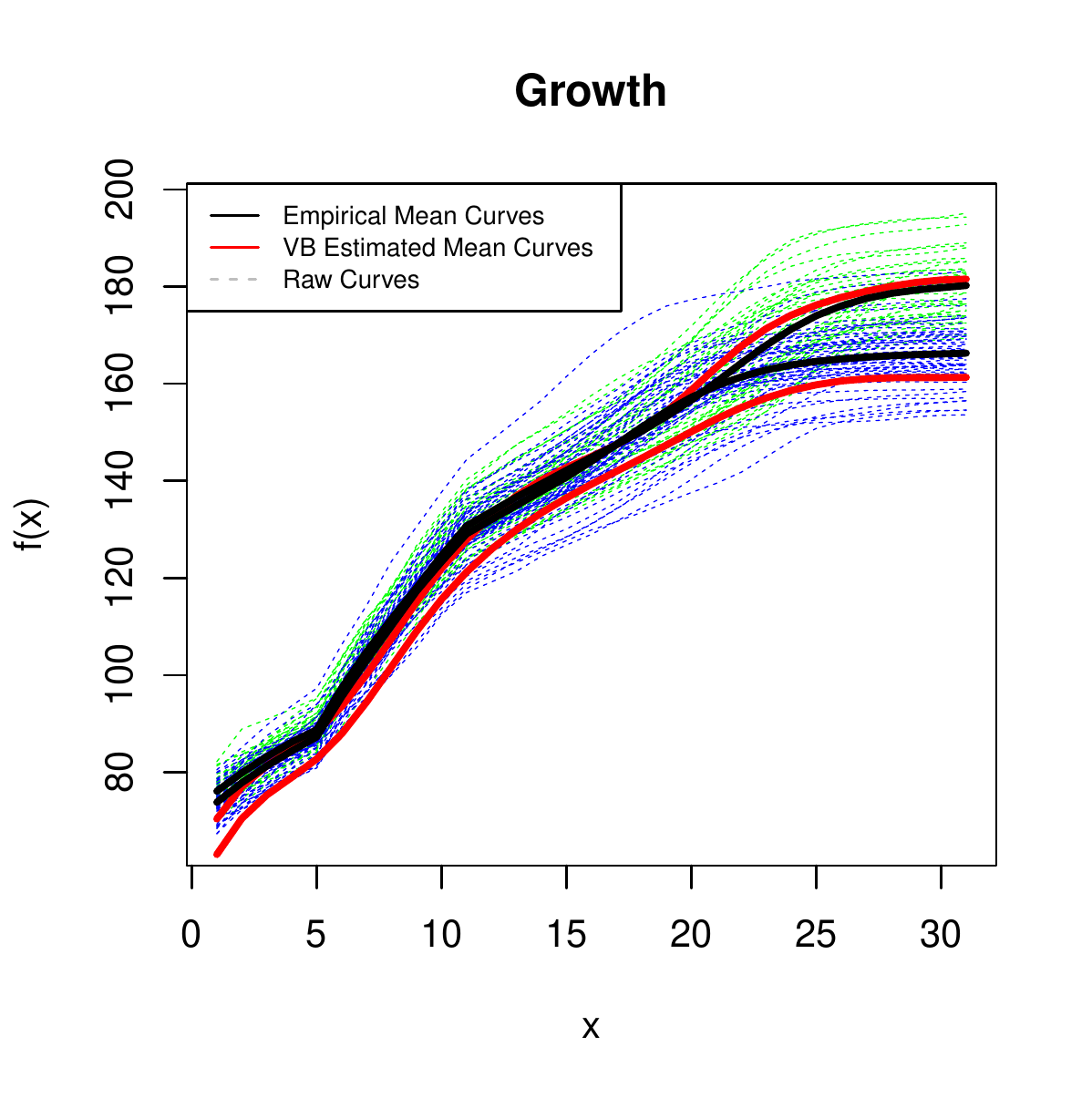}
\end{subfigure} 
\caption{Raw curves (dashed curves) from the Growth dataset where green curves refer to the boys' heights while the blue ones are for the girls', with empirical mean curves (in solid black) and our VB estimated mean curves (in solid red). The left graph is resulted from Model 1 while the right is from Model 2.}
\label{Growth}
\end{figure}

The Growth data \citep{Tuddenham_1954} includes heights (in cm) of the 93 children over 31 unevenly spaced time points from the age of one to eighteen. Raw curves without any smoothing are shown in Figure \ref{Growth}, where the green curves correspond to boys and blue curves to girls. In this case, we apply our proposed method to the growth curves considering two clusters and compare the inferred cluster assignments (boys or girls) to the true ones.  

The Canadian weather data (raw data are presented in Figure \ref{Canadian_raw} in Appendix B) contains the daily temperature at 35 different weather stations (cities) in Canada, averaged out from the year of 1960 to 1994. However, unlike the growth data, we do not know the true number of clusters in the weather data. Therefore, in order to find the best number of clusters, we apply the deviance information criterion (DIC) \citep{Spiegelhalter_2002} for model comparison. DIC is built to balance the model fitness and complexity under a Bayesian framework, and lower DIC indicates a better model. In our Model 1 setting, the DIC can be obtained as follows:
\begin{eqnarray}
 DIC=-4\E_{q^*} \big[\log p(\vecY \vert \vect{Z},\negr{\pi},\negr{\phi},\negr{\tau}) \big] + 2\overline{D},
 \label{eq:dic}
\end{eqnarray} 

\noindent where $\E_{q^*} \big[ \log p(\vecY \vert\vect{Z},\negr{\pi},\negr{\phi},\negr{\tau}) \big]$ can be computed after the convergence of our proposed VB algorithm based on the  ELBO. The term  $\overline{D}$ corresponds to the log-likelihood $\log p(\vecY \vert\vect{Z},\negr{\pi},\negr{\phi},\negr{\tau})$ evaluated at the expected value of each parameter posterior. For example, when we calculate the term $\log \tau_{k}$ in $\log p(\vecY \vert\vect{Z},\negr{\pi},\negr{\phi},\negr{\tau})$, we replace it by $\log \,(\E_{q^*(\tau_k)}(\tau_{k}))$.

In our experiments, ten and six B-spline basis functions are considered for clustering the Growth and Canadian weather curves, respectively. The ELBO convergence threshold is 0.001. It is important to note that we do not have a strong prior knowledge of these real datasets but still need to provide appropriate priors for the VB algorithm. As a solution, we randomly select one underlying curve in each dataset and fit a B-spline regression to obtain a vector of coefficients  which is then modified across different clusters resulting in the prior mean vectors $\vect{m}_k^0$ for  $k=1,...,K$. We set $s^0=0.1$  as the prior variance of these coefficients. For the Dirichlet prior distribution of $\negr{\pi}$, we use $\vect{d}^0=(1/K,...,1/K)$. For the Gamma prior distribution of the precision, $\tau_k=1/\sigma_k^2$, we set $a^0 = 2000$ and $r^0 = 100$ for the growth data, and $a^0 = 1000$ and $r^0 = 800$ for the weather data.

Since we know there are two clusters (boys and girls) in the growth dataset, $K=2$ is preset for the clustering procedure. We apply the proposed VB algorithms under Models 1 and 2 to cluster the growth curves with 50 runs corresponding to 50 different initializations. The classical $k$-means method is also applied to the raw curves for performance comparison purposes. Figure \ref{Growth} presents the estimated mean curves for each cluster corresponding to the the best VB run (the one with maximum ELBO after convergence) along with the empirical mean curves from both models (left graph for Model 1 while right for Model 2). The empirical mean curves are calculated by considering the true clusters and calculating their corresponding point-wise mean at each time point. Some difference between the estimated and the empirical curves can be observed for the girls due to a potential outlier. Regarding clustering performance, the mean mismatch rates for the VB algorithms under Model 1 and Model 2, and $k$-means are 33.33\%, 20.47\% and 34.41\%, respectively. V-measure is more sensitive to misclassification than mismatch rate and, therefore, we obtain low mean V-measure values of 7.75\% for VB under Model 1, 33.75\% for VB under Model 2, and  6.37\% for $k$-means. We can see the clustering performance significantly improved after adding a random intercept to each curve. Compared with Model 1, the mean mismatch rate from Model 2 is lower by 12.86\% , and the mean V-measure is higher by 26\%.

\begin{figure}[!ht]
\centering
\begin{subfigure}{.5\textwidth}
  \centering
  \includegraphics[height = 4.8cm, width = 5cm]{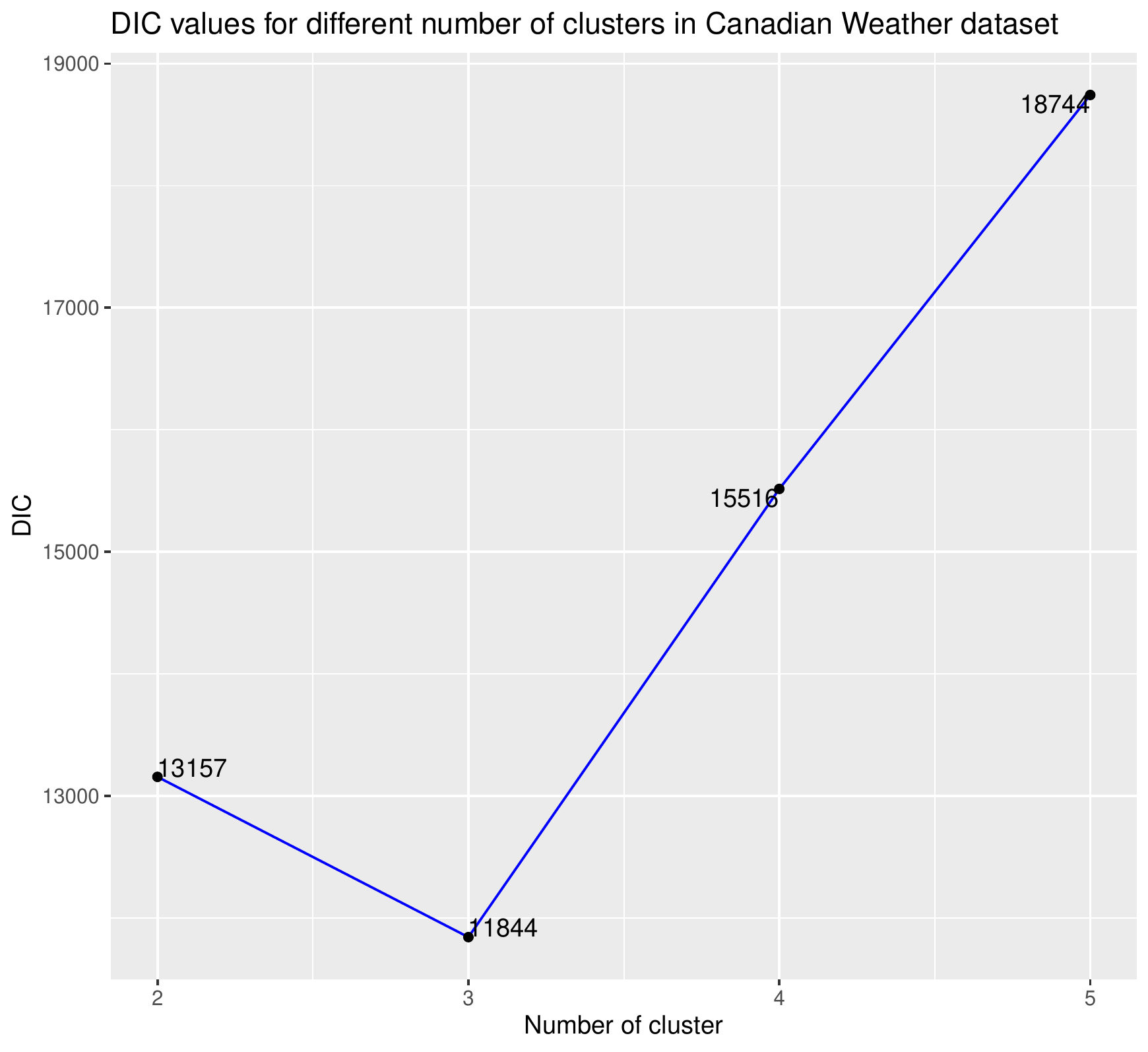}
\end{subfigure}%
\begin{subfigure}{.5\textwidth}
  \centering
  \includegraphics[height =5.8cm, width = 6cm]{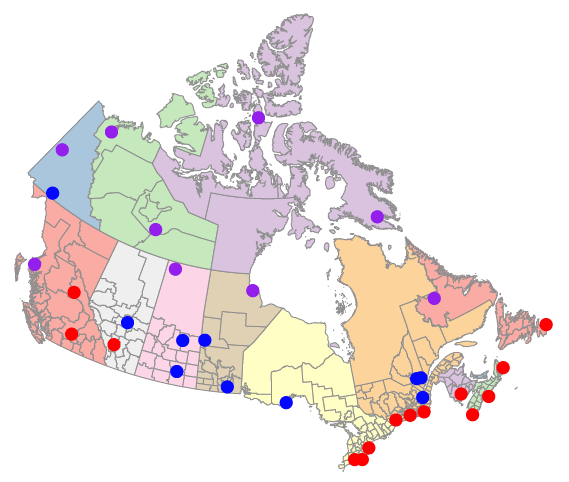}
\end{subfigure}
\caption{Left: DIC values for different clusters ($K=2,3,4,5$) in Canadian weather data. The best number of clusters is three which has the smallest DIC. Right: Clustering results under Model 1 (cities with same color are predicted in the same cluster) for Canadian weather data with preset three clusters ($K=3$).}
\label{CanadianWeather}
\end{figure}

For the Canadian weather dataset analysis, we considered temperature data from all stations except those located in Vancouver and Victoria because they present relatively flat temperature curves compared to other locations. We applied the proposed VB algorithm under Model 1 to the weather data. The left plot in Figure \ref{CanadianWeather} shows the DIC values for different possible numbers of clusters ($K=2,3,4,5$). We can observe that the best number of clusters for separating the Canadian weather data is three, which corresponds to the smallest DIC. Finally, we present the clustering results with $K=3$ on a map of Canada in the right plot in Figure \ref{CanadianWeather}. As can be seen, when $K=3$, we have three resulting groups in three different colors. In general, most of the weather stations in purple are located in northern Canada. In contrast, stations in southern Canada are separated into two groups color-coded in blue and red on the map of Canada. Although some stations may be incorrectly clustered, we can still see a potential pattern that makes sense geographically. 

\section{Conclusion and Discussion}\label{sec.con.dis}
This paper develops a new model-based algorithm to cluster functional data via Bayesian variational inference. We first provide an overview of variational inference, a method used to approximate the posterior distribution under the Bayesian framework through optimization. We then derive a mean-field Variational Bayes (VB) algorithm. Next, the coordinate ascent variational inference is applied to update each term in the variational distribution factorization until convergence of the evidence lower bound. Finally, each observed curve is assigned to the cluster with the largest posterior probability. 

We build our proposed VB algorithm under two different models. In Model 1, we assume the errors are independent, which may be a strong assumption. Therefore, we extended our approach to Model 2, which includes more complex variance-covariance structures by adding a random intercept for each curve.

Simulations and real data analysis suggest that the proposed VB algorithm performs better than the classical $k$-means method in functional data with different patterns. In simulation studies, we also carried out a sensitivity analysis to show that our proposed VB algorithm still has a satisfactory performance across different prior settings. If we know the exact number of groups for real data, a fixed value can be set for $K$ (e.g., $K=2$ in the growth data) before running the algorithm. For cases where we do not know the exact number of clusters, the deviance information criterion (DIC) is adopted to guide the selection of the best number of clusters. 

The main advantage of our proposed VB algorithm is that we model the raw data and obtain clustering assignments and cluster-specific smooth mean curves simultaneously. In other words, compared to some previous methods where researchers first smooth the data and then cluster the data using only the information after smoothing (e.g., the coefficients of B-spline basis functions), our method directly uses the raw data as input, performing smoothing and clustering simultaneously. In addition, we can measure the uncertainty of our proposed clustering using the obtained cluster assignment posterior probabilities.

\backmatter

\bmhead{Acknowledgments}

This research is supported by the Natural Sciences and Engineering Research Council of Canada (NSERC) and the São Paulo Research Foundation (FAPESP).

\section*{Declarations}

\textbf{Conflict of interest} The authors declare that they have no conflict of interest.
\begin{appendices}

\section{VB algorithm for Model 1}\label{secA1}
\subsection{Main steps}
This section describes the main steps of the VB algorithm for inferring $\negr{Z}$, $\bphi$, $\negr{\pi}$ and $\negr{\tau}$ in Model 1 in Section \ref{sec:model:one}, which is summarized in Algorithm \ref{VBalgorithm}. 

\noindent\textbf{\textit{1. VD factorization:}}
\begin{eqnarray}
  q(\vect{Z},\negr{\pi},\negr{\phi},\negr{\tau}) = \prod_{i=1}^N q(Z_i) \times \prod_{k=1}^K q(\bphi_k) \times \prod_{k=1}^K q(\tau_k) \times q(\negr{\pi})  \label{eq:factorization}
\end{eqnarray}

\noindent\textbf{\textit{2. Complete data log-likelihood:}}
\begin{eqnarray}
 \log p(\vect{Y},\vect{Z},\negr{\pi},\negr{\phi},\negr{\tau}) 
&=& \log p(\vecY\vert\vect{Z},\negr{\phi},\negr{\tau}) \; + \; \log p(\vect{Z} \vert \negr{\pi})  \nonumber \\
&& + \log p(\negr{\phi}) \;+\; \log p(\negr{\tau}) \;+\; \log p(\negr{\pi}).
\label{eq:complete}
\end{eqnarray}

\noindent\textbf{\textit{3. Update equations:}}

\textit{i) Update equation for $q(\negr{\pi}$})

Since only the second term, $\log p(\vect{Z} \vert \negr{\pi})$, and the last term, $\log p(\negr{\pi})$, in (\ref{eq:complete}) depend on $\negr{\pi}$, the update equation $q^*(\negr{\pi})$ can be derived as follows.
\begin{eqnarray}
&& {\log q^*(\negr{\pi})} \nonumber \\
  &\addeq& \E_{-\negr{\pi}} \big( \log p(\vect{Y},\vect{Z},\negr{\pi},\negr{\phi},\negr{\tau}) \big) \nonumber \\
 &\addeq & \E_{-\negr{\pi}} \big( \log p(\vect{Z}\vert \negr{\pi}) \big)\; +\; \E_{-\negr{\pi}} \big(\log p(\negr{\pi})\big)  \nonumber \\
 &= & \E_{-\negr{\pi}} \Big[ \sum_{i=1}^N \sum_{k=1}^{K} \I(Z_i =k) \log \pi_k \Big]+  \log p(\negr{\pi})  \nonumber \\
 &\addeq& \sum_{k=1}^{K} \log \pi_k \Big[\sum_{i=1}^N \EZi \Big] + \sum_{k=1}^{K} [d^0_k -1 ]\log \pi_k\nonumber \\
 &=& \sum_{k=1}^{K} \log \pi_k \Big[ \Big( \sum_{i=1}^N \EZi + d^0_k   \Big) -1  \Big] .\nonumber 
\end{eqnarray}
Therefore, $q^*(\negr{\pi})$ is a Dirichlet distribution with parameters $\vect{d}^*=(d_1^*,\ldots,d_K^*)$, where  

\begin{eqnarray}d^{*}_{k}= d^0_k + \sum_{i=1}^N \EZi. 
\label{Eq:qstarPi}
\end{eqnarray}

\vspace{0.4cm}

ii) \textit{Update equation for $q(Z_i)$} 
\begin{eqnarray}
\log q^*(Z_i)  \addeq \E_{-Z_i} \big( \log p(\vect{Y},\vect{Z},\negr{\pi},\negr{\phi},\negr{\tau}) \big)
\label{eq:logqZi}
\end{eqnarray}

When taking the expectation above we just need to consider the first term,  $\log p(\vecY\vert\vect{Z},\negr{\phi},\negr{\tau})$, and the second term, $\log p(\vect{Z} \vert \negr{\pi})$, in (\ref{eq:complete}). Note that we can write $\log p(\vecY\vert\vect{Z},\negr{\phi},\negr{\tau})$ and $\log p(\vect{Z} \vert \negr{\pi})$ into two parts, one that depends on $Z_i$ and one that does not. 
\begin{eqnarray}
\log p(\vecY\vert\vect{Z},\negr{\phi},\negr{\tau})&=& \sum_{k=1}^K \I(Z_i = k) \log p(\vecY_i \vert Z_i =k, \bphi_k,\tau_k) \nonumber \\
&& +\,\sum_{l:l\neq i} \sum_{k=1}^K \I(Z_l=k) \log p(\vecY_l \vert Z_l=k,\negr{\phi}_k,\tau_k)  \nonumber
\end{eqnarray}
\begin{eqnarray}
\log p(\vect{Z} \vert \negr{\pi}) &=& \sum_{k=1}^K \I (Z_i=k) \log \pi_k + \sum_{l:l\neq i} \sum_{k=1}^K \I (Z_l = k) \log \pi_k  \nonumber
\end{eqnarray}
Now when taking the expectation in (\ref{eq:logqZi}) the parts that do not depend on $Z_i$ in $\log p(\vecY\vert\vect{Z},\negr{\phi},\negr{\tau})$ and $\log p(\vect{Z} \vert \negr{\pi})$ in (\ref{eq:complete}) will be added as a constant in the expectation. So, we obtain
\begin{eqnarray}
\log q^*(Z_i) &\addeq& \sum_{k=1}^K I(Z_i =k)\Big \{ \frac{n}{2}\E_{q^*(\tau_k)}(\log \tau_k)   
 \nonumber \\
&& -\frac{1}{2}\E_{q^*(\tau_k)}(\tau_k)\E_{q^*(\negr{\phi}_k)}\big[ (\vecY_i - \matr{B}\negr{\phi}_k)^T(\vecY_i - \matr{B}\negr{\phi}_k) \big]   \nonumber \\
&& + \,\E_{q^*(\negr{\pi})} (\log \pi_k)
\Big\}  \nonumber     
\end{eqnarray}

Therefore, $q^*(Z_i)$ is a categorical distribution with parameters 
\begin{eqnarray}
    p^*_{ik} = \frac{e^{\alpha_{ik}}}{\sum_{k=1}^Ke^{\alpha_{ik}}},
    \label{eq:pikstar}
\end{eqnarray} 

\noindent where 
\begin{eqnarray}
 \alpha_{ik} &=& \frac{n}{2}\E_{q^*(\tau_k)}(\log \tau_k)  -\frac{1}{2}\E_{q^*(\tau_k)}(\tau_k)\E_{q^*(\negr{\phi}_k)}\big[ (\vecY_i \; - \, \matr{B}\negr{\phi}_k)^T(\vecY_i - \matr{B}\negr{\phi}_k) \big] 
 \nonumber \\ 
 &&+ \,\E_{q^*(\negr{\pi})} (\log \pi_k). \nonumber
\end{eqnarray} 

\vspace{0.5cm}
\textit{iii) Update equation for $q(\negr{\phi}_k)$} 

Note that only the first term, $\log p(\vecY\vert\vect{Z},\negr{\phi},\negr{\tau})$, and the third term, $\log p(\negr{\phi})$, in (\ref{eq:complete}) depend on $\bphi_k$. Similarly to the previous case for $q^*(Z_i)$, we can write $\log p(\vecY\vert\vect{Z},\negr{\phi},\negr{\tau})$ and $\log p(\negr{\phi})$ in two parts, one that depends on $\bphi_k$ and the other that does not. Therefore, we obtain
\begin{eqnarray}
\log q^*(\bphi_k)  &\addeq& \E_{-\bphi_k} \big( \log p(\vect{Y},\vect{Z},\negr{\pi},\negr{\phi},\negr{\tau}) \big) \nonumber \\
&\addeq& \frac{n}{2}\E_{q^*(\tau_k)}(\log \tau_k) \sum_{i=1}^N \E_{q^*(Z_i)}[\I(Z_i=k)] \nonumber \\
&& \; - \; \frac{1}{2} \E_{q^*(\tau_k)}(\tau_k)\sum_{i=1}^N \nonumber \\
&&  \Big\{\E_{q^*(Z_i)}[\I(Z_i=k)] (\vecY_i \; - \, \matr{B}\negr{\phi}_k)^T(\vecY_i - \matr{B}\negr{\phi}_k)\Big\} \label{eq:quad1} \\
&& -\frac{M}{2} \log v^0 \; - \;\frac{1}{2}v^0 (\bphi_k -\vect{m}_k^0)^T(\bphi_k -\vect{m}_k^0)
\label{eq:quad2}
\end{eqnarray}

All expectations will be later defined, but note that, for example, $\E_{q^*(Z_i)}[\I(Z_i=k)] = p^*_{ik}$. First, we will focus on the quadratic forms that appear in (\ref{eq:quad1}) and (\ref{eq:quad2}).
\begin{eqnarray}
&& -\frac{1}{2} \E_{q^*(\tau_k)}(\tau_k) \sum_{i=1}^N p^*_{ik} (\vecY_i \; - \, \matr{B}\negr{\phi}_k)^T(\vecY_i - \matr{B}\negr{\phi}_k) \nonumber \\
&& - \;\frac{1}{2}v^0 (\bphi_k -\vect{m}_k^0)^T(\bphi_k -\vect{m}_k^0) = \nonumber \\
&& - \; \frac{1}{2} \E_{q^*(\tau_k)}(\tau_k) \sum_{i=1}^N p^*_{ik}\big[ \vecY_i^T \vecY_i - 2\vecY_i^T\matr{B}\bphi_k + \bphi_k^T\matr{B}^T\matr{B}\bphi_k \big] \nonumber \\
&& -\frac{1}{2}v^0\big[ \bphi^T_k\bphi_k  - 2(\vect{m}_k^0)^T\bphi_k + (\vect{m}_k^0)^T\vect{m}_k^0 \big] \addeq \nonumber \\
&& -\frac{1}{2} \bphi_k^T \Big[v^0\matr{I} + \E_{q^*(\tau_k)}(\tau_k) \sum_{i=1}^N p^*_{ik} \matr{B}^T\matr{B} \Big] \bphi_k \nonumber \\
&& -\,\frac{1}{2}(-2)\Big [v^0(\vect{m}_k^0)^T + \E_{q^*(\tau_k)}(\tau_k) \sum_{i=1}^N p^*_{ik} \vecY_i^T \matr{B}  \Big]\bphi_k 
\label{eq:quad_form}
\end{eqnarray}

Now let 
\begin{eqnarray}
    \matr{\Sigma}^*_k = \Big[v^0\matr{I} + \E_{q^*(\tau_k)}(\tau_k) \sum_{i=1}^N p^*_{ik} \matr{B}^T\matr{B}   \Big]^{-1}.
    \label{eq:sigma_star}
\end{eqnarray} 

\noindent We can then rewrite (\ref{eq:quad_form}) as

$$ -\frac{1}{2} \bphi_k^T \matr{\Sigma}^{*-1}_k \bphi_k -\frac{1}{2}(-2)\Big [v^0(\vect{m}_k^0)^T + \E_{q^*(\tau_k)}(\tau_k) \sum_{i=1}^N p^*_{ik} \vecY_i^T \matr{B} \Big]\matr{\Sigma}^*_k\matr{\Sigma}^{*-1}_k \bphi_k.$$  
\noindent Therefore, $q^*(\bphi_k)$ is $MVN(\vect{m}^*_k,\matr{\Sigma}^*_k)$ with  $\matr{\Sigma}^*_k$ as in (\ref{eq:sigma_star}) and mean vector

\begin{eqnarray}
 \vect{m}^*_k = \big [v^0(\vect{m}_k^0)^T + \E_{q^*(\tau_k)}(\tau_k) \sum_{i=1}^N p^*_{ik} \vecY_i^T \matr{B}  \big]\matr{\Sigma}^*_k. 
 \label{eq:mkstar}
\end{eqnarray}

\textit{iv) Update equation for $q(\tau_k)$}

Similarly to the calculations in i) and ii) we can write
\begin{eqnarray}
 \log q^*(\tau_k) &\addeq& \frac{n}{2}\log \tau_k \sum_{i=1}^N p^*_{ik} -\frac{1}{2}\tau_k\sum_{i=1}^N p^*_{ik}\E_{q^*(\bphi_k)}\Big[ (\vecY_i \; - \, \matr{B}\negr{\phi}_k)^T(\vecY_i - \matr{B}\negr{\phi}_k)\Big]  \nonumber \\
&&+ \; (a^0 -1)\log \tau_k - r^0\tau_k  \nonumber
\end{eqnarray}
Therefore, $q^*(\tau_k)$ is a Gamma distribution with parameters
\begin{eqnarray}
A^*_k = a^0 + \frac{n}{2} \sum_{i=1}^N p^*_{ik} 
\label{eq:Akstar}
\end{eqnarray}
\noindent and
\begin{eqnarray}
R^*_{k} = \Big( r^0 + \frac{1}{2} \sum_{i=1}^N p^*_{ik} \E_{q^*(\bphi_k)}\Big[ (\vecY_i \; - \, \matr{B}\negr{\phi}_k)^T(\vecY_i - \matr{B}\negr{\phi}_k)\Big] \Big).
\label{eq:Rkstar}
\end{eqnarray}

\noindent\textbf{\textit{4. Expectations:}}

Next, we calculate the expectations in the update equations for each component in the VD.
Let  $ \negr{\Psi}$ be the digamma function defined as
\begin{eqnarray}
 \negr{\Psi}(x)=\frac{d}{dx}\log \Gamma(x), 
 \label{eq:digamma}
\end{eqnarray}
\noindent which can be easily calculated via numerical approximation. The values of the expectations taken with respect to the approximated distributions are given as follows.
\begin{eqnarray}\E_{q^*(Z_i)}[\I(Z_i=k)] = p^*_{ik}
\label{eq:EqZstar}
\end{eqnarray}
\begin{eqnarray}\E_{q^*(\tau_k)}(\tau_{k}) = \frac{A^{*}_{k}}{R^{*}_{k}}
\label{eq:Eqtaustar}
\end{eqnarray}
\begin{eqnarray}
\E_{q^*(\tau_k)}(\log \tau_{k}) =  \negr{\Psi}(A^{*}_k) - \log R^{*}_{k}
\label{eq:Eqlogtau}
\end{eqnarray}
\begin{eqnarray}
\E_{q^*(\negr{\pi})}(\log \pi_{k}) =  \negr{\Psi}(d^{*}_{k}) -  \negr{\Psi}\Big(\sum_{k=1}^K d^{*}_{k}\Big)
\label{eq:Eqlogpi}
\end{eqnarray}

\noindent In addition, using the fact that $\E(\vecX^T \vecX) = \mbox{trace}[\mbox{Var}(\vecX)] + \E(\vecX)^T\E(\vecX)$, we obtain
\begin{eqnarray}
 && \E_{q^*(\bphi)}\Big[ (\vecY_i \; - \, \matr{B}\negr{\phi}_k)^T(\vecY_i - \matr{B}\negr{\phi}_k)\Big] \nonumber \\ &=&  \mbox{trace}\big(\matr{B} \matr{\Sigma}^*_k \matr{B}^T \big) +\, (\vecY_i - \matr{B}\vect{m}^*_k)^T(\vecY_i - \matr{B}\vect{m}^*_k).
 \label{eq:Eqphistar}
\end{eqnarray}

\subsection{ELBO calculation}\label{sec:elbo_derivation}

In this section, we show how to calculate the ELBO, which is the convergence criterion of our proposed VB algorithm and will be updated at the end of each iteration until it converges. Equation (\ref{Eq:elbo}) gives the ELBO:

$$
    \mbox{ELBO}(q) = \E_{q^*} \big[ \log p(\vecY ,\vect{Z},\negr{\pi},\negr{\phi},\negr{\tau}) \big] - \E_{q^*} \big[ \log q(\vect{Z},\negr{\pi},\bphi,\negr{\tau}) \big],
$$

\noindent where 
\begin{eqnarray}
    \E_{q^*} \big[ \log p(\vecY ,\vect{Z},\negr{\pi},\negr{\phi},\negr{\tau}) \big] = 
    \E_{q^*} \big[ \log p(\vecY \vert \vect{Z},\negr{\pi},\negr{\phi},\negr{\tau}) \big] + 
    \E_{q^*} \big[ \log p(\vect{Z} \vert \negr{\pi} \big)] + \nonumber \\
    \E_{q^*} \big[ \log p(\negr{\phi})] + 
    \E_{q^*} \big[ \log p(\negr{\tau})] + 
    \E_{q^*} \big[ \log p(\negr{\phi})],  \nonumber
\end{eqnarray}
\noindent and 
\begin{eqnarray}
    \E_{q^*} \big[ \log q(\vect{Z},\negr{\pi},\bphi,\negr{\tau}) \big] &=&  \E_{q^*} \big[ \log q(\vect{Z}) \big] +  
    \E_{q^*} \big[ \log q(\bphi) \big]  \nonumber \\
&& +\, \E_{q^*} \big[ \log q(\negr{\pi}) \big] + 
    \E_{q^*} \big[ \log q(\negr{\tau}) \big]  \nonumber
\end{eqnarray}
\noindent Therefore, we can write the ELBO as the summation of 5 terms: 
\begin{eqnarray}
    \mbox{ELBO}(q) &=& \E_{q^*} \big[ \log p(\vecY \vert \vect{Z},\negr{\pi},\negr{\phi},\negr{\tau}) \big] + diff_{\vect{Z}} +  diff_{\negr{\phi}} \nonumber \\
    && + \,diff_{\negr{\tau}} + diff_{\negr{\pi}}  
    \label{eq:elbo_calc}
\end{eqnarray}
\noindent where, 
$$diff_{\vect{Z}} = \E_{q^*}\big[\log p (\vect{Z} \vert\negr{\pi})\big ]-\E_{q^*}\big[\log q(\vect{Z})\big].$$
\noindent Specifically,
\begin{eqnarray}
 diff_{\vect{Z}} \equiv \sum_{i=1}^{N}\sum_{k=1}^K p^{*}_{ik} \E_{q^*(\negr{\pi})}(\log \pi_k) - \sum_{i=1}^{N}\sum_{k=1}^K p^{*}_{ik} \log p^{*}_{ik}.
 \label{eq:diff_Z}
\end{eqnarray}
The other terms in (\ref{eq:elbo_calc}) are calculated as follows:
$$diff_{\negr{\phi}} \equiv -\frac{1}{2}\sum_{k=1}^K v_k^0\{\mbox{trace}\big(\matr{\Sigma}^*_k \big) + (\vect{m}^*_k-\vect{m}^0_k)^T(\vect{m}^*_k-\vect{m}^0_k)\} +\frac{1}{2}\sum_{k=1}^K \log \vert\matr{\Sigma}^*_k\vert,$$
\begin{eqnarray}
    diff_{\negr{\tau}} &\equiv& \sum_{k=1}^K \{(a^0-1)\E_{q^*(\tau_k)}(\log \tau_{k})-r^0\E_{q^*(\tau_k)}(\tau_{k})\} \nonumber\\
    &&-\, \sum_{k=1}^K\{A^*_k(\log R_k^{*}-1) - \log \Gamma(A^*_k)
\nonumber\\
    && +\,(A^*_k-1)\E_{q^*(\tau_k)}(\log \tau_{k})\},
\label{diff.tau}
\end{eqnarray}
$$diff_{\negr{\pi}} \equiv \sum_{k=1}^K (d_k^0-d_k^{*})\E_{q^*(\negr{\pi})}(\log \pi_{k}),$$

\noindent and 
\begin{eqnarray}
\E_{q^*} \big[ \log p(\vecY \vert\vect{Z},\negr{\pi},\negr{\phi},\negr{\tau}) \big]&=&\sum_{i=1}^{N}\sum_{k=1}^K p^{*}_{ik}\Big\{\frac{n}{2}\E_{q^*(\tau_k)}(\log \tau_{k}) -\frac{1}{2}\frac{A_k^{*}}{R_k^{*}} \nonumber \\
&& \E_{q^*(\bphi)}\Big[ (\vecY_i \; - \, \matr{B}\negr{\phi}_k)^T(\vecY_i - \matr{B}\negr{\phi}_k)\Big]\Big\}.    
\end{eqnarray}
\noindent Therefore, at iteration $c$, we calculate $\mbox{ELBO}^{(c)}$ using all parameters obtained at the end of iteration $c$. Convergence of the algorithm is achieved if $\mbox{ELBO}^{(c)}-\mbox{ELBO}^{(c-1)}$ is smaller than a given threshold. It is important to note that we use the fact that $\displaystyle \lim_{p^{*}_{ik} \to 0} p^{*}_{ik}\log p^{*}_{ik}=0$ to avoid numerical issues when calculating (\ref{eq:diff_Z}). 
\vspace{0.2cm}

\begin{algorithm}[H]
  \KwData{$N$ original curves with $n$ evaluation points; number of clusters $K$; values of hyperparameters: $\vect{d}^0$, $\vect{m}_k^0, k=1,...,K$, $s^0$, $a^0$, $r^0$; convergence threshold and maximum number of iterations}
  \KwResult{VB estimated mean curves for each cluster and the cluster index for each original curve}
  \textbf{Initialization}: initialize $R_{k}^*$ with arbitrary values (e.g., $R_{k}^*= r^0$) and $p^{*}_{ik}$ from $k$-means, and set $c=0$\;
  \While{$c<$ maximum number of iterations and difference of ELBO $>$ convergence threshold}{
    \Repeat{maximum iteration is achieved or the ELBO converges}{
      $c=c+1$\;
      update $A^{*(c)}_k$ using $p^{*(c-1)}_{1k},\ldots,p^{*(c-1)}_{Nk}$ with equation (\ref{eq:Akstar})\;
      update $\matr{\Sigma}^{*(c)}_k$ using $A^{*(c)}_k$, $R_{k}^{*(c-1)}$ and $p^{*(c-1)}_{1k},\ldots,p^{*(c-1)}_{Nk} $ with equations (\ref{eq:sigma_star}) and (\ref{eq:Eqtaustar})\;
      update $\vect{m}^{*(c)}_k$ using $\matr{\Sigma}^{*(c)}_k$, $A^{*(c)}_k$, $R_{k}^{*(c-1)}$ and $p^{*(c-1)}_{1k},\ldots,p^{*(c-1)}_{Nk} $ with equations (\ref{eq:mkstar}) and (\ref{eq:Eqtaustar})\;
      update $R_k^{*(c)}$ using $\vect{m}^{*(c)}_k$, $\matr{\Sigma}^{*(c)}_k$ and $p^{*(c-1)}_{1k},\ldots,p^{*(c-1)}_{Nk} $ with equations (\ref{eq:Rkstar}) and (\ref{eq:Eqphistar})\;
      update $\negr{d}^{*(c)}$ using $p^{*(c-1)}_{1k},\ldots,p^{*(c-1)}_{Nk} $ with equations (\ref{Eq:qstarPi}) and (\ref{eq:EqZstar})\;
      update $p^{*(c)}_{1k},\ldots,p^{*(c)}_{Nk} $ using $R_k^{*(c)}$ , $\negr{d}^{*(c)}$, $\vect{m}^{*(c)}_k$ and $\matr{\Sigma}^{*(c)}_k$ with equations (\ref{eq:pikstar}), (\ref{eq:Eqtaustar}), (\ref{eq:Eqlogtau}), (\ref{eq:Eqlogpi}) and (\ref{eq:Eqphistar})\;
      calculate the current ELBO, $\text{ELBO}^{(c)}$ using formulas in section \ref{sec:elbo_derivation}\;
      calculate the difference of ELBO $=\text{ELBO}^{(c)}-\text{ELBO}^{(c-1)}$\;
    }
  }
  \caption{Clustering functional data via variational inference}
  \label{VBalgorithm}
\end{algorithm}

\section{Plots}\label{secA2}

\begin{figure}[!ht]
\centering
\begin{subfigure}{.5\textwidth}
  \centering
  \includegraphics[height = 4.5cm, width = 5.5cm]{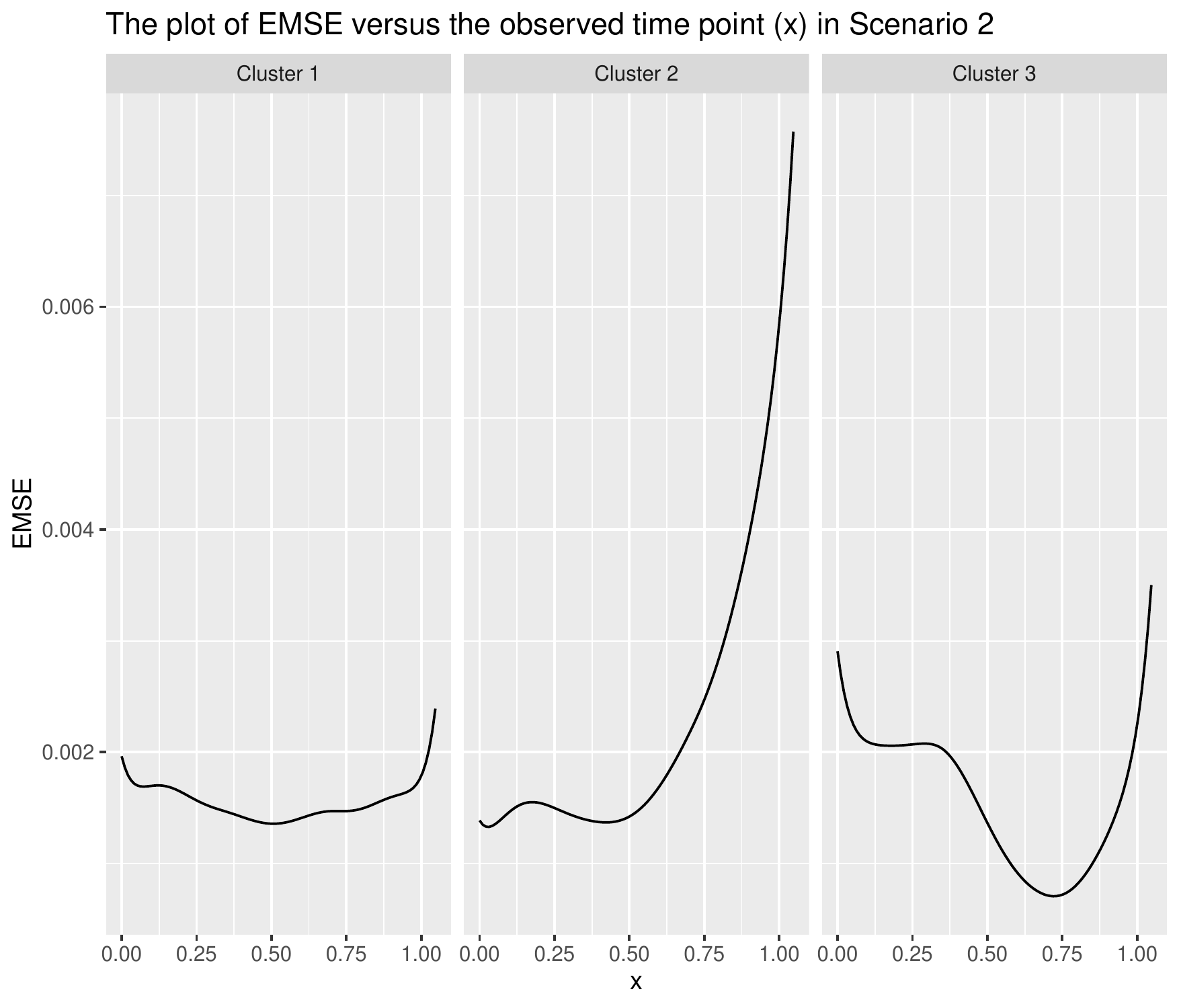}
\end{subfigure}%
\begin{subfigure}{.5\textwidth}
  \centering
  \includegraphics[height = 4.5cm, width = 5.5cm]{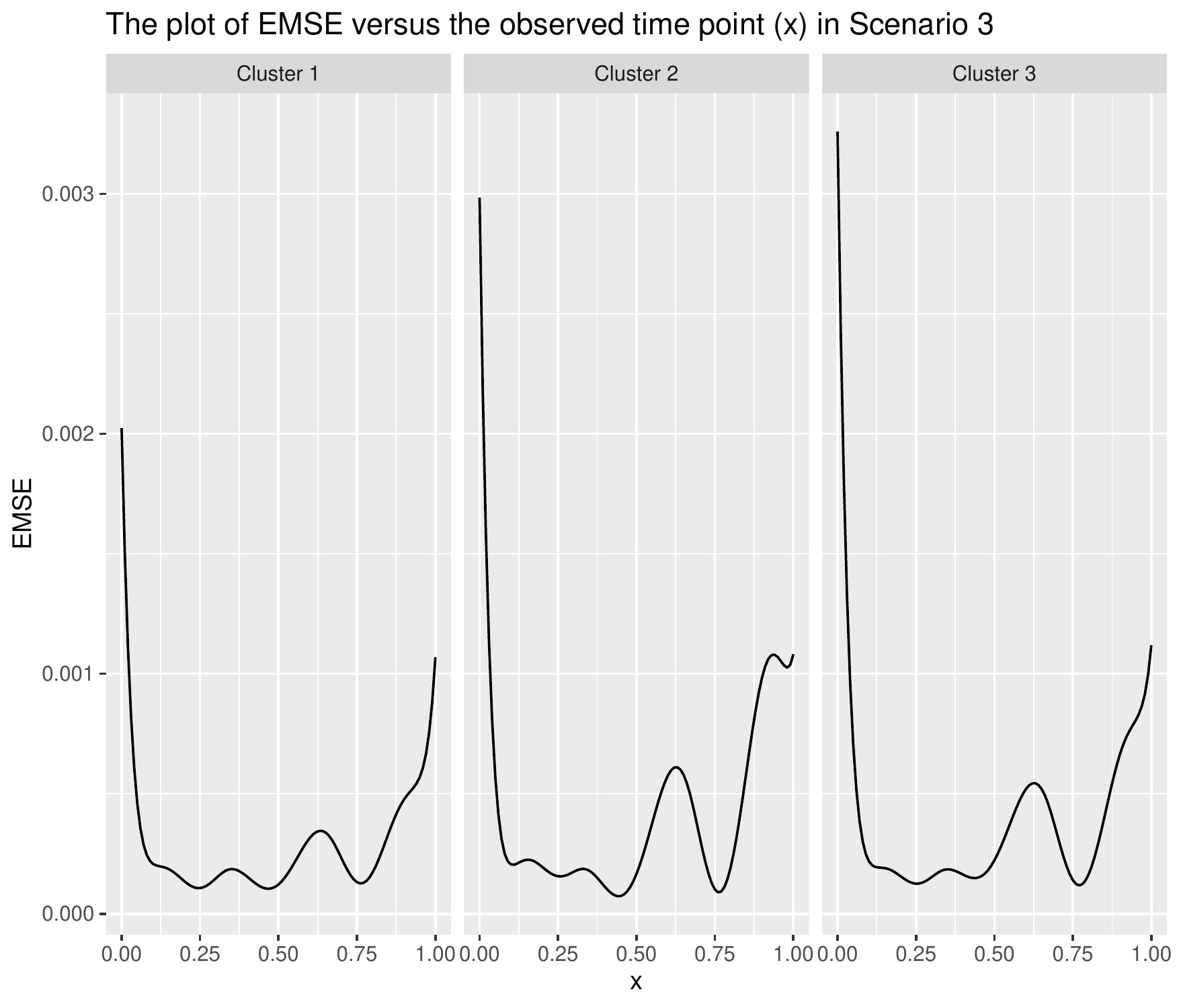}
\end{subfigure} 
\medskip
\begin{subfigure}{.5\textwidth}
  \centering
  \includegraphics[height = 4.5cm, width = 5.5cm]{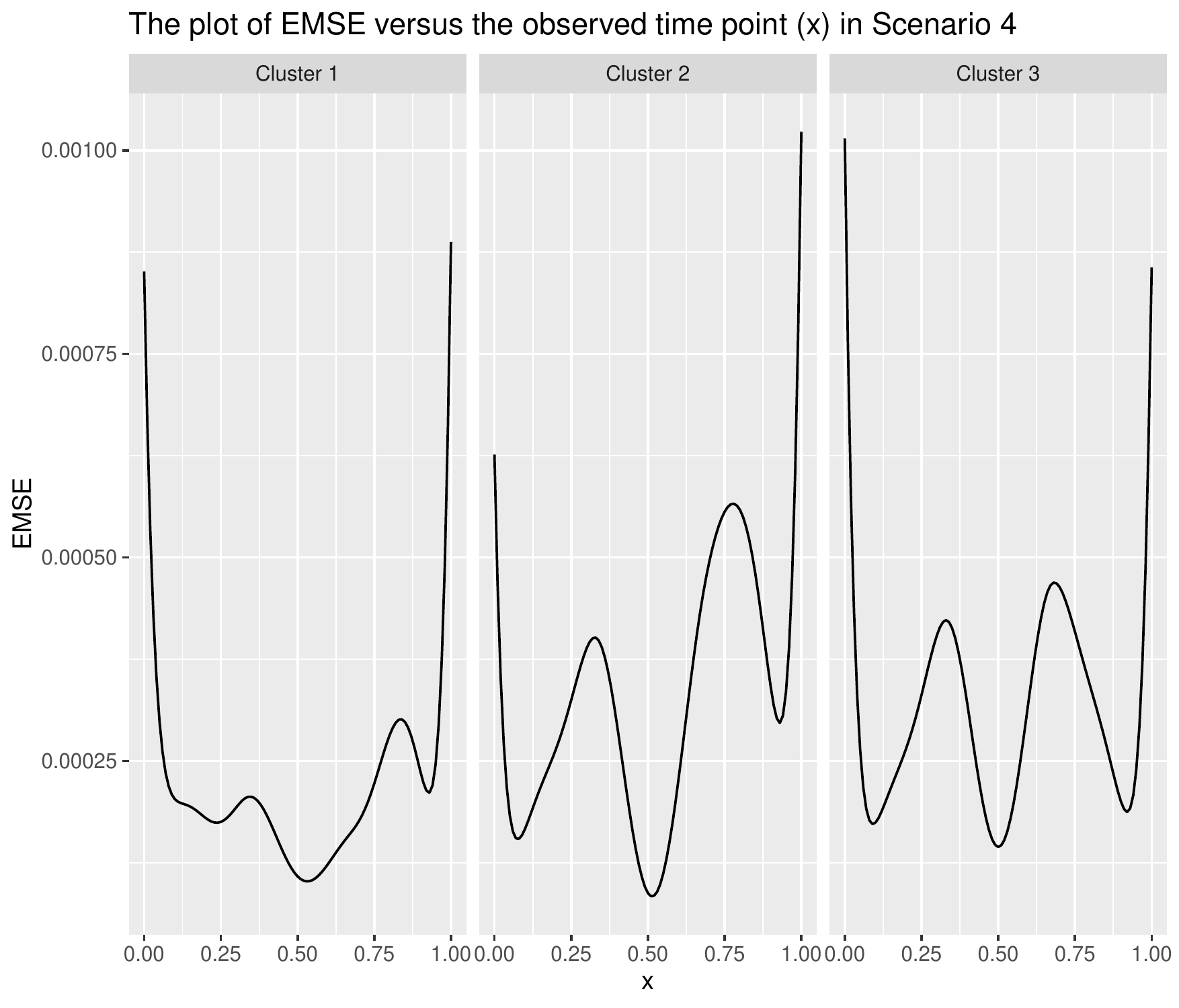}
\end{subfigure}%
\begin{subfigure}{.5\textwidth}
  \centering
  \includegraphics[height = 4.5cm, width = 5.5cm]{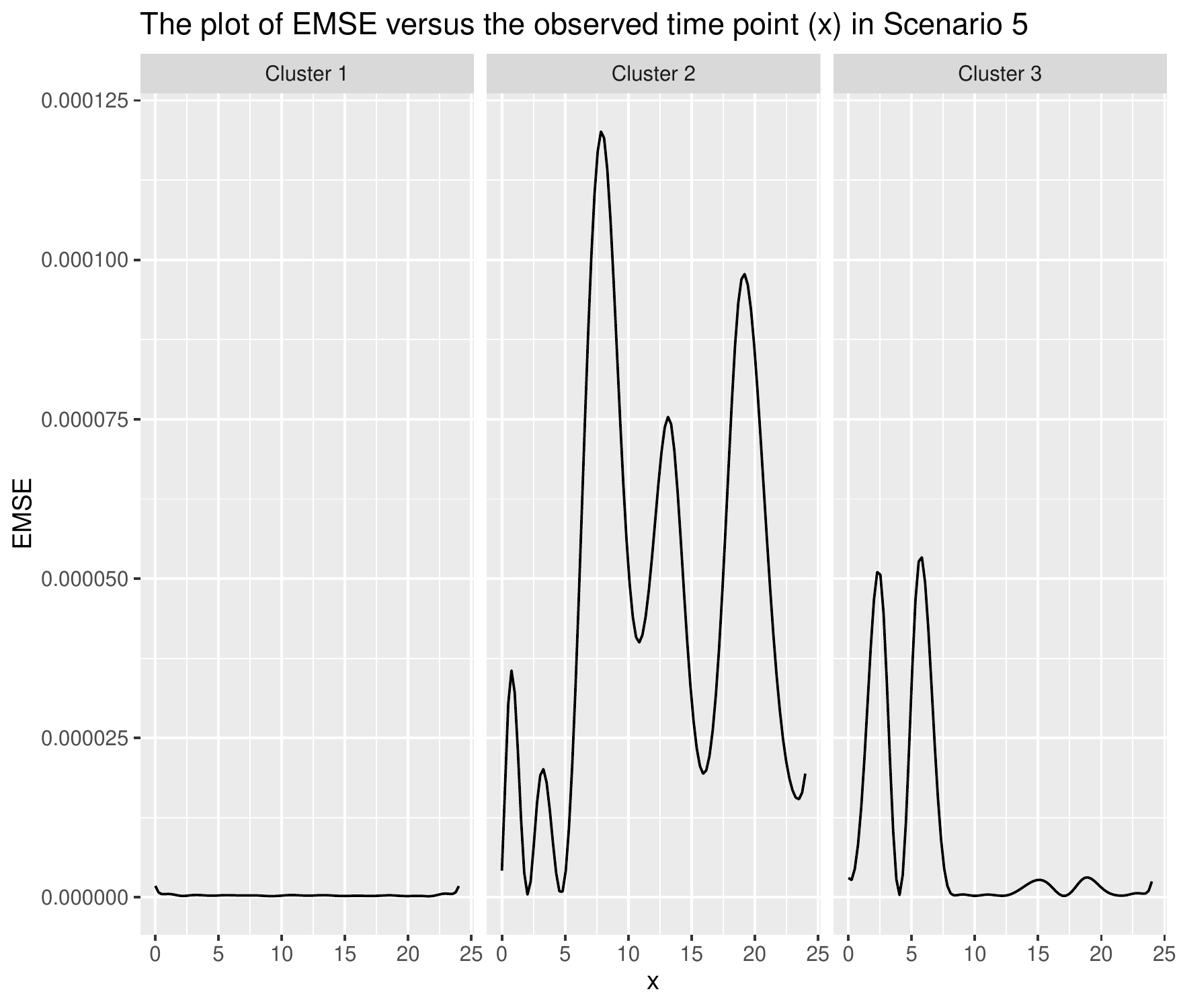}
\end{subfigure} 
\medskip
\begin{subfigure}{.5\textwidth}
  \centering
  \includegraphics[height = 4.5cm, width = 5.5cm]{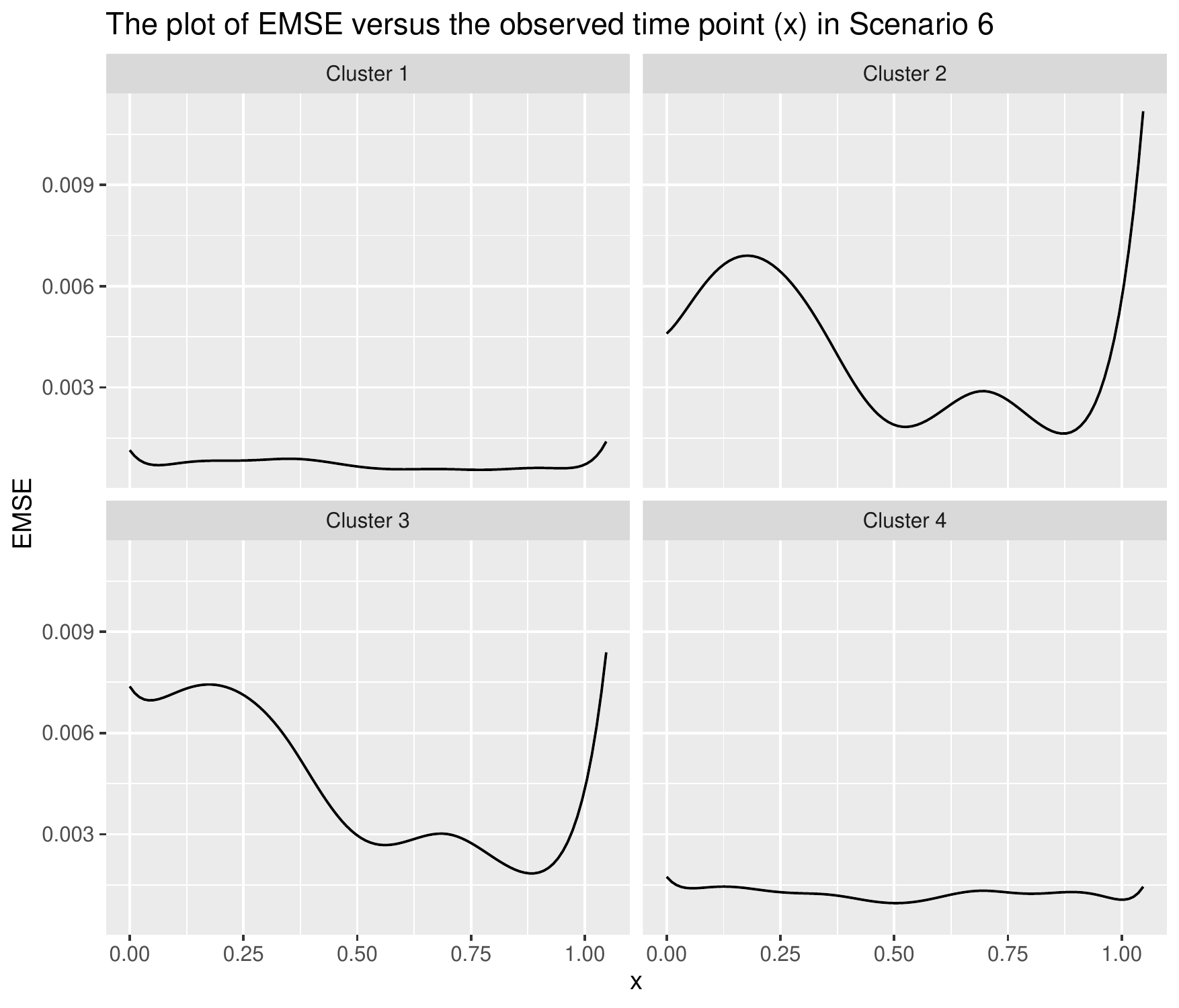}
\end{subfigure}%
\begin{subfigure}{.5\textwidth}
  \centering
  \includegraphics[height = 4.5cm, width = 5.5cm]{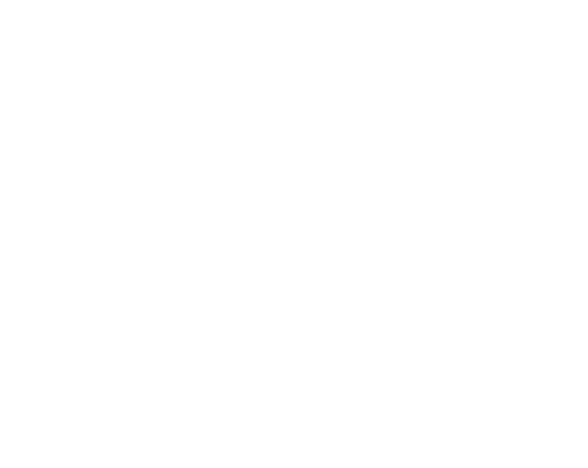}
\end{subfigure}%
\caption{EMSE versus the observed evaluation point for each cluster in Scenarios 2, 3, 4, 5 and 6. In Scenario 5, the straight line in cluster one does not mean there is no EMSE. This is because compared to cluster two and three, the EMSE in cluster one is very small (the median is $1.41\times 10^{-11}$).}
\label{EMSE_others}
\end{figure}

\begin{figure}[!ht]
\centering
\begin{subfigure}{.5\textwidth}
  \centering
  \includegraphics[height = 5.6cm, width = 6cm]{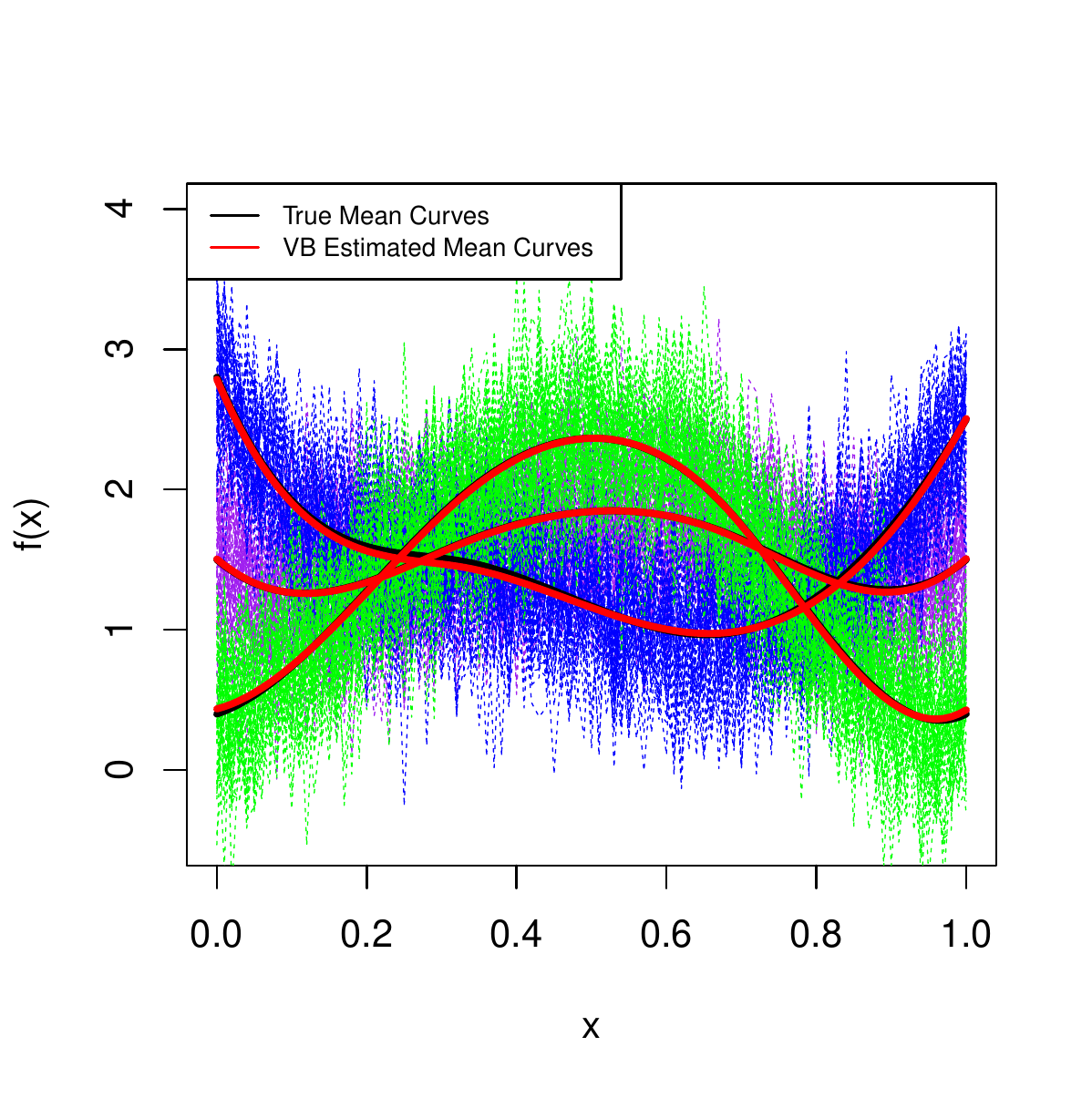}
\end{subfigure}%
\begin{subfigure}{.5\textwidth}
  \centering
  \includegraphics[height = 5.4cm, width = 5.8cm]{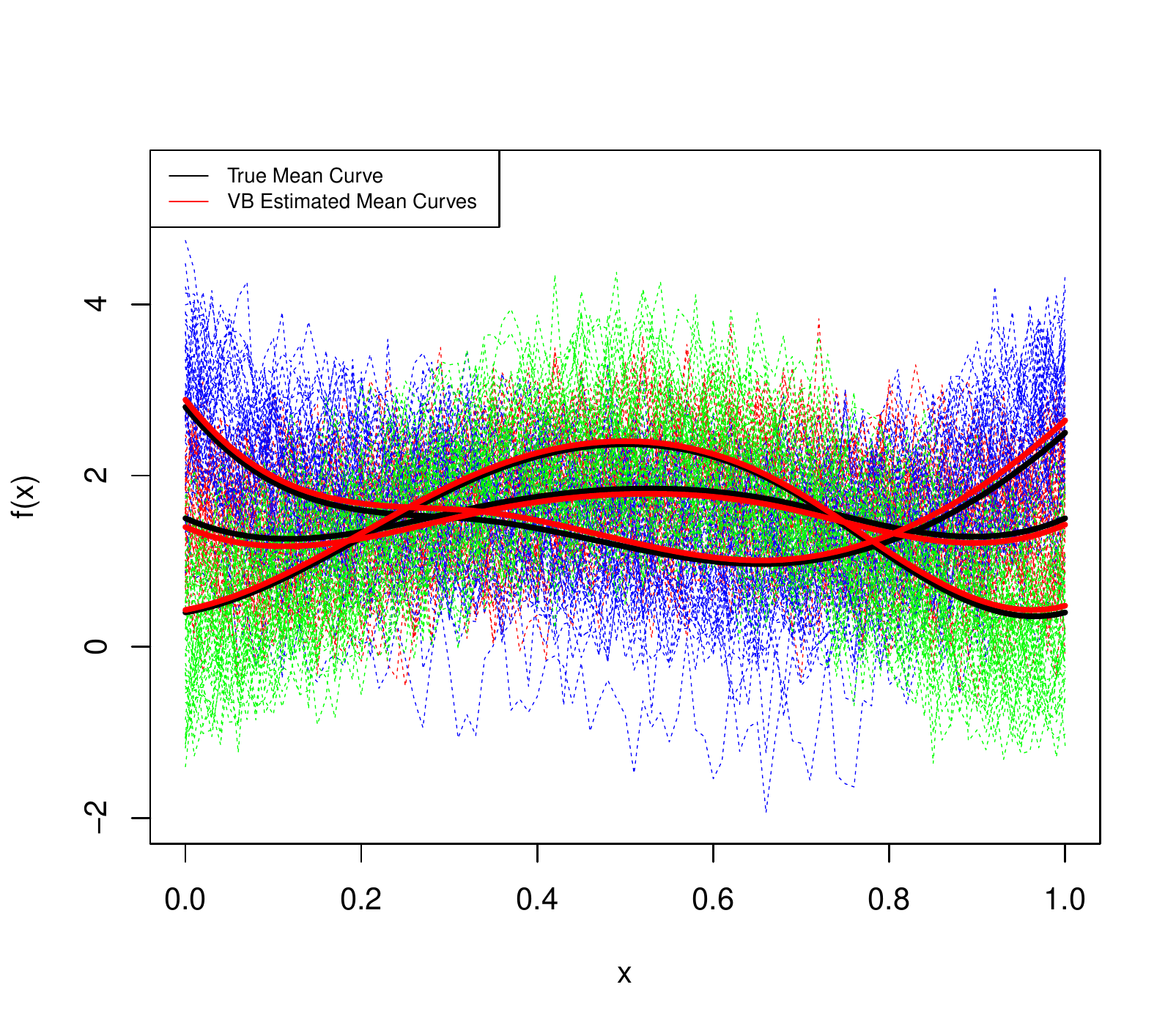}
\end{subfigure} 
\caption{Example of simulated data under Scenario 8 (left) and Scenario 10 (right) for Model 2. Raw curves (different colors correspond to different clusters), cluster-specific true mean curves (in black) and corresponding estimated mean curves (in red).}
\label{Scenario.8.10}
\end{figure}

\begin{figure}[!ht]
\centering
\begin{subfigure}{.5\textwidth}
  \centering
  \includegraphics[height = 4.5cm, width = 5.5cm]{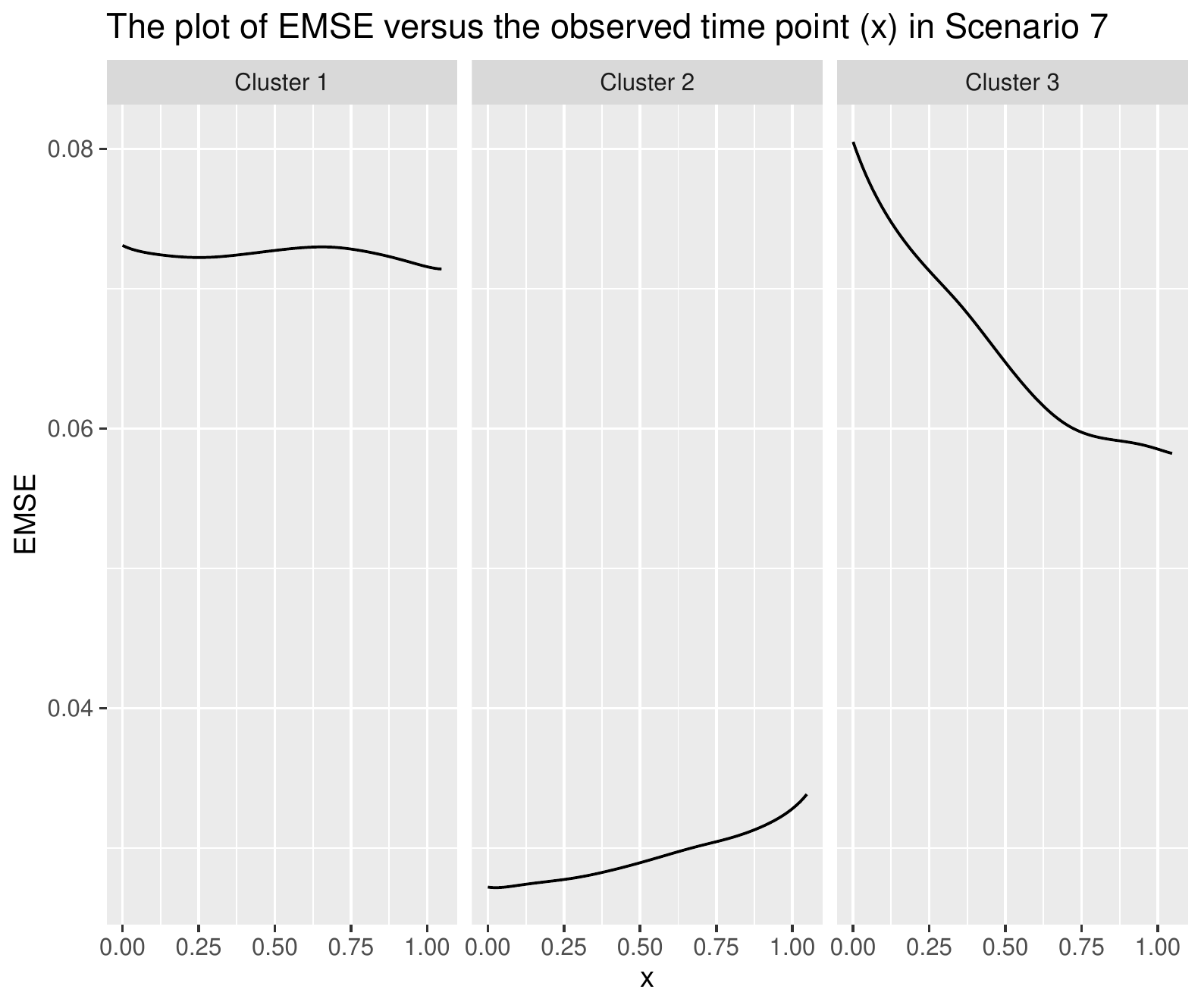}
\end{subfigure}%
\begin{subfigure}{.5\textwidth}
  \centering
  \includegraphics[height = 4.5cm, width = 5.5cm]{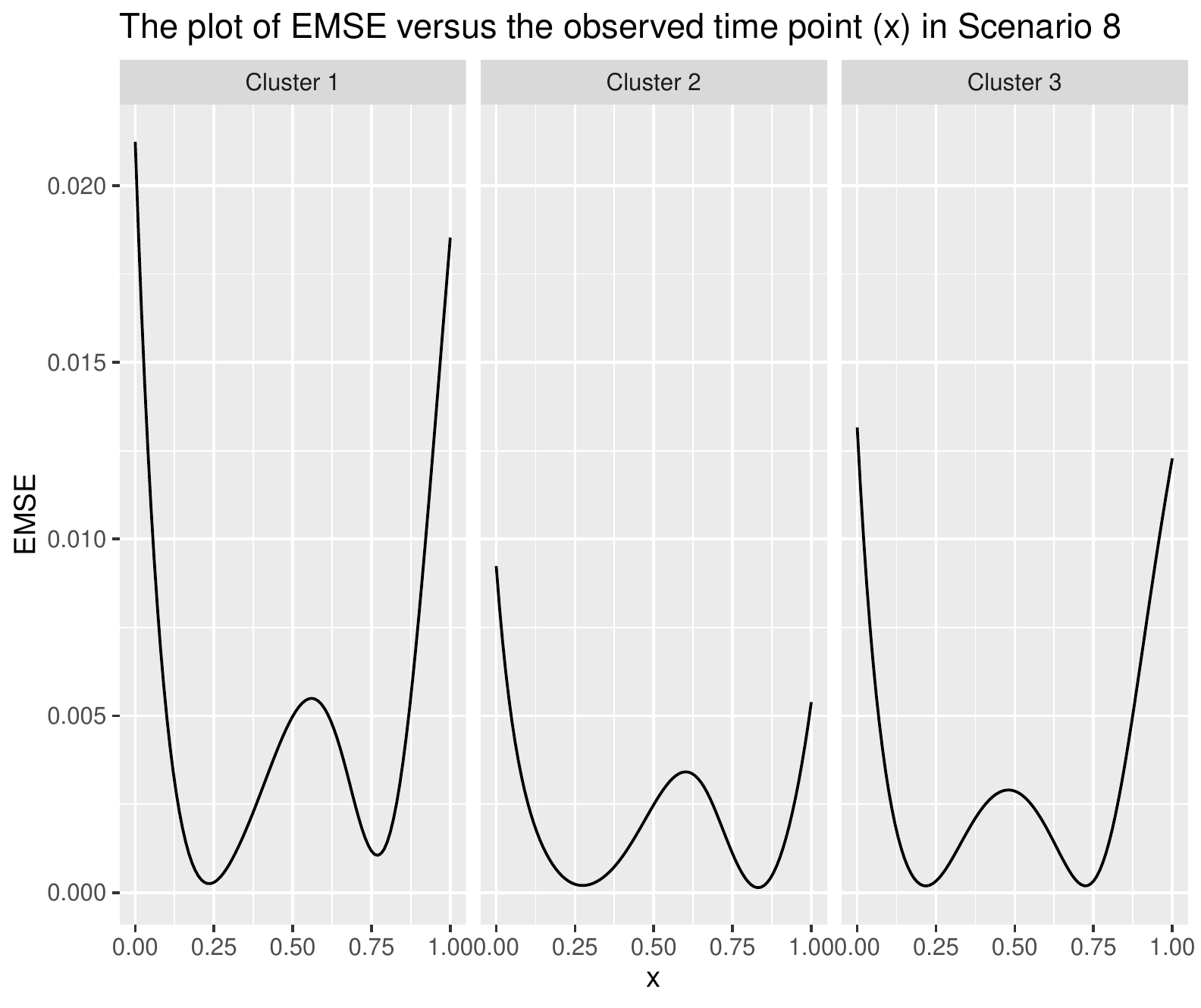}
\end{subfigure} 
\medskip
\begin{subfigure}{.5\textwidth}
  \centering
  \includegraphics[height = 4.5cm, width = 5.5cm]{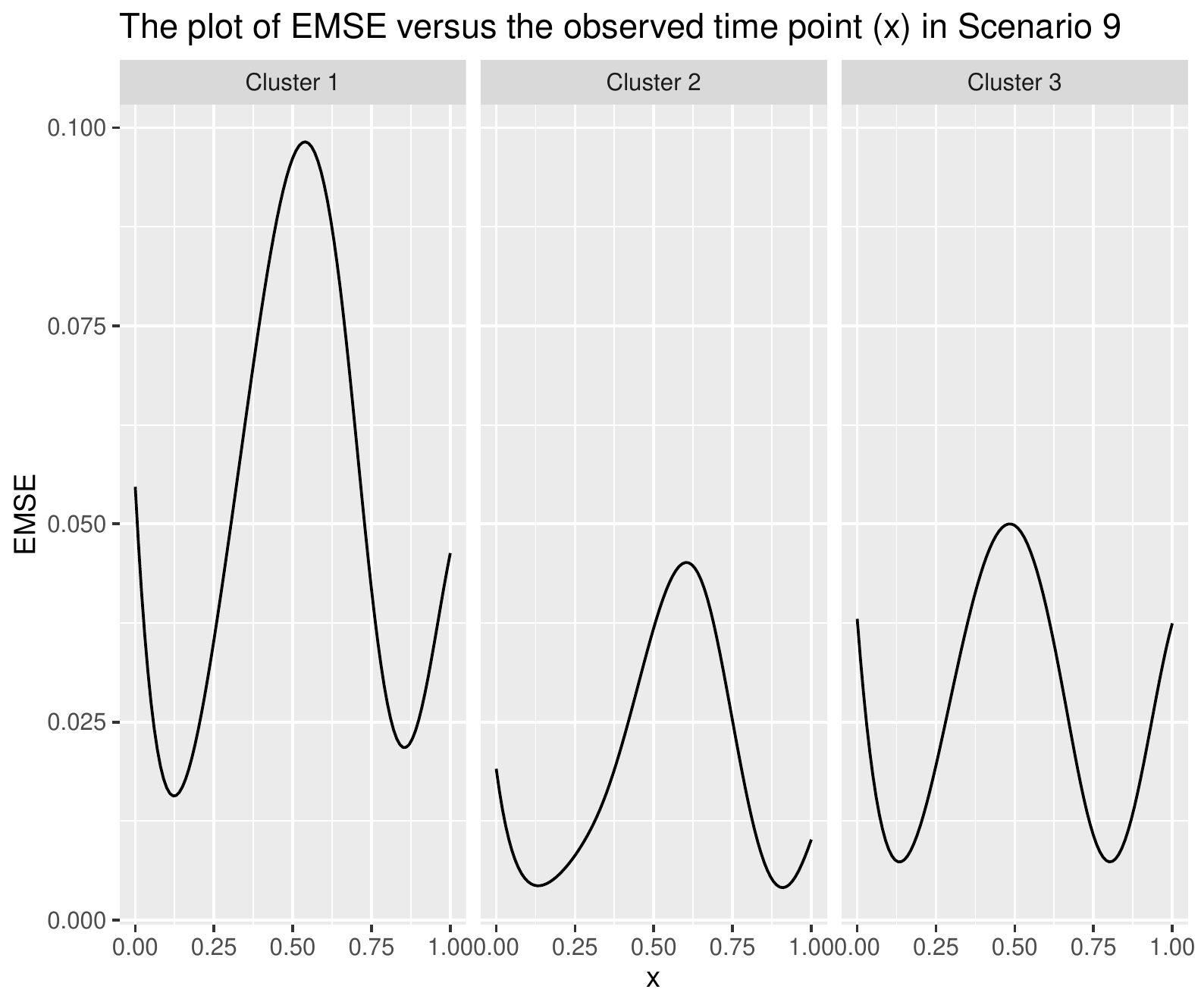}
\end{subfigure}%
\begin{subfigure}{.5\textwidth}
  \centering
  \includegraphics[height = 4.5cm, width = 5.5cm]{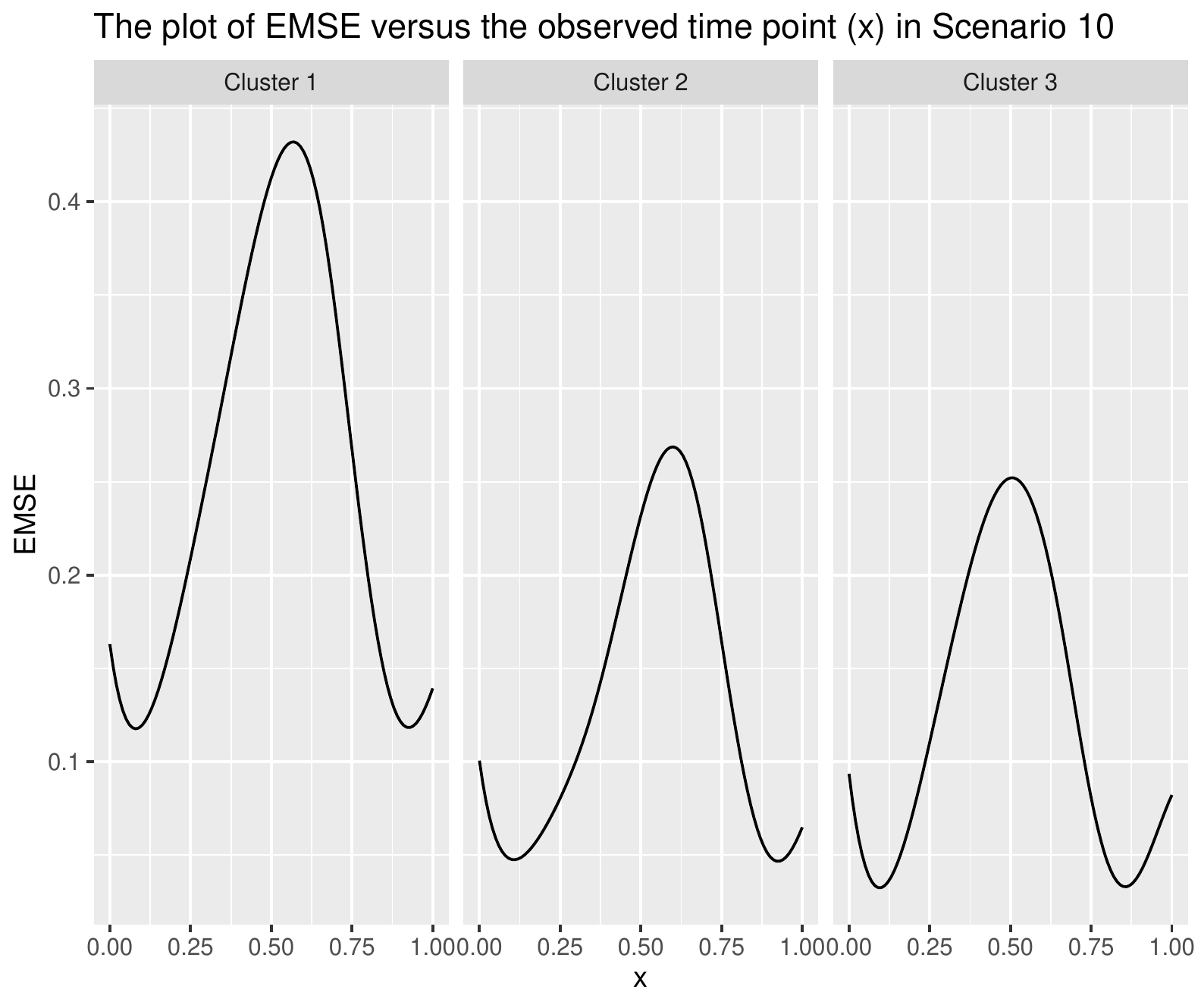}
\end{subfigure} 
\caption{EMSE versus the observed evaluation point for each cluster in Scenarios 7, 8, 9 and 10.}
\label{EMSE_others_ext}
\end{figure}

\begin{figure}[!ht]
\centering
  \includegraphics[height = 8cm, width = 10cm]{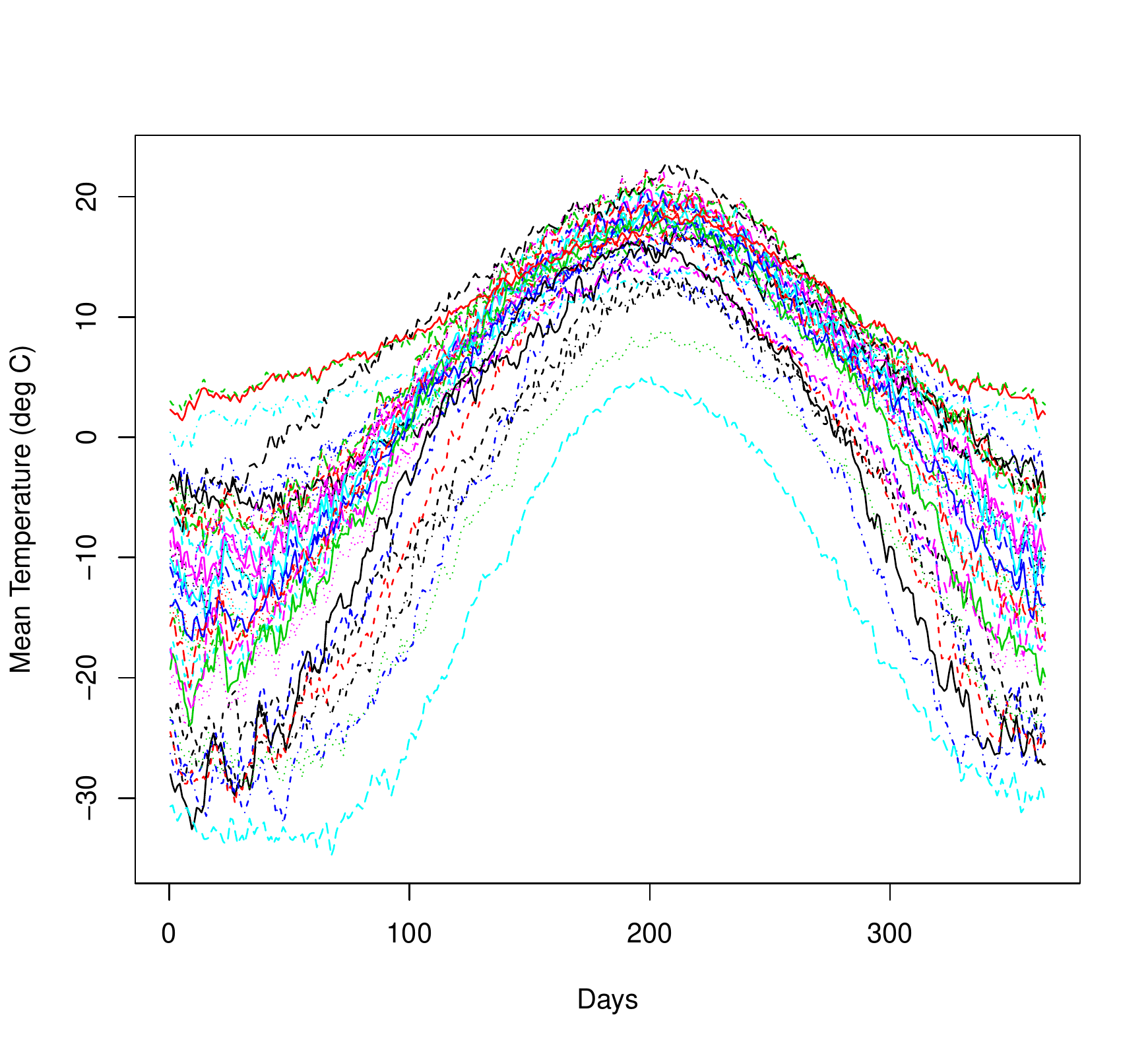}
\caption{Raw curves of the Canadian weather data. Different curves have different colors.}
\label{Canadian_raw}
\end{figure}



\end{appendices}

\newpage
\,
\newpage
\bibliography{sn-bibliography}



\end{document}